\documentclass[a4paper,fleqn,usenatbib]{mnras}

\usepackage[T1]{fontenc}
\usepackage{ae,aecompl}

\usepackage{float}
\usepackage{graphicx}	
\usepackage{graphicx}
\usepackage{txfonts}
\usepackage{pdflscape}
\usepackage{natbib}
\usepackage{threeparttable, tablefootnote}
\usepackage{aas_macros}

\def\equationautorefname~#1\null{equation~(#1)\null}

\title[Kinematics of the SMC]{Kinematics of stellar substructures in the Small Magellanic Cloud}

\author[El Youssoufi et al.]{Dalal El Youssoufi,$^{1,2}$
Maria-Rosa L. Cioni,$^{1}$ Nikolay Kacharov,$^{1}$ 
Cameron P. M. Bell,$^{1}$
\newauthor Gal Matjevi\'{c},$^{1}$
Kenji Bekki,$^{3}$
Richard de Grijs,$^{4,5}$
Valentin D. Ivanov,$^{6}$
Jacco Th. van Loon$^{7}$
\\
$^{1}$Leibniz-Institut f\"ur Astrophysik Potsdam, An der Sternwarte 16, D-14482 Potsdam, Germany\\
$^{2}$Institut f\"{u}r Physik und Astronomie, Universit\"{a}t Potsdam, Haus 28, Karl-Liebknecht-Str. 24/25, D-14476 Golm (Potsdam), Germany\\
$^{3}$ICRAR, M468, The University of Western Australia, 35 Stirling highway, Crawley, WA 6009, Australia\\
$^{4}$School of Mathematical and Physical Sciences, Macquarie University, Balaclava Road, Sydney, NSW 2109, Australia\\
$^{5}$Research Centre for Astronomy, Astrophysics and Astrophotonics, Macquarie University, Balaclava Road, Sydney, NSW 2109, Australia\\
$^{6}$European Southern Observatory, Karl-Schwarzschild-Str. 2, D-85748 Garching bei M\"{u}nchen, Germany\\
$^{7}$Lennard-Jones Laboratories, Keele University, ST5 5BG, UK\\
}

\date{Accepted 2023 April 27. Received 2023 April 24; in original form 2022 November 17}

\pubyear{2022}

\begin{document}
\label{firstpage}
\pagerange{\pageref{firstpage}--\pageref{lastpage}}
\maketitle

\begin{abstract}
We present a kinematic analysis of the Small Magellanic Cloud  using 3700 spectra extracted from the European Southern Observatory archive. We used data from Gaia and near-infrared photometry to select stellar populations and discard Galactic foreground stars. The sample includes main-sequence, red giant branch and red clump stars, observed with Fibre Large Array Multi Wavelength Spectrograph. The spectra have a resolving power $\lambda/\Delta\lambda$ from 6500 to 38\,000. We derive radial velocities by employing a full spectrum ﬁtting method using a penalised pixel ﬁtting routine. We obtain a mean radial velocity for the galaxy of 159$\pm$2 km s$^{-1}$, with a velocity dispersion of 33$\pm$2 km s$^{-1}$. Our velocities agree with literature estimates for similar (young or old) stellar populations. The radial velocity of stars in the Wing and bar-like structures differ as a consequence of the dynamical interaction with the Large Magellanic Cloud. The higher radial velocity of young main-sequence stars in the bar compared to that of supergiants can be attributed to star formation around 40 Myr ago from gas already inﬂuenced by tidal stripping. Similarly, young main-sequence stars in the northern part of the bar, resulting from a prominent star forming episode 25 Myr ago, have a higher radial velocity than stars in the southern part. Radial velocity diﬀerences between the northern and southern bar overdensities are also traced by giant stars. They are corroborated by studies of the  cold gas and proper motion indicating stretching/tidal stripping of the galaxy.
\end{abstract}



\section{Introduction}\label{section1}

The Small Magellanic Cloud (SMC) is a dwarf irregular galaxy, located at $\sim$60~kpc \citep{deGrijs2015} and characterised by a gas rich and low metallicity environment. Together with the Large Magellanic Cloud (LMC), it represents the nearest interacting pair of dwarf galaxies to the Milky Way (MW). The pair has been heavily influenced by dynamical interactions of tidal and/or ram pressure nature and has likely experienced a collision $\sim$200 Myr ago leading to the formation of the Magellanic Bridge  \citep[e.g.,][]{Yoshizawa2003,Diaz2012,Kallivayalil2013,Zivick2019,Schmidt2020} and 
$\sim$2~Gyr ago leading to the formation of the Magellanic Stream \citep[e.g.,][]{Mathewson1974,Nidever2008,D'Onghia2016}. 

The structure of the SMC reflects its complex dynamical H\,{\sc i} story where different stellar populations show different features along the line of sight. The galaxy is characterized by an elongated bar-like (bar hereafter) structure along the North East--South West (NE--SW) axis \citep[e.g.,][]{DeVaucouleurs1972,Subramanian2012,Scowcroft2016,Jacyszyn-Dobrzeniecka2016,Ripepi2017} and an eastern Wing towards the Magellanic Bridge \citep{Shapley1940}. Young stellar populations follow the irregular and asymmetric distribution typical of the H\,{\sc i} gas \citep[e.g.,][]{Stanimirovic2004,DiTeodoro2019} while older stellar populations depict an elliptical/spheroidal distribution \citep[e.g.,][]{Cioni2000a,Zaritsky2000,Rubele2015,ElYoussoufi2019}.
 
 \cite{ElYoussoufi2019,ElYoussoufi2021} used photometric data from the Visible and Infrared Survey Telescope for Astronomy (VISTA; \citealp{Sutherland2015}) survey of the Magellanic Clouds system (VMC; \citealp{Cioni2011}) and the VISTA Hemisphere Survey (VHS; \citealp{McMahon2013}) to obtain a comprehensive morphological view of the Magellanic Clouds and their periphery by tracing stellar populations of different median ages.
\cite{ElYoussoufi2019} provided the highest spatial resolution maps of the main bodies of the Magellanic Clouds in the near-infrared (NIR) to date.  They found that young main sequence stars in the SMC delineate a broken bar while red clump (RC) as well as red giant branch (RGB) stars display signatures of elongations towards the Magellanic Bridge, also detected by \cite{Belokurov2017} and \cite{Muraveva2018}, due to a LMC--SMC interaction $\sim$200 Myr ago. Irregular central features, showing that the inner SMC has been affected by tidal disruption, were also evidenced from the distribution of intermediate-age/old stars (1--7 Gyr old); they include two main overdensities along the NE--SW axis. 

 \cite{Pieres2017} identified an overdensity (SMCNOD) eight degrees from the centre of the SMC, mainly comprised of intermediate-age (a few Gyr old) stars, which were probably removed from the SMC disc by tidal stripping.   \cite{ElYoussoufi2021} discovered another substructure  in the vicinity of SMCNOD, Northern Substructure 2, lying at a similar distance as the SMC and located seven degrees NE of the centre; the substructure is probably associated with the ellipsoidal shape of the galaxy and is also comprised of intermediate-age stars. The shell substructure, located at about two degrees from the centre of the SMC in the NE direction is mainly composed of young ($\sim$ 150~Myr old) stars suggesting it formed during a recent star formation event as no evidence was found of a tidal origin \citep{MartinezDelgado2019}. 
 
 The outskirts of the SMC appear highly disturbed with tails connected to the Old Bridge, a tail of stars offset from the gaseous Bridge \citep{Belokurov2017,Mackey2018,ElYoussoufi2021}. 
 The SMC is also characterised by a line-of-sight depth that can reach $\sim$30~kpc depending on the type of stellar tracer \citep[e.g.,][]{Subramanian2012,Jacyszyn-Dobrzeniecka2016,Jacyszyn-Dobrzeniecka2017,Ripepi2017,Muraveva2018}. Furthermore, a foreground population manifesting itself as a bi-modality in RC distances has been measured $\sim$11 kpc in front of the SMC's main body \citep[e.g.,][]{Nidever2010,Subramanian2017,Omkumar2020,ElYoussoufi2021}. This is likely a perturbed population formed at smaller SMC radii and driven outwards \citep{Cullinane2023}.
 The complex nature of the structure of the SMC is also emphasized by filaments, arcs and shells in the distribution of H\,{\sc i} gas \citep[e.g.,][]{Stanimirovic2004,DiTeodoro2019}. \cite{Pingel2022} provided
the most sensitive and detailed view of  H\,{\sc i} emission associated
with the SMC, showing details on scales from 10 pc to 5 kpc.

Spectroscopic observations have made significant contributions to the measurement of stellar motions leading to a greater understanding of galaxy kinematics, formation and evolution. Our knowledge of radial velocities (RVs) of SMC stars has been limited to modest stellar samples from a few  types of tracers such as luminous supergiants \citep{Ardeberg1977,Ardeberg1979}, Cepheids \citep{Mathewson1988}, carbon stars \citep{Hardy1989,Hatzidimitriou1997,Kunkel2000}, RC stars \citep{Hatzidimitriou1993} and RGB stars \citep{Suntzeff1986}. More recently, the kinematics of RC and RGB stars in the periphery of the galaxy have been mapped by the Magellanic Edges Survey \citep{Cullinane2020}.

\cite{Harris2006} provided the first large spectroscopic survey of RGB stars in the SMC and derived a velocity gradient of 8.3 km\,s$^{-1}$\,deg$^{-1}$. They also concluded that the SMC is supported by its velocity dispersion rather than by rotation. \cite{Evans2008} undertook a large spectroscopic survey of massive stars in the SMC. They obtained a local velocity dispersion similar to that of RGB stars ($\sigma$ = 30~km\,s$^{-1}$), however unlike \cite{Harris2006}, they found evidence of rotation with a mean velocity gradient of 26.3 $\pm$ 1.6  km\,s$^{-1}$\,deg$^{-1}$ at PA = 126.2 $\pm$ 3.9 \,deg. \cite{Dobbie2014} used a large sample of RGB stars to study the kinematics of the SMC, and detected signatures of rotation with an observed rotation curve between 20--40 km\,s$^{-1}$ at similar position angles (120--130 deg) to \cite{Evans2008}. Additionally, \cite{Dobbie2014} found evidence of tidal stripping in the outer SMC. \cite{Deleo2020} used a sample of red giants and found significant tidal disruption in the inner 2~kpc of the SMC using a combination of proper motion (PM) and RV measurements, but with no obvious signature of rotation. \cite{Stanimirovic2004} obtained one of the first rotation measurements in H\,{\sc i} using RVs with a maximum velocity of 50 km s$^{-1}$, a mean PA of 40 deg and a velocity dispersion of about 22 km s$^{-1}$. These results were confirmed by \cite{DiTeodoro2019} using spatially  higher resolution observations of H\,{\sc i} in the SMC. Differences in velocity dispersion of stars from different stellar populations result also from chemodynamical simulations of the SMC, LMC and the Galaxy (e.g., \citealp{Bekki2009}). In a first instance, the apparent gradient in RV is of the same order of magnitude as the velocity dispersion and the rotation amplitude which means that bulk motion and rotation cannot be inferred independently from RV data alone.

Until recently, advances made by spectroscopic studies were mostly constrained to one dimensional RVs, highlighting the importance of PMs for a comprehensive understanding of the three-dimensional (3D) kinematics and dynamics of the galaxies. The \textit{Hipparcos} satellite provided one of the first PM measurements in the SMC \citep{Kroupa1997}, the sample however was  limited to eleven massive stars and the uncertainties were of the order of $\sim$65\%. Observations with the \textit{Hubble Space Telescope} (HST) by \cite{Kallivayalil2006} measured the PM of the SMC to an accuracy of 15\%  allowing the determination of the PM of the centre of mass of the galaxy and showcasing that the Magellanic Clouds are most likely on their first infall towards the MW or are on a long eccentric orbit. Subsequently,  \cite{Kallivayalil2013} reduced the PM uncertainties to 1--2\% by increasing the time baseline with additional HST observations, but no conclusion could be reached about the rotation of the galaxy perhaps due to the small number of targeted fields. No evident rotation in the plane of the sky was also found by \cite{VanderMarel2016} using PMs for eight stars from the Tycho--Gaia Astrometric Solution (TGAS) Catalog, which contains PMs only for stars in common between \textit{Gaia} Data Release 1 (DR1;  \citealp{Brown2016}) and the {\it Hipparcos} Tycho--2 Catalog. 

\cite{Oey2018} used \textit{Gaia} DR2 PMs in combination with RVs from the  Runaways and Isolated O-Type Star Spectroscopic Survey of the SMC (RIOTS4; \citealp{Lamb2016}) and concluded that the bar and Wing of the SMC are kinematically distinct features, with the Wing having a consistent transverse motion along the Bridge towards the LMC. Results by \cite{Murray2019} showed that the 3D kinematics of a sample of young massive stars (their RV follows that of the H\,{\sc i} gas) is inconsistent with disc rotation in the SMC. \cite{Niederhofer2018} used ground-based VMC observations to present the first stellar resolved PM map of the SMC. This map revealed a non-uniform velocity pattern indicative of a tidal feature behind the main body of the galaxy and a flow of stars along the line-of-sight. \cite{Zivick2019} and \cite{Schmidt2020} studied the kinematics of the Magellanic Bridge and confirmed that old and young stellar populations are moving away from the SMC towards the LMC. Additionally, \cite{Niederhofer2021} suggested a dynamical stretching of the galaxy with ordered motion of intermediate-age/old stars from the SMC towards the Old Bridge, as well as a stellar motion in the North which might be related to the Counter Bridge \citep{Diaz2012}. \cite{Grady2021} found evidence of tidal stripping in the SMC both in PM and metallicity space. The density and velocity flow of stars from the SMC to the LMC along the Bridge is traced by \textit{Gaia} Early Data Release 3 (EDR3) data globally and separately for young and evolved stellar populations \citep{Luri2021}. A best-fit model of the SMC kinematics, based on Gaia DR2, highlights the need to distinguish among the different stellar populations once sufficiently populated and widely distributed RV samples become available \citep{Zivick2021}.

The aim of this paper is to use available spectra from the European Southern Observatory (ESO) Science Archive Facility (SAF) to increase the sample of RVs and characterise the kinematics of the morphological features belonging to the SMC and its vicinity. The paper is organised as follows: in Section \ref{section2} we describe the used data sets and sample selection; we outline the method and steps involved in obtaining RVs in Section \ref{section3} and we describe our results in Section \ref{section4}, while in Section \ref{section6} we summarise our main conclusions and future prospects.


\section{Sample Selection}\label{section2}

 To select a sample of stars for which to search the ESO SAF for available spectra, we used NIR photometry from the VMC survey and VHS. We distinguished between stars belonging to different stellar populations as in \cite{ElYoussoufi2019}, using regions in the colour-magnitude diagram (CMD) with different median ages and evolutionary stages. Additionally, we used Two Micron All Sky Survey (2MASS; \citealp{Skrutskie2006}) photometry for objects that are too bright for the VISTA surveys.
 The ESO SAF provides access to data acquired with ESO telescopes. Raw, processed and catalogued data can be queried  via different web query interfaces. In our work, we used the spectral query form\footnote{\url{https://Archive.eso.org/wdb/wdb/adp/phase3_spectral/form}} that gives access to reduced spectral ESO phase 3 data products. In the following subsections, we outline our selection criteria for both the photometric and spectroscopic data sets as well as describe the parameters encompassed by the spectroscopic sample.

\subsection{Photometric data}\label{section2.1}

The VMC survey and VHS are multi-band photometric surveys carried out using the VISTA infrared camera (VIRCAM; \citealp{Dalton2006, Emerson2006}). We used observations in the $J$ and $K_\mathrm{s}$ bands within a 10 deg radius from the SMC's optical centre, Right Ascension (RA)=$13.05$ deg and Declination (Dec)=$-72.83$ deg at the epoch J2000 \citep{DeVaucouleurs1972}. The VMC data were acquired between February 2010 and November 2016 and were released as part of Data Release 5 (DR5) of the VMC survey\footnote{\url{https://www.eso.org/sci/publications/announcements/sciann17232.html}}. VMC observations reach 5$\sigma$ point source limits of $J$ = 22~mag and $K_\mathrm{s}$ = 21.5~mag (in the Vega system) across an area of $\sim 43$ deg$^2$ on the SMC component of the survey. The median full width at half maximum (FWHM) of point-like sources in the VMC images corresponds to 1 arcsec in $J$ and 0.93 arcsec in $K_\mathrm{s}$. Beyond this area, VHS observations reach 5$\sigma$ limits of $J$=19.3 mag and $K_\mathrm{s}$=18.5~mag (also in the Vega system).
The median FWHM of point-like sources in the VHS images corresponds to 0.99 arcsec in $J$ and 0.91 arcsec in $K_\mathrm{s}$. VHS observations obtained until 30th March 2017 were released as part of DR5 of the VHS survey\footnote{\url{https://www.eso.org/sci/publications/announcements/sciann17290.html}}, whereas for VHS observations obtained until 30th September 2017 only raw images are publicly available. VMC and VHS images were initially processed by the Cambridge Astronomical Survey Unit (CASU) using the VISTA Data Flow System \citep{Emerson2006} and further processed by the Wide Field Astronomy Unit (WFAU) as well as stored and made available to the community at the VISTA Science Archive (VSA; \citealp{Cross2012}). We used the aperture corrected magnitudes calculated within a 2 arcsec diameter \texttt{(\texttt{jAperMag3} and \texttt{ksAperMag3})}, as these are the most reliable magnitudes for point sources, from the \textit{vmcsource} and \textit{vhssource} tables. These magnitudes were calibrated as outlined by \cite{Gonzalez-Fernandez2018}.
VMC and VHS observations overlap in the area of the Magellanic Bridge and around the SMC. We eliminated duplicate sources by keeping the VMC sources because VMC data are deeper than the VHS data. Our initial selection criteria refer to detections in both $J$ and $K_\mathrm{s}$ bands for objects that have at least a 70\% probability of being stars \texttt{(flag \texttt{mergedClass} = $-1$ or $-2$)}, as well as with  photometric uncertainties better than $0.1$~mag in both bands. Furthermore, we only selected sources with minor quality issues \texttt{(flag \texttt{ksppErrBits} $\leq$ $256$ and \texttt{jppErrBits} $\leq$ $256$)}.
Applying these criteria, we obtained 3\,407\,966 sources.

2MASS data across the SMC were obtained from a dedicated 1.3 metre telescope at the Cerro Tololo Inter-American Observatory. They correspond to a sensitivity of $J$=15.8 mag and $K_\mathrm{s}$=14.3 mag (in the Vega system) at a signal-to-noise ratio (SNR) of 10. The median FWHM of the 2MASS images corresponds to  2.9 arcsec in both $J$ and $K_\mathrm{s}$. However, multiple dithered observations improved the spatial resolution of the images. 2MASS observations were acquired between June 1997 and February 2001. The data were downloaded from the Gaia@AIP portal\footnote{\url{https://gaia.aip.de/}}. We required that the objects were detected in both bands, with photometric uncertainties smaller than 0.05~mag, and quality flags applied to the $J$, $H$ and $K_\mathrm{s}$ bands (we included the $H$ band to obtain a more reliable sample), as follows: photometric quality flag \texttt{(flag ph\_qual) = AAA}, reflecting detections where valid measurements were made with a SNR>10 and photometric errors below 0.1 mag, Read flag \texttt{(flag rd\_flg) = 222} indicating that the  default magnitude is derived from a profile-fitting measurement made on six individual 1.3 s exposures covering the sources, blend flag \texttt{(flag rd\_flg) = 111}, indicating that one component was fit when estimating the brightness of the source, contamination and confusion flag \texttt{(flag cc\_flg) = 000}, where only sources unaffected by known artifacts are kept.  In total, 27\,716 sources were selected, excluding sources in common with the VISTA sample. We note that the difference between the filter systems (e.g., \citealp{Cioni2011,Gonzalez-Fernandez2018}) does not influence our study.

We combined data from \textit{Gaia} EDR3 \citep{Brown2021} with the NIR samples 
because the astrometric solution (RA and Dec -- position on the sky, $\omega$ -- parallax, $\mu_\mathrm{RA}$ and $\mu_\mathrm{Dec}$ -- proper motion) allowed us to reduce the number of MW foreground stars. \textit{Gaia} EDR3 data are based on observations collected during the first 34 months of the mission which started in December 2013. They include objects with a limiting magnitude of $G$= 21~mag (in the Vega system).
We performed the cross-correlation between \textit{Gaia} EDR3, VMC and VHS data
to retain sources within a distance of 1 arcsec. The cross-match of \textit{Gaia} EDR3 and 2MASS is provided by the \textit{Gaia} Data Processing and Analysis Consortium. The selection criteria we applied to maximise the number of SMC stars are the following: $\omega\geq0.2$ mas, $-4\leq \mu_\mathrm{Dec} \leq 2$~mas~yr$^{-1}$ and $-2.5\leq \mu_\mathrm{RA} \leq 4.5$ mas~yr$^{-1}$. The parallax criterion is the same as in \cite{ElYoussoufi2021} whereas the proper motion criteria encompass also the diffuse component around the galaxy. These criteria remove bright main-sequence and giant stars of the MW (e.g., \citealp{Vasiliev2018}), but due to the uncertainties of parallax and proper motion at the distance of the SMC do not allow us to distinguish MW stars among the faint sources. In total, 1\,931\,462 sources were removed from our sample based on these criteria.

Furthermore, we used the Optical Gravitational Lensing Experiment (OGLE) III \citep{Soszynski2011} and IV \citep{Soszynski2015,Soszynski2016} data to exclude Cepheids, RR Lyrae stars and Long Period Variables from our sample because their corresponding ESO SAF spectra will also include a variation due to the intrinsic stellar pulsation velocity which is difficult to disentangle with single-epoch spectra. 
The exclusion of variable stars was performed using the Tool for OPerations on Catalogues And Tables
(TOPCAT\footnote{http://www.star.bris.ac.uk/~mbt/topcat/}) and a maximum cross-matching distance of 1 arcsec. In total, 30\,668 variable sources were removed from our sample.

\begin{figure*}
	\centering
	\includegraphics[scale=0.163]{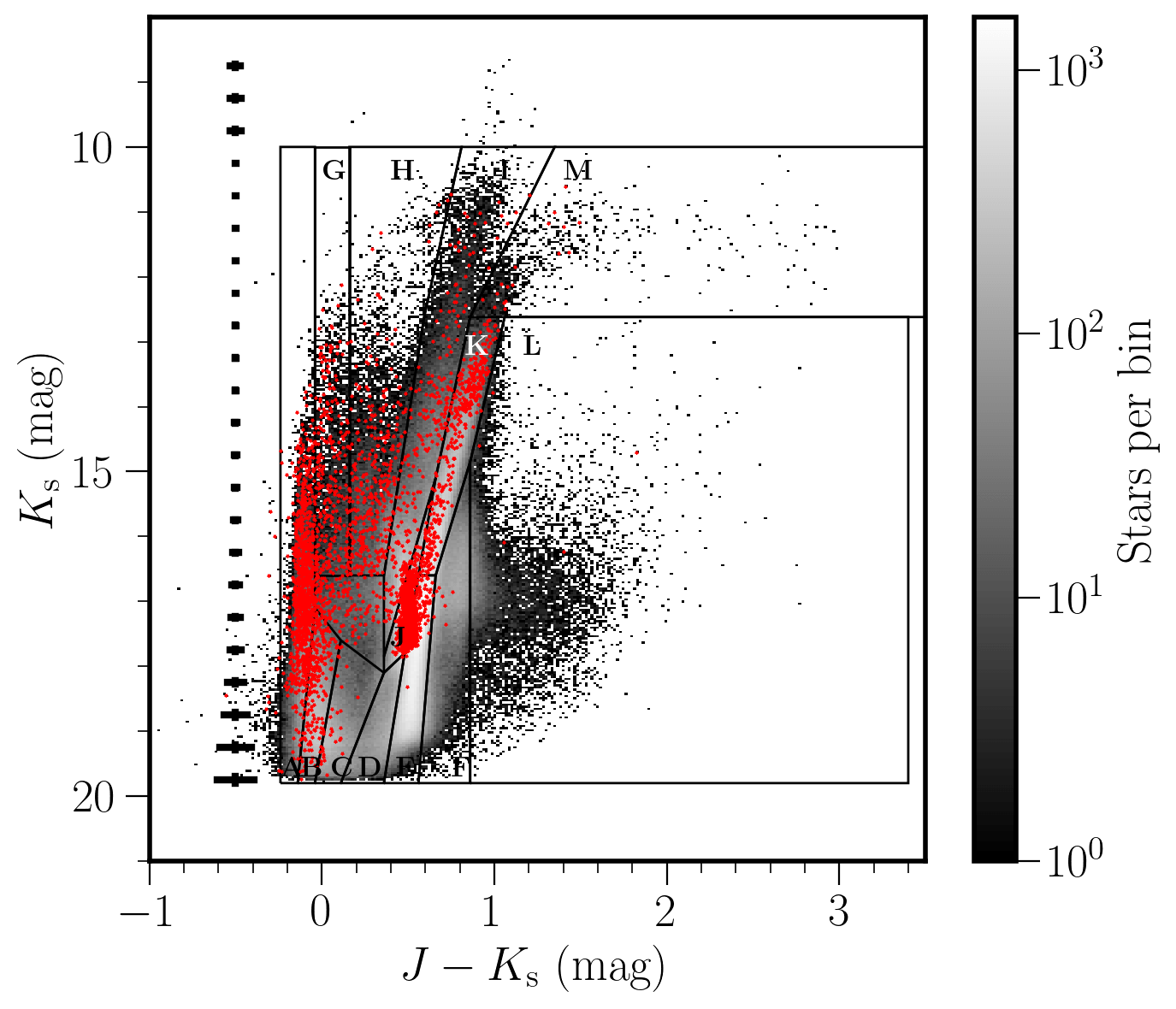}
	\includegraphics[scale=0.181]{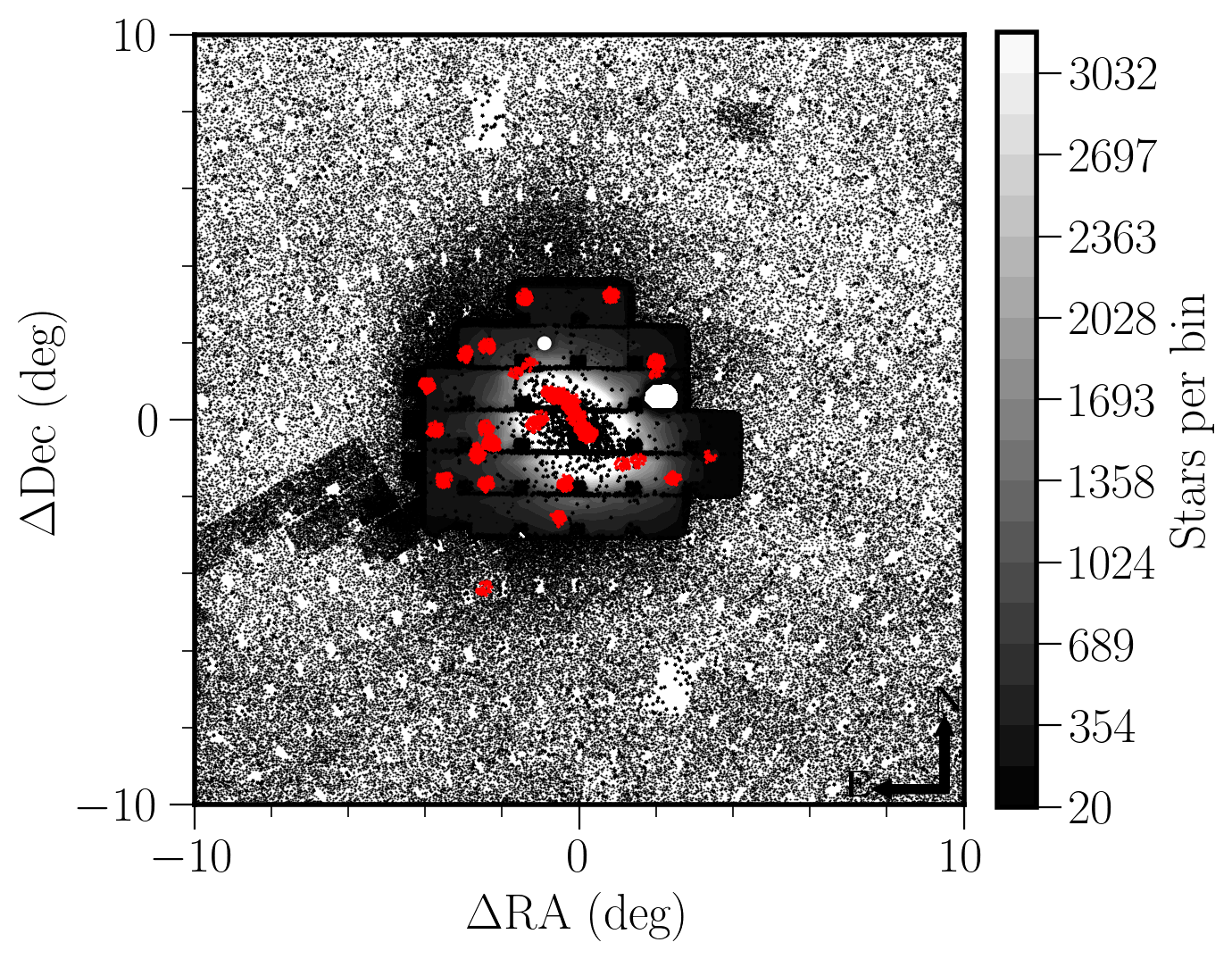}\\
	
	\includegraphics[scale=0.165]{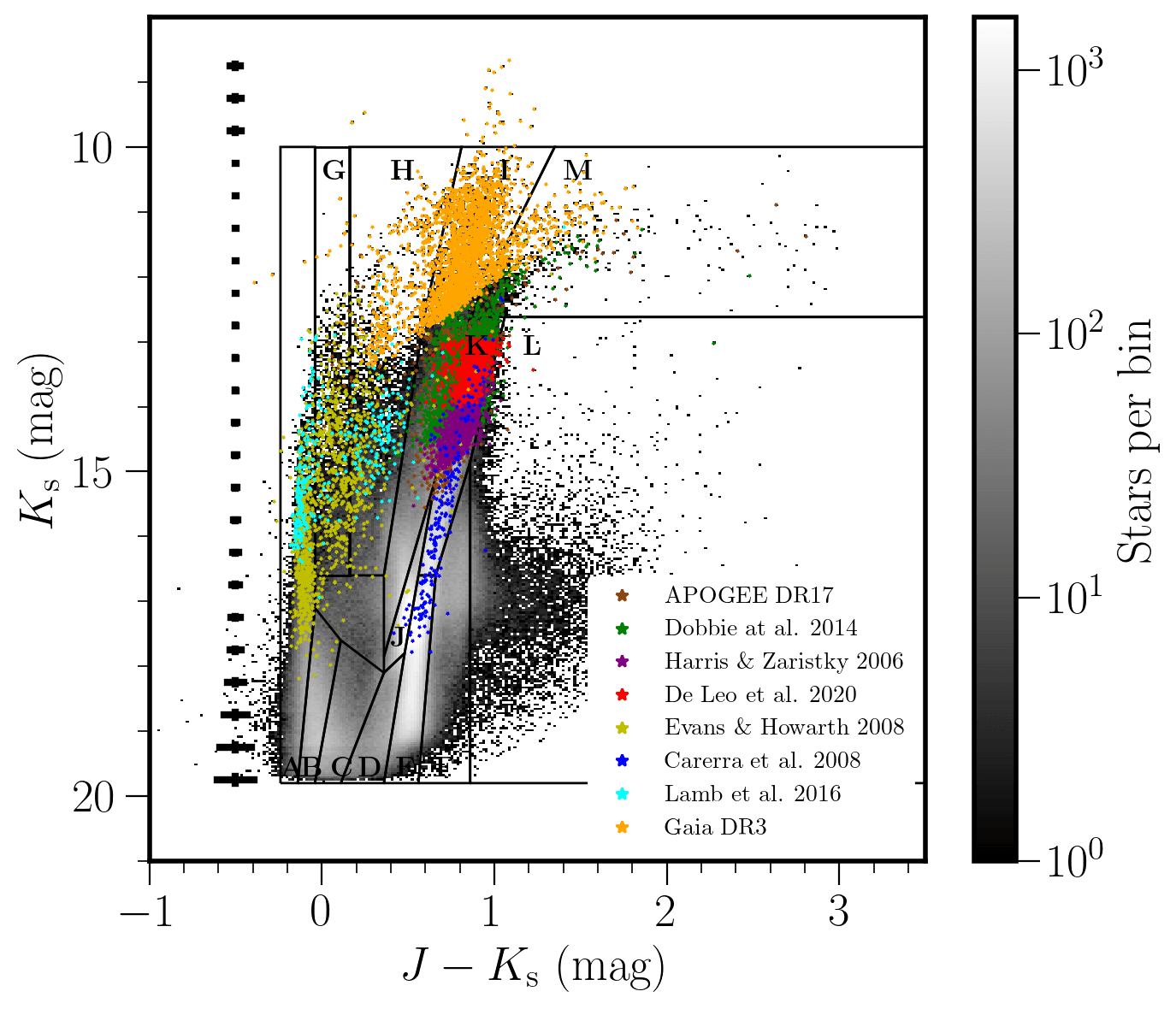}
	\includegraphics[scale=0.174]{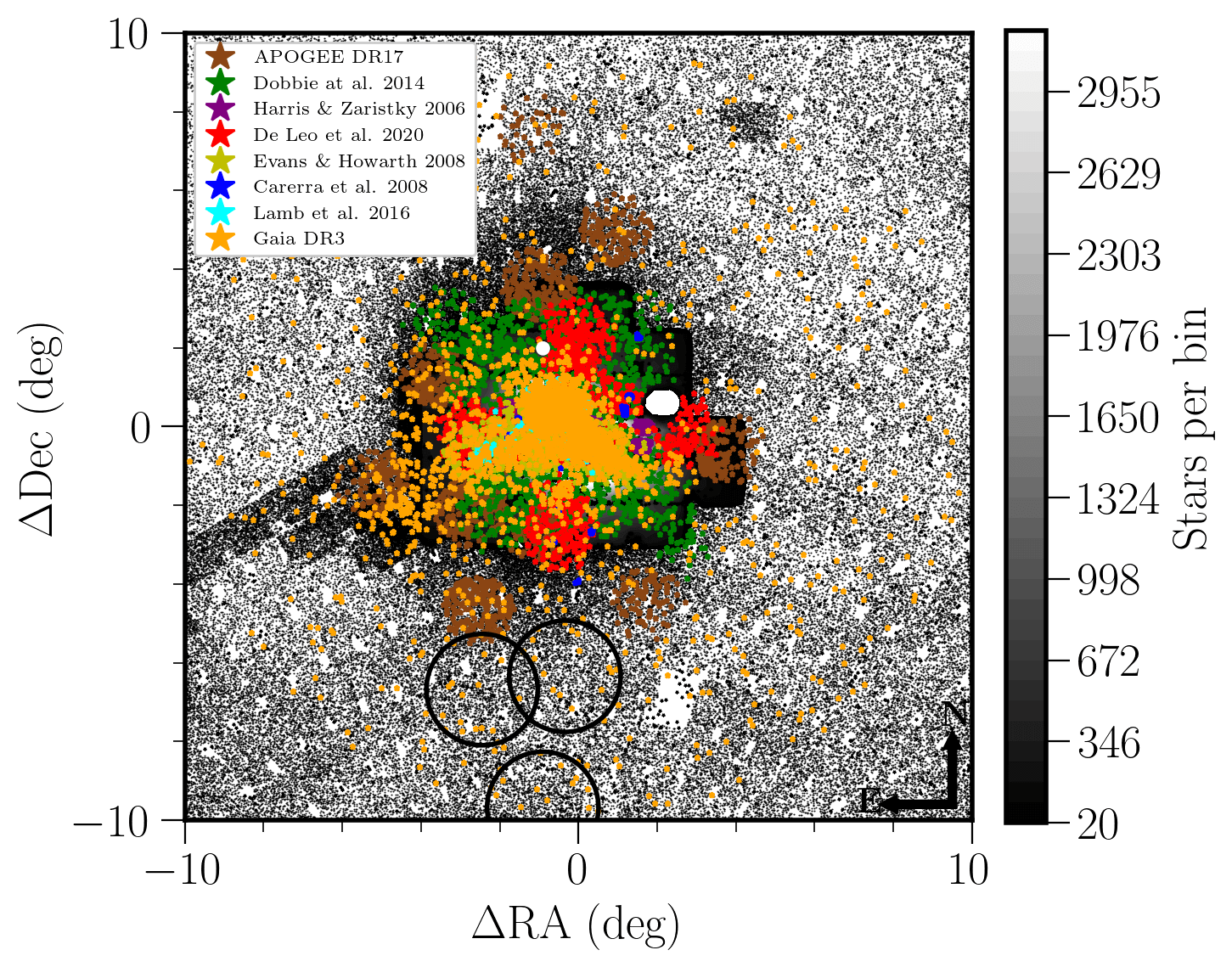}

 	\caption{(top left) NIR ($J-K_\mathrm{s}$, $K_\mathrm{s}$) Hess diagrams of the SMC stars after the cross-match with \textit{Gaia} EDR3 data. The grey scale indicates the stellar density on a logarithmic scale whereas horizontal bars show the photometric uncertainties as a function of magnitude. VMC, VHS and 2MASS photometry is used. Points in red refer to stars with available spectra in the ESO archive, after applying the selection criteria discussed in Sect.\ref{section2}. Boxes outline the stellar population regions as in \protect\cite{ElYoussoufi2019}. (top right) Spatial distribution of SMC stars. The stellar clusters 47 Tuc and NGC 362, at ($\Delta$RA, $\Delta$Dec) of (2.1 deg, 0.6 deg) and ($-$0.9 deg, 1.96 deg) respectively, have been excluded using circular masks. Rectangular regions of high stellar density represent the location of VMC tiles. The overdensity at ($\Delta$RA, $\Delta$Dec) of (5 deg, 8 deg) corresponds to a tile in the Magellanic Stream component of the VMC survey. The gaps at the bottom of each VISTA tile refer to the excluded detector \#16. The grey scale indicates the stellar density per deg$^2$, with (red) and without available spectra in the ESO archive. (bottom left and right) As figures in the upper panels but with points in colour referring to stars with available spectra in different literature studies. Circles to the extreme outskirts encompass the regions studied using data from the Magellanic Edges Survey \citep{Cullinane2023}.}
	\label{fig:CMD_SD_ESO_Archive}
	
	
\end{figure*}

The resulting photometric data set 
contained 1\,504\,220  sources. Their colour--magnitude diagram and spatial distribution are displayed in Figure \ref{fig:CMD_SD_ESO_Archive}.

\subsection{Spectroscopic data} \label{section2.2}

We initially selected from the ESO SAF all objects with 1D spectra within a 10 deg radius from the SMC optical centre. The query form offers access to various parameters and the most relevant to us are outlined in Table \ref{table:spectra_eso_Archive}.
The ARCFILE (column 1) represents the name under which the data product is stored in the ESO SAF where the timestamp (ADP.timestamp) specifies the time of the data archival. The coordinates RA and Dec (columns 2 and 3) are given in the FK5 J2000 coordinate system which are equivalent to equatorial coordinates and coordinates in the International Celestial Reference System (ICRS); the instrument and spectral grating used for the observations (columns 4 and 5) are also indicated. Column 6 represents the wavelength coverage $\lambda$, column 7 the resolving power {\it R} ($\lambda/\Delta\lambda$) and column 8 the  median SNR of the spectra. Columns 9 and 10 show the Programme ID associated with the spectra and the observing date on which the spectra were obtained, respectively. Column 11 contains the product version. For single observing blocks where more than just one exposure is available, the individual spectra are stacked and are referred to as product version 2. However this is only available for science spectra with identical instrument set ups belonging to the same observing block. Otherwise, single spectra are referred to as product version 1.

\begin{figure*}
	\centering
	\includegraphics[scale=0.07]{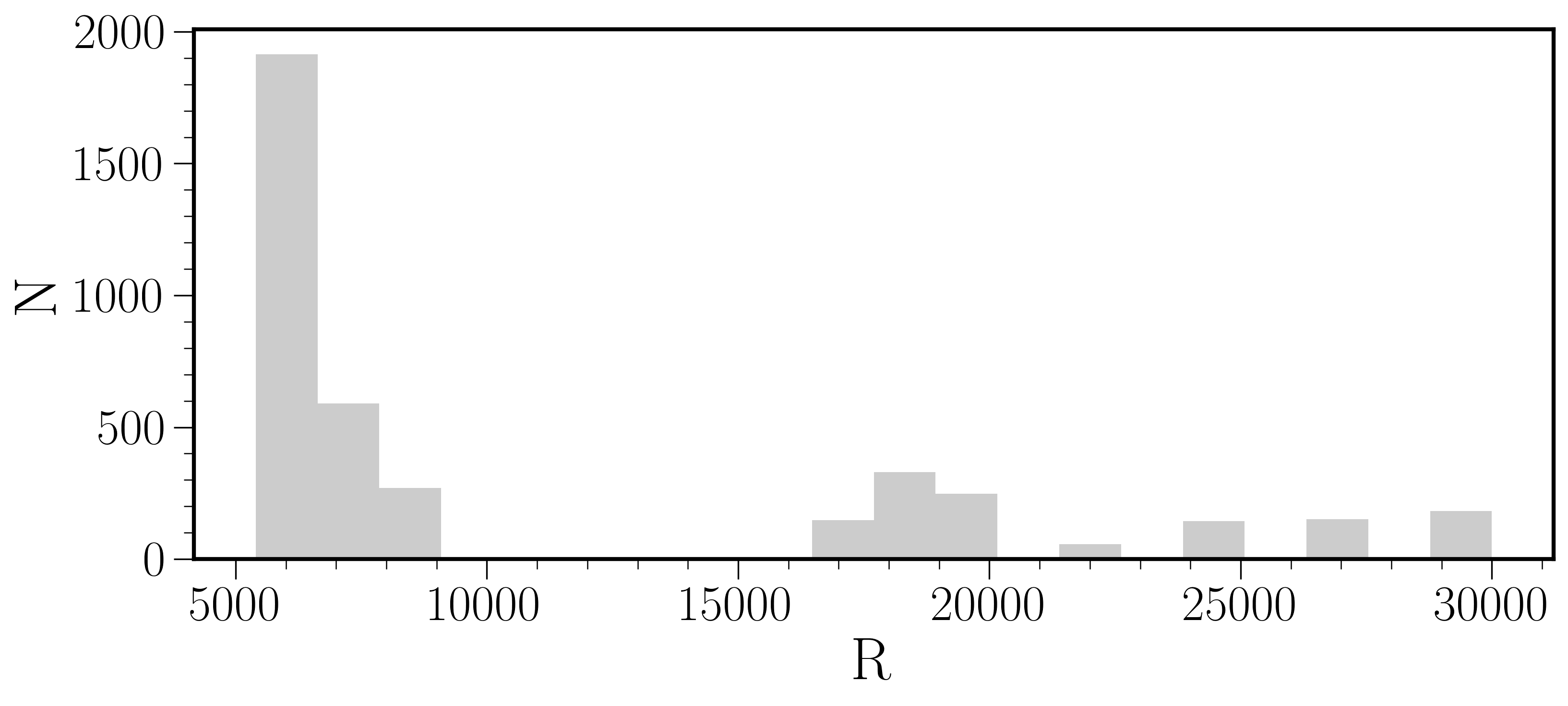}
	\includegraphics[scale=0.07]{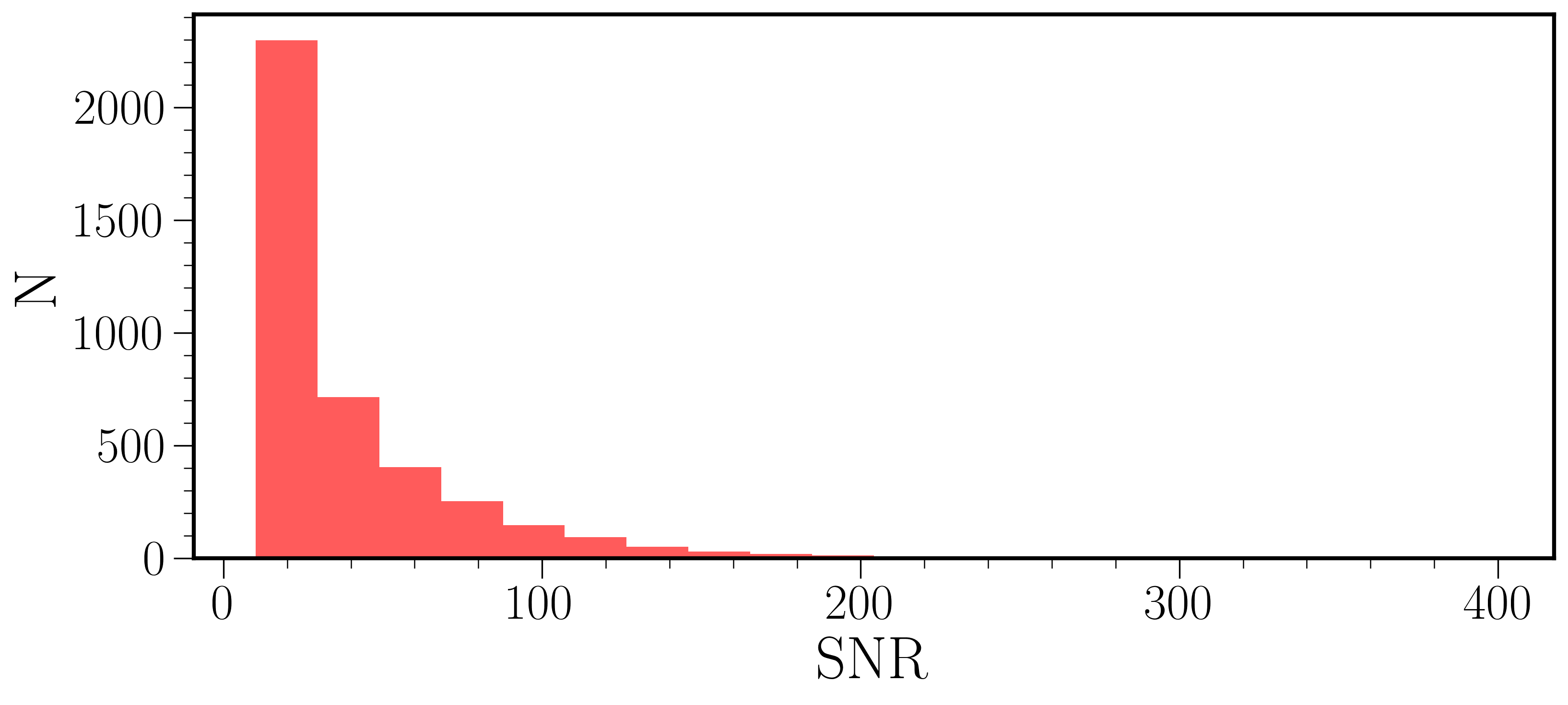}
	\includegraphics[scale=0.07]{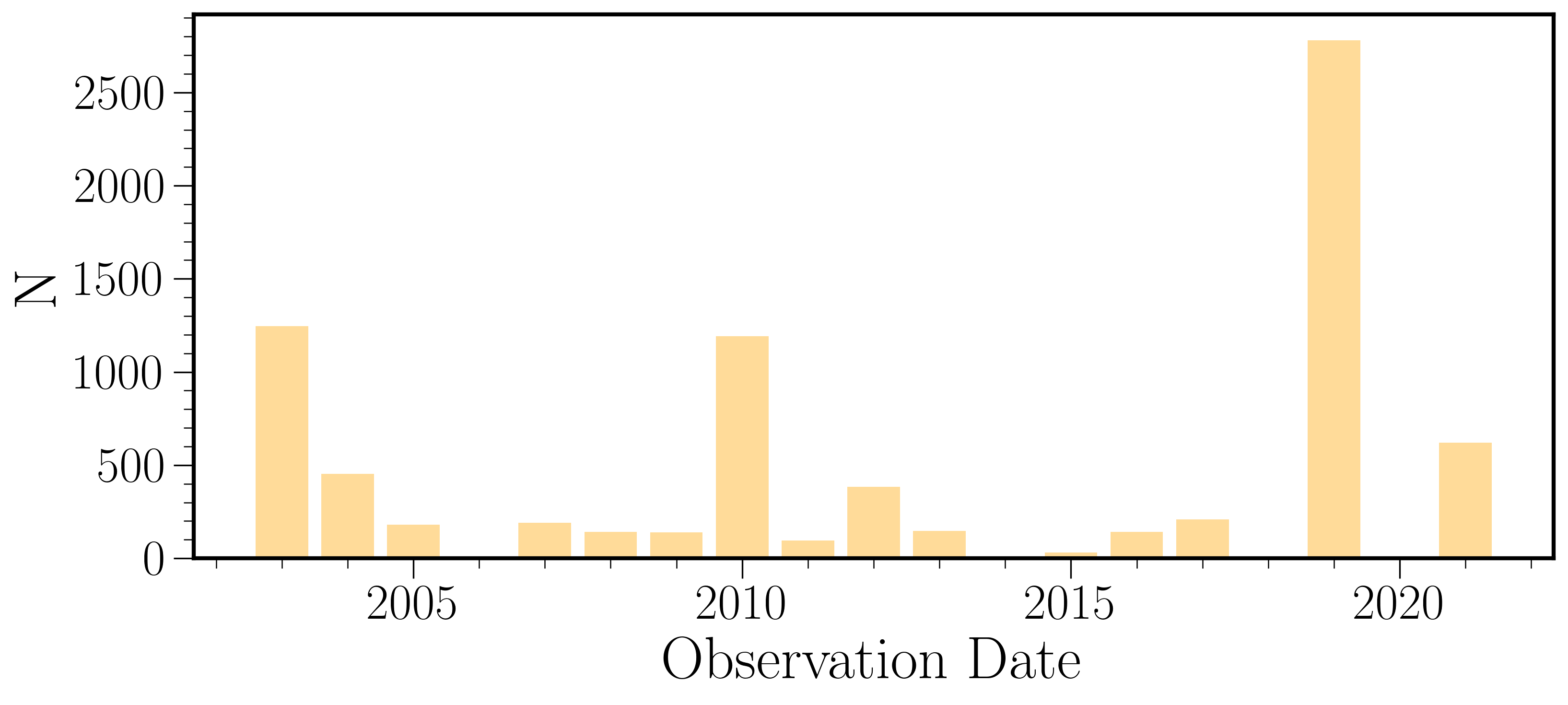}
	\includegraphics[scale=0.07]{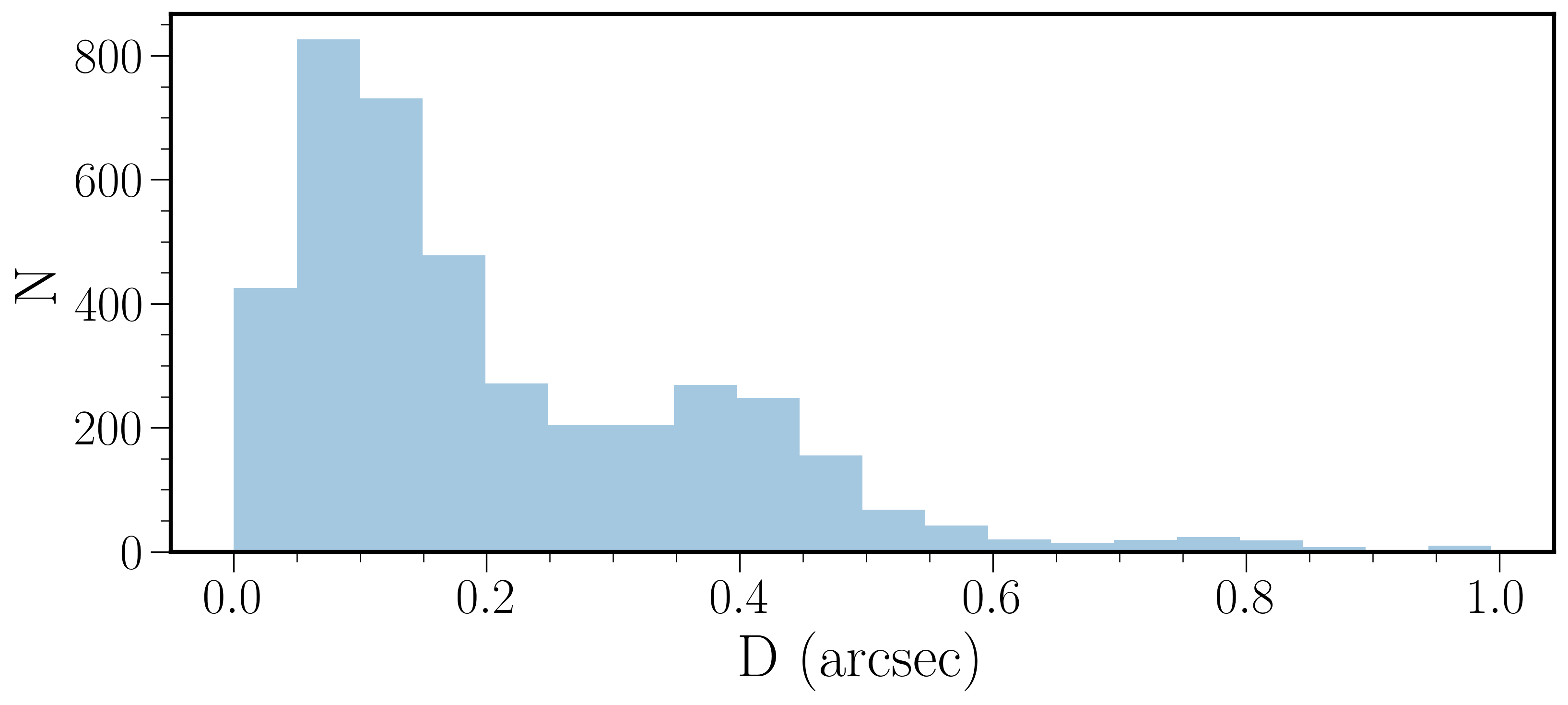}
	\includegraphics[scale=0.07]{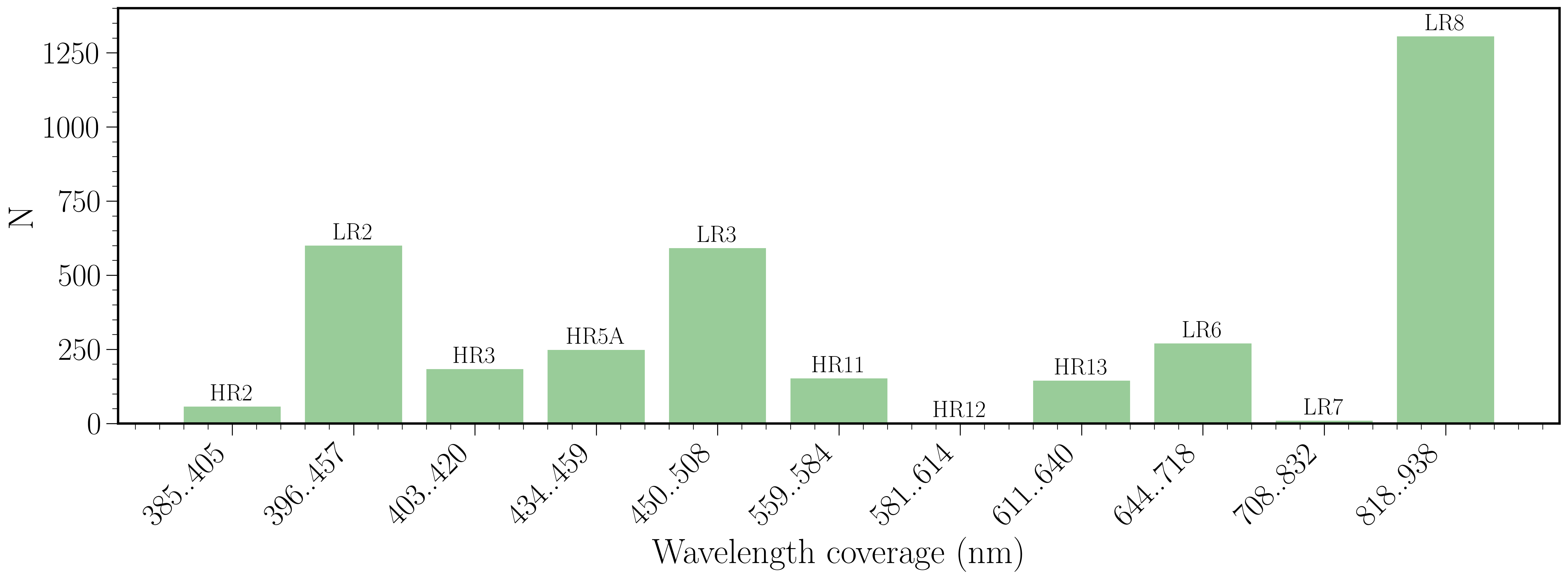}
	\caption{Histograms showing the parameter range of the spectra used in this study in terms of resolving power {\it R} (top-left), signal-to-noise ratio SNR (top-right), year of observation (middle-left), the separation between the coordinates of the photometric and spectroscopic data (middle-right), and the wavelength coverage (bottom).}
	\label{fig:AAA}
\end{figure*}

In order to find photometric counterparts for the selected spectra, we used TOPCAT and a maximum cross-matching distance of 1 arcsec (Fig.~\ref{fig:AAA}). The cross-match was done using ICRS coordinates at the J2000 epoch. For each row in the spectroscopic catalogue only the best (nearest) match from the photometric catalogue will appear as a result, but rows from the photometric catalogue may appear multiple times. This allowed us to group data by their source identifier (source ID) and identify objects with more than one spectrum. Not every object of the spectroscopic catalogue had a match in the photometric catalogue because MW stars and variable stars were excluded from our photometric catalogue using the selection criteria described in Section \ref{section2.1}. We found a photometric counterpart for 24\,609 spectra. 

As stars can have multiple observations, we decided to use only the best spectrum of each star in terms of {\it R} and SNR. The first step was to group data by source ID and identify sources with more than one spectrum. We then proceeded to select spectra from each data group with the highest value of {\it R}; at this stage there can still be multiple spectra as more than one spectrum can have the same {\it R}. 
From this list of high resolution spectra, we proceeded to select spectra with the highest value of SNR, which can also result in multiple spectra. We found that only 22 sources still have multiple spectra and we proceeded to pick a random spectrum from among those available. Furthermore, we applied one last positional cut to discard spectra belonging to the MW globular clusters 47 Tuc and NGC 362. We masked two areas centred at ($\Delta$RA= 2.1 deg, $\Delta$Dec= 0.6 deg) and ($\Delta$RA=$-$0.9 deg, $\Delta$Dec = 1.96 deg) with radii of 0.22 deg and 0.17 deg, respectively. Finally, we selected only spectra with SNR $\geq$ 10. After applying these selection criteria we obtain a sample comprised of 3814 spectra, of which about 97\% (3700) refer to observations with the GIRAFFE spectrograph of the Fibre Large Array Multi Element Spectrograph (FLAMES). We decided to focus only on this sample because the remaining 114 spectra were obtained with other instruments and are insufficient to quantify possible systematic uncertainties. The GIRAFFE spectra in our sample have not been associated with journal publications as is visible in the ESO SAF. However, during the revision of our work an analysis of 206 RGB stars members of the SMC appeared in \cite{Mucciarelli2023}. 
The distribution of the 3700 sources on the colour--magnitude diagram and across the SMC is shown in Fig.~\ref{fig:CMD_SD_ESO_Archive}.

 	\begin{table*}
	\setlength{\tabcolsep}{3pt}
		\caption{Parameters of the spectra used in this study. The table is published in its entirety as
        supporting material with the electronic version of the article.}                    
		\label{table:spectra_eso_Archive}      
		
		\begin{tabular}{cccccccrcc}
			\hline
			ARCFILE & RA & Dec & Instrument & Grating & $\lambda$ & {\it R} & SNR & Program ID& Product version \\
			&(deg)&(deg)&(nm)&&&&&&\\
			\hline
			ADP.2015-04-13T10:11:06.923 & 13.713166 & -72.637166 & GIRAFFE & LR3 & 450$-$508 & 7500 & 15.8 & 386.D-0541 & 1 \\
            ADP.2015-04-13T10:11:07.283 & 13.797333 & -72.285638 & GIRAFFE & LR3 & 450$-$508 & 7500 & 60.6 & 386.D-0541 & 1 \\
            ADP.2015-04-13T10:11:13.337 & 14.055874 & -72.652472 & GIRAFFE & LR3 & 450$-$508 & 7500 & 42.2 & 386.D-0541 & 1 \\
            ADP.2015-04-13T10:11:16.837 & 14.412333 & -72.511472 & GIRAFFE & LR3 & 450$-$508 & 7500 & 12.9 & 386.D-0541 & 1 \\
            ADP.2015-04-13T10:11:26.400 & 13.348958 & -72.481388 & GIRAFFE & LR3 & 450$-$508 & 7500 & 29.7 & 386.D-0541 & 1 \\
            ADP.2015-04-13T10:11:53.937 & 13.948374 & -72.658611 & GIRAFFE & LR3 & 450$-$508 & 7500 & 138.0 & 386.D-0541 & 1 \\
            ADP.2015-04-13T10:12:02.720 & 13.237999 & -72.484055 & GIRAFFE & LR3 & 450$-$508 & 7500 & 74.8 & 386.D-0541 & 1 \\
            ADP.2015-04-13T10:12:15.860 & 14.158791 & -72.609666 & GIRAFFE & LR3 & 450$-$508 & 7500 & 112.0 & 386.D-0541 & 1 \\
            ADP.2015-04-13T10:12:19.657 & 13.607041 & -72.393305 & GIRAFFE & LR3 & 450$-$508 & 7500 & 110.4 & 386.D-0541 & 1 \\
			\hline
		\end{tabular} 
	\end{table*}

FLAMES is a multi-object spectrograph mounted on the Unit Telescope 2 (UT2; Kueyen) of the Very Large Telescope (VLT) facility at the ESO Paranal Observatory. It gives access to targets across a field-of-view of 25 arcmin in diameter and feeds two spectrographs, GIRAFFE and UVES. GIRAFFE covers the visible range, 370--950 nm, with three types of feeding fibre systems: MEDUSA, IFU, ARGUS. We limited our study to the MEDUSA mode as we are interested in 1D spectra of individual stars. In this mode, GIRAFFE provides intermediate to high resolving power ({\it R} from $\sim$ 5000 to 38\,100) spectra and can observe up to 132 targets (including sky fibres) at a given time. Each fibre has an aperture of 1.2 arcsec on the sky.
The spectra obtained by GIRAFFE are cleaned from cosmic rays as a pre-processing step and consecutively have most of their instrument signature removed; they have been de-biased, flat fielded, extracted, and wavelength calibrated. Their wavelength scale has been corrected to the  heliocentric reference system using barycentric, heliocentric and geocentric corrections. However they are neither flux calibrated\textcolor{green}{,} nor sky subtracted. Additional information about the spectra is available in the ESO Phase 3 Data Release Description document\footnote{\url{https://www.eso.org/rm/api/v1/public/releaseDescriptions/73}}.

\begin{figure}
	\centering
	\includegraphics[scale=0.1]{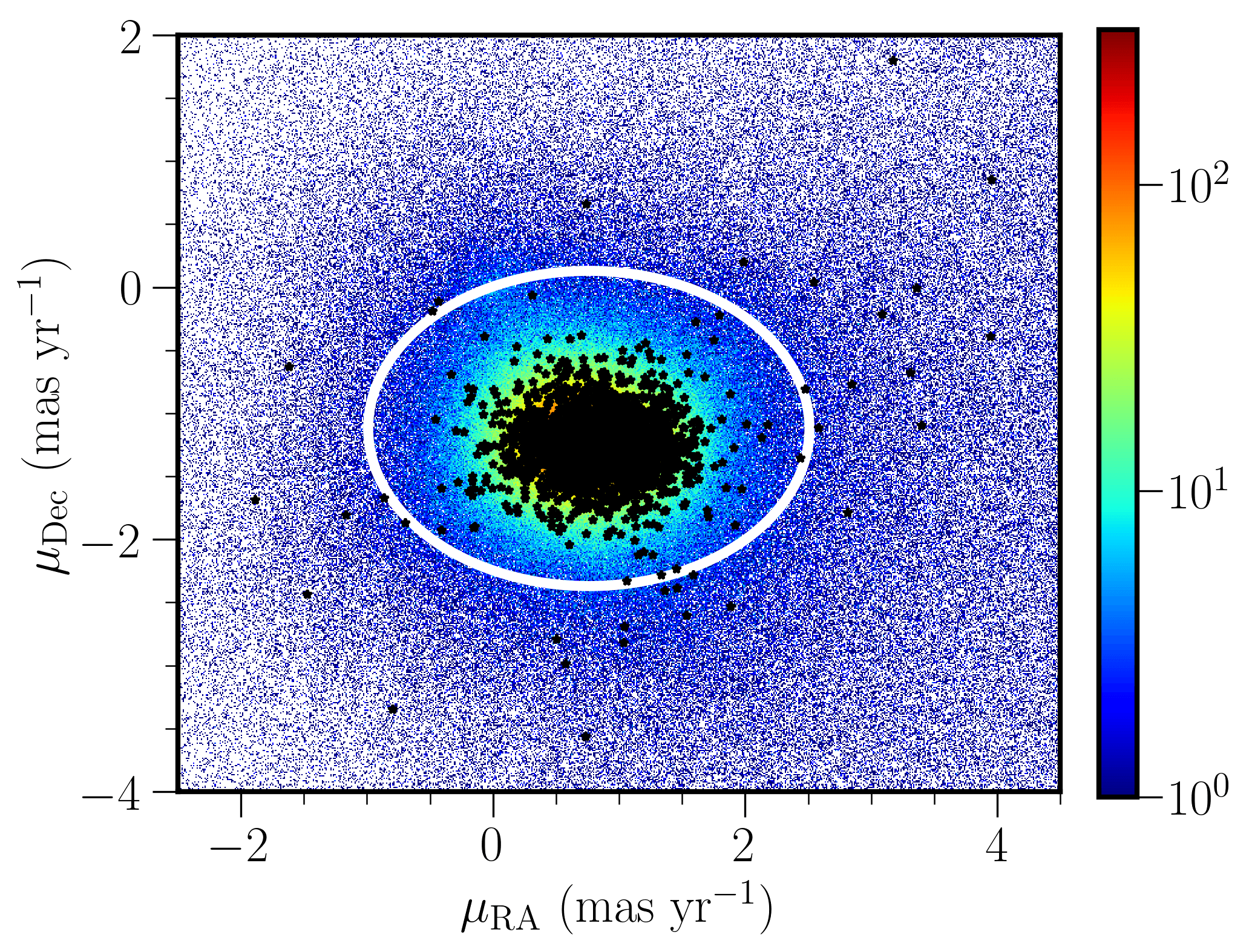}
	\includegraphics[scale=0.1]{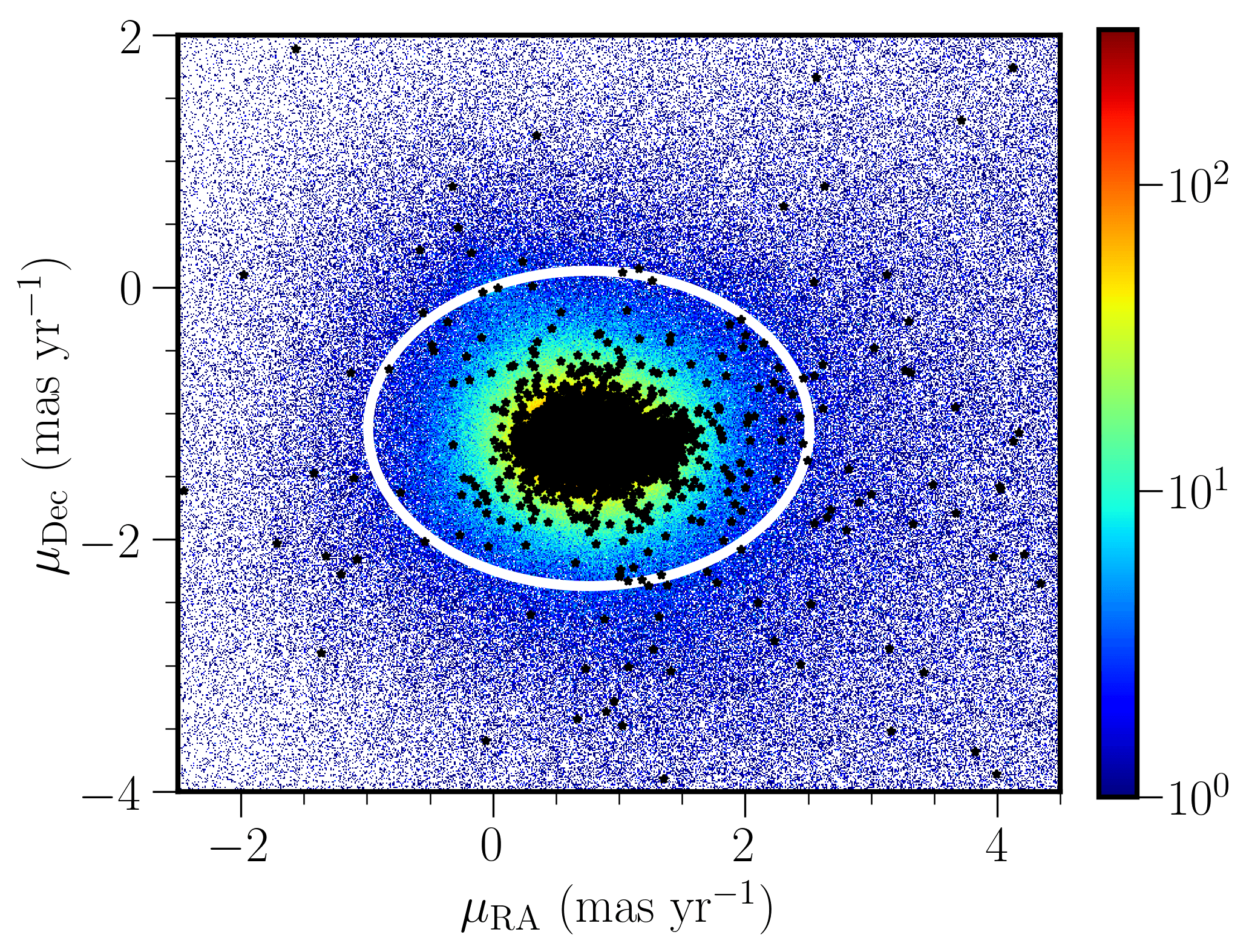}
	\caption{Stellar density of objects up to 10 deg from the SMC centre in proper motion space. The colour bars represent the number of stars. The ellipse encloses a region dominated by SMC stars. The black points show objects with available spectra in our sample (top) and in the literature sample (bottom).}
    \label{fig:BBB}
\end{figure}

Figure \ref{fig:AAA} shows the parameter ranges of the spectra used in our study. The spectra have {\it R} from 6500 to 38\,000, with 68 percent having {\it R}$\le$10\,000 and the SNR ranges from 10 to 200, with 75 percent of the spectra having a SNR $\le$50. Most of the spectra were obtained during the years 2003, 2010 and 2019. The distance between photometric and spectroscopic counterparts ranges from 0 to 1 arcsec with 61 percent of the sources having a distance $\le$0.02 arcsec. Figure \ref{fig:BBB} shows the distribution of the sources in PM space. Most of the  sources with available spectra in the ESO SAF (black points) are within the main stellar locus of the SMC, which is outlined by a white ellipse. The photometric properties of the 3700 sources such as the NIR survey of provenance, their NIR magnitudes in the $J$ and $K_\mathrm{s}$ bands, their optical magnitudes in the $G$, $G_\mathrm{BP}$ and $G_\mathrm{RP}$ bands, as well as the respective photometric uncertainties are given in Table \ref{table:photometric_properties}.
 
 
 	\begin{table*}
 	\setlength{\tabcolsep}{3pt}
		\caption{Photometric characteristics of the spectra. The table is published in its entirety as
        supporting material with the electronic version of the article.}                        
		\label{table:photometric_properties}      
		
		\begin{tabular}{ccccccccccccccccc}
			\hline
			NIR source ID & NIR survey &\textit{Gaia} EDR3 source ID & $G$ & $\sigma_G$ $^{a}$ & $G_\mathrm{BP}$ & $\sigma_{G_\mathrm{BP}}$ $^{a}$ & $G_\mathrm{RP}$ & $\sigma_{G_\mathrm{RP}}$ $^{a}$ & $J$ & $\sigma_J$& $K_\mathrm{s}$ & $\sigma_{K_\mathrm{s}}$ \\
			&&& (mag)&(mag)&(mag)&(mag)&(mag) &(mag)&(mag)&(mag)&(mag)&(mag)\\
			\hline
			558358499858 & VMC & 4688987114793104384 & 18.665 & 0.003 & 18.477 & 0.015 & 18.640 & 0.036 & 18.775 & 0.038 & 18.799 & 0.072 \\
            558370021706 & VMC & 4689011166567250944 & 15.344 & 0.003 & 15.283 & 0.006 & 15.429 & 0.006 & 15.408 & 0.004 & 15.145 & 0.007 \\
            558358869977 & VMC & 4685986066534948864 & 15.755 & 0.003 & 15.661 & 0.004 & 15.883 & 0.005 & 16.099 & 0.006 & 16.297 & 0.012 \\
            558370088420 & VMC & 4688995498568136704 & 19.036 & 0.004 & 18.696 & 0.039 & 18.715 & 0.051 & 18.675 & 0.027 & 18.466 & 0.082 \\
            558369977606 & VMC & 4689004986121150464 & 17.156 & 0.005 & 16.922 & 0.006 & 17.106 & 0.008 & 17.274 & 0.011 & 17.369 & 0.034 \\
            558358846869 & VMC & 4685985933366169600 & 13.271 & 0.003 & 13.289 & 0.003 & 13.207 & 0.004 & 13.158 & 0.001 & 13.108 & 0.003 \\
            558369967505 & VMC & 4689005054840926208 & 14.645 & 0.004 & 14.648 & 0.021 & 14.602 & 0.028 & 14.615 & 0.003 & 14.561 & 0.006 \\
            558358892671 & VMC & 4685986483121884160 & 13.567 & 0.003 & 13.511 & 0.003 & 13.661 & 0.004 & 13.828 & 0.002 & 13.865 & 0.004 \\
            558370002756 & VMC & 4689007322583053312 & 13.844 & 0.003 & 13.794 & 0.003 & 13.883 & 0.005 & 13.988 & 0.002 & 13.956 & 0.004 \\
			\hline
		\end{tabular} 
		\vspace{1pt}
	\begin{flushleft}
		$^{a}$ The standard error of $G$, $G_\mathrm{BP}$, and $G_\mathrm{RP}$ mean magnitudes were computed as a simple error propagation on the fluxes, according to the formula: $\sigma_G=\sqrt{(-2.5/\mathrm{ln}(10)*\sigma_{F_G}/F_G)^2 +\sigma_{G_{0}}^2}$, where $F_G$ is the mean flux in the $G$, $G_\mathrm{BP}$, or $G_\mathrm{RP}$ bands respectively, $\sigma_{F_G}$ is the error on the mean flux, while $\sigma_{G_{0}}$ is the zero point uncertainty.
	\end{flushleft}
	\end{table*}



\section{Analysis}\label{section3}

After applying the selection criteria outlined in the previous section and obtaining a final sample of 3700 sources, we proceed to analyse the spectra by initially performing the sky subtraction followed by full spectrum fitting to obtain line-of-sight velocities.

\subsection{Sky subtraction}
\label{sky}
The ESO pipeline recipes for GIRAFFE do not perform any sky subtraction, therefore reduced GIRAFFE spectra obtained as ESO phase 3 data products still have their sky signatures. However when available, all fibres with a sky signal (as defined by the observer) are collected in the associated MOSSKY ancillary product files, allowing for optimal sky subtraction. Usually 10 to 20 sky fibres are assigned by the observer to obtain sky spectra. We proceed to obtain a median sky spectrum for each observing block, comprising the observation of many targets within a given plate configuration, by median combining the sky spectra available at the MOSSKY ancillary product files. Since the object spectra are already shifted to the heliocentric RV system, while the sky spectra are not, we correct the median sky spectra to the heliocentric rest frame that has been applied to the science spectra by using the header keyword HELICORR. The corrected wavelength is defined by $\lambda_\mathrm{cor}=(1+v_\mathrm{helio}/c) \times \lambda_\mathrm{uncor}$ with {\it c} being the speed of light and $v_\mathrm{helio}$ the projected heliocentric velocity shift that was calculated from the ESO pipeline for each science spectrum. We follow a similar methodology for the sky subtraction as in \cite{Koch2007}, where the sky spectrum is modeled as a second-order Legendre polynomial plus the median sky. Additionally, we allow a small velocity shift in the sky spectrum so it is wavelength calibrated to our science spectra. We employed a Markov Chain Monte Carlo (MCMC) approach to determine the three parameters of the second order Legendre polynomial and the small additional velocity correction using the logarithmic likelihood function given as:
\begin{equation}
\log \mathcal{L}=\sum_{i=1}^N{\lbrack \log \left(\frac{1}{\sqrt{2\pi} \sigma}\right) -\frac{(F_{\rm obj}-F_{{\rm sky},i})^2}{2\sigma^2}}\rbrack 
\end{equation}
$N$ is the number of tracers, $F_{\mathrm{obj}}$ is the flux of the object spectrum and $F_{\mathrm{sky}, i}$ is the modeled sky spectrum defined as the median sky in addition to the second order polynomial: 
\begin{equation}
F_{\mathrm{sky}, i}= a\lambda^2 + b\lambda + c + F_\mathrm{\rm median~sky}
\end{equation}
where $a$, $b$ and $c$ are the parameters to be determined.
The uncertainty $\sigma$ results from the uncertainties on the flux of the object and sky spectra as:
\begin{equation}
\sigma^2 =\sigma_{\rm obj}^2+\sigma_{\rm median~sky,i}^2.
\end{equation}



\subsection{pPXF method}
\label{method}
In order to determine line-of-sight velocities, we are performing a full spectrum fitting using the penalised pixel fitting routine (\texttt{pPXF}; \citealp{Cappellari2004,Cappellari2017}). \texttt{pPXF} fits an observed spectrum with a best-fitting linear combination of differently weighted stellar templates in pixel space. It recovers the line-of-sight velocity distribution through moments of a Gauss--Hermite series and the parameters of the distribution are optimised by a $\chi^2$ minimisation by direct comparison with the observed spectrum. The main advantage of \texttt{pPXF} is that it makes optimal usage of the entire spectrum, rather than focusing on a few spectral lines. \texttt{pPXF} has been extensively used to study kinematics of galaxies \citep[e.g.,][]{Boardman2017}, however its implementation is independent of the system in question. The main improvement introduced in \cite{Cappellari2017} resulted in accurate velocities regardless of the velocity dispersion $\sigma$, which in our case is simply the instrumental broadening of the spectra. The \texttt{pPXF} method approximates the observed spectrum using the following parameterisation:

\begin{equation}
\label{eq:model}
G_{\rm mod}(x) =
\sum_{n=1}^{N} w_n \left\{\left[T_n(x) \ast \mathcal{R}_n(c x)\right]
\sum_{k=1}^{K} a_k \mathcal{P}_k(x)\right\} \nonumber + \sum_{l=0}^{L} b_l \mathcal{P}_l(x),
\end{equation}
where \textbf{$\mathcal{R}_n$} are the line-of-sight velocity distributions, $T_n(x)$ represent the templates, $\mathcal{P}_k(x)$ and $\mathcal{P}_l(x)$ are multiplicative and additive orthogonal polynomials of degrees $k$ and $l$ respectively, while $w_n$, $a_k$ and $b_i$ are weights to be solved for. 
The implementation of \texttt{pPXF} requires several inputs including a set of spectral templates. The noise spectrum of the observations and the starting value for the velocity and line broadening are also required. The input and noise spectra of the object to be measured are normalized and logarithmically rebinned while conserving the flux, using the \texttt{log\_rebin} routine from \texttt{pPXF}.

Even after performing the sky subtraction, some of our spectra still suffer from contamination of sky emission lines and residual bad pixels, which can affect the fit severely. In order to further clean our spectra we use the following methodology. We median smooth the log re-binned spectrum by 7 pixels and then subtract the smoothed spectrum from the not-smoothed spectrum. Using this residual-subtracted spectrum, we proceed to use an iterative sigma clipping that selects all pixels that deviate from the mean by more than 3$\sigma$; the clipping is iterative until convergence is achieved (i.e., until no pixels are removed). We then expand the binning mask by $\pm$2 pixels, we obtain the indices of the good pixels and feed them to \texttt{pPXF} by using the \texttt{goodpixels} keyword; only these spectral pixels are included in the fit.


The \texttt{pPXF} method combines the set of input template spectra to best fit the observed spectrum. These templates can either be synthetic or empirical. Considering the resolving power and wavelength range of our spectra (Figure \ref{fig:AAA}), we chose the X-shooter Spectral Library (XSL; \citealp{Arentsen2019,Gonneau2020}). This library contains empirical spectra with a resolving power close to $R=10\,000$ for  stars across a large range of spectral types and chemical compositions, which account for the different stellar populations we have in our sample. The spectra cover the following ranges of stellar parameters: $2\,600\leq T_\mathrm{eff}\leq 38\,000$ K, $0.0\leq\log(g)\leq5.7$, and $-2.5\leq$ [Fe/H] $\leq+1.0$~dex. They are corrected for RV and provided in the rest frame (with wavelength in air), logarithmically sampled in wavelength and normalised. 

The full spectrum fit is conducted using fourth order additive and multiplicative Legendre polynomials in order to account and correct for any potential low frequency differences in shape between the target's and the template's spectra continuum shape. 
Since \texttt{pPXF} uses a local minimisation algorithm, reasonably close starting values ($V_\mathrm{start}$) of the velocity and line broadening are required in order to avoid the algorithm to converge prematurely and hence being stuck in a local minimum. For $V_\mathrm{start}$ values ranging from 0 to 200 km\,s$^{-1}$ with a step of 25 km\,s$^{-1}$ are used; as to the line broadening, since we are dealing with stellar spectra, we set the starting value as the quadratic difference between the stellar spectra and templates' instrumental resolutions.
We re-sample the stellar spectra and their associated uncertainties onto the arbitrary wavelength grid of the templates, while preserving the integrated flux.
In order to evaluate whether the results represent a good fit, a reduced $\chi^2$ as close as possible to 1 should be returned; furthermore we visually inspect the fit. We assess any potential template mismatch by analysing the residual of the fit where the standard deviation of the residuals should be similar to their mean, for a good template match. Note that pPXF assigns a linear combination of templates to each source (usually between 6 and 10) so there is no single template star corresponding to each source. In addition, our templates are not set on a regular grid and since we did not apply regularisation of the fits they are not expected to cluster around specific parameter values. In the Appendix we show two examples. This work is focused on determining line-of-sight velocities whereas stellar parameters (temperature, surface gravity and metallicity) will be explored in a subsequent study.

\subsection{Systematic uncertainties}

\begin{figure*}
	\centering
		\includegraphics[scale=0.12]{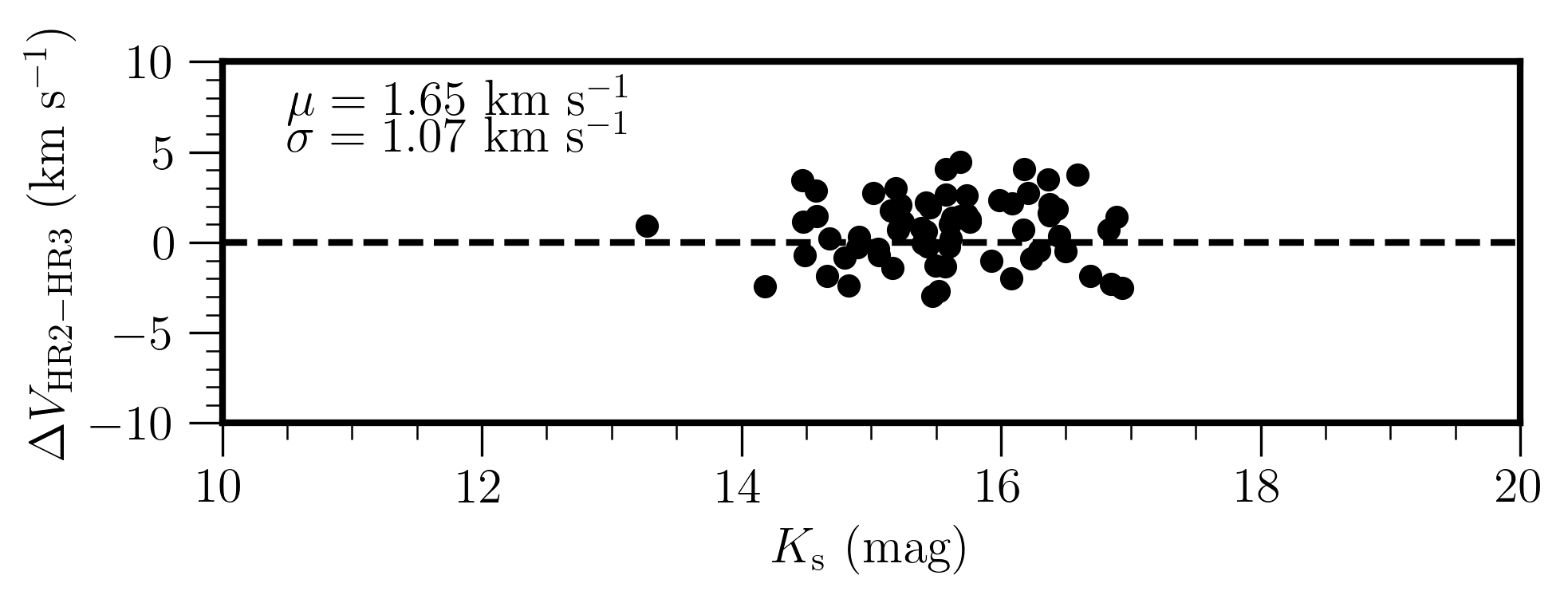}
		\includegraphics[scale=0.12]{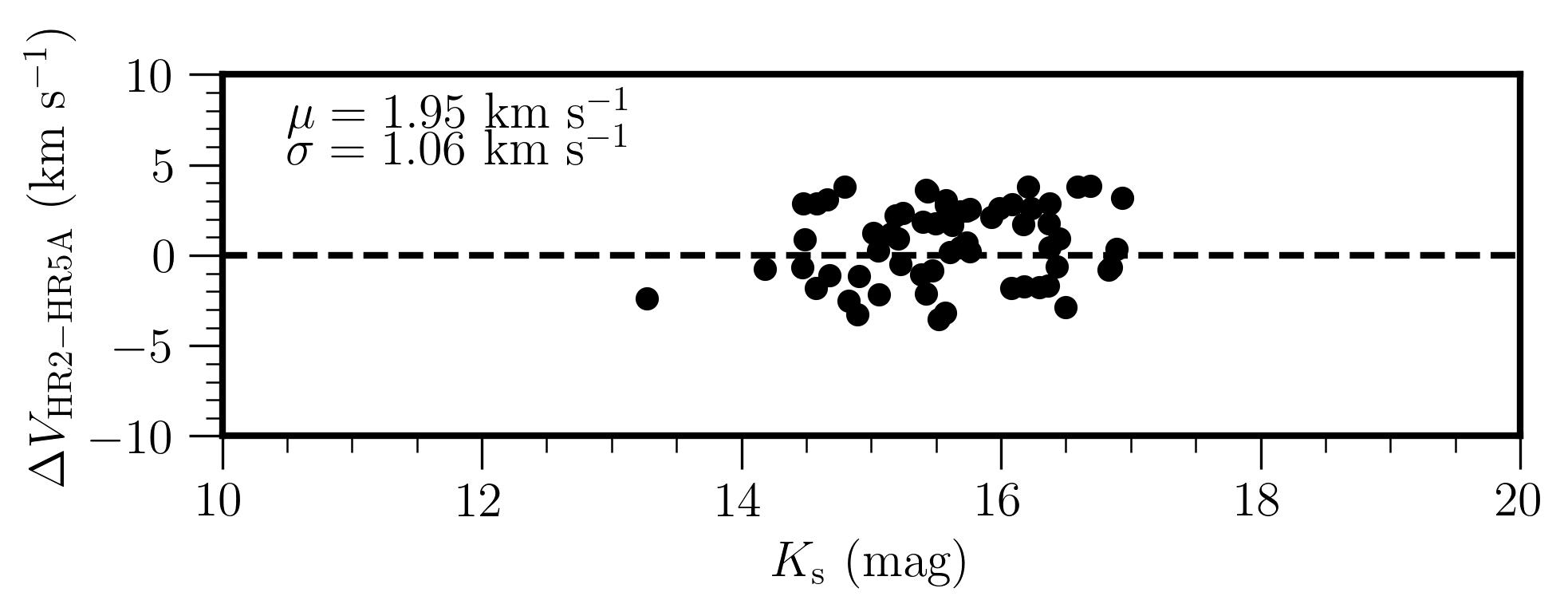}
		\includegraphics[scale=0.12]{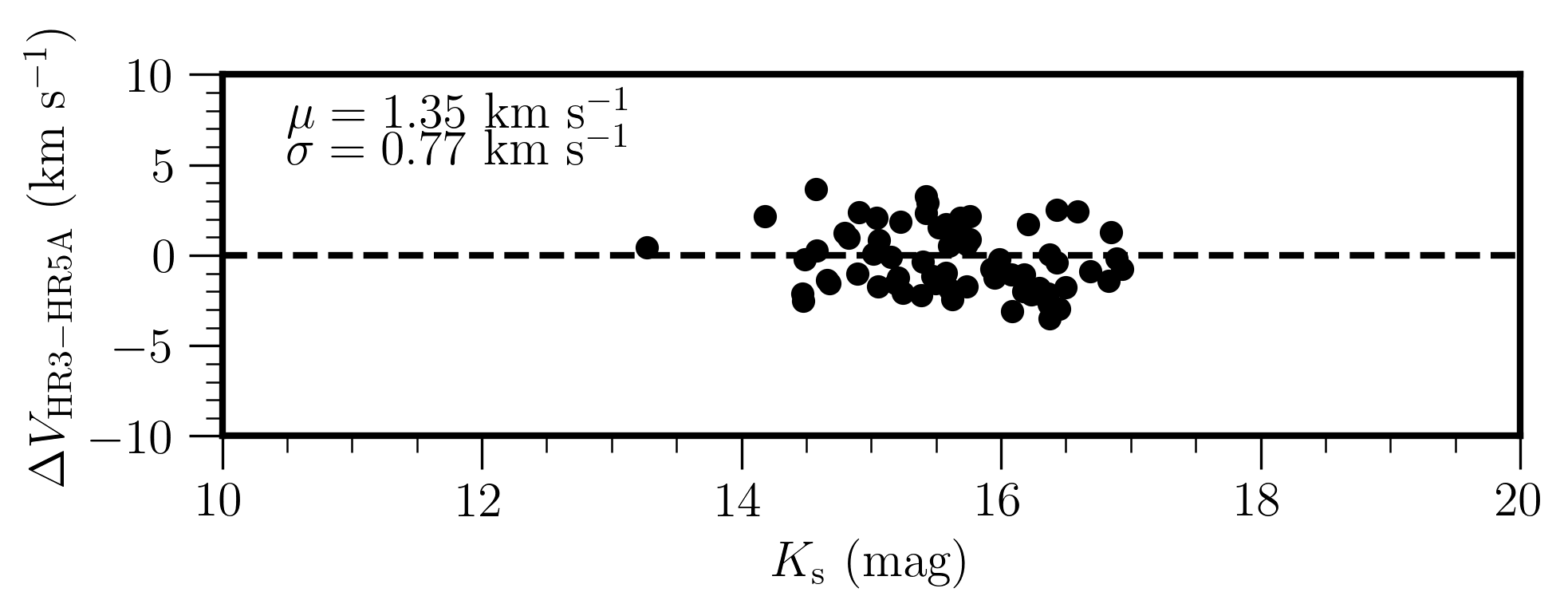}
		\includegraphics[scale=0.12]{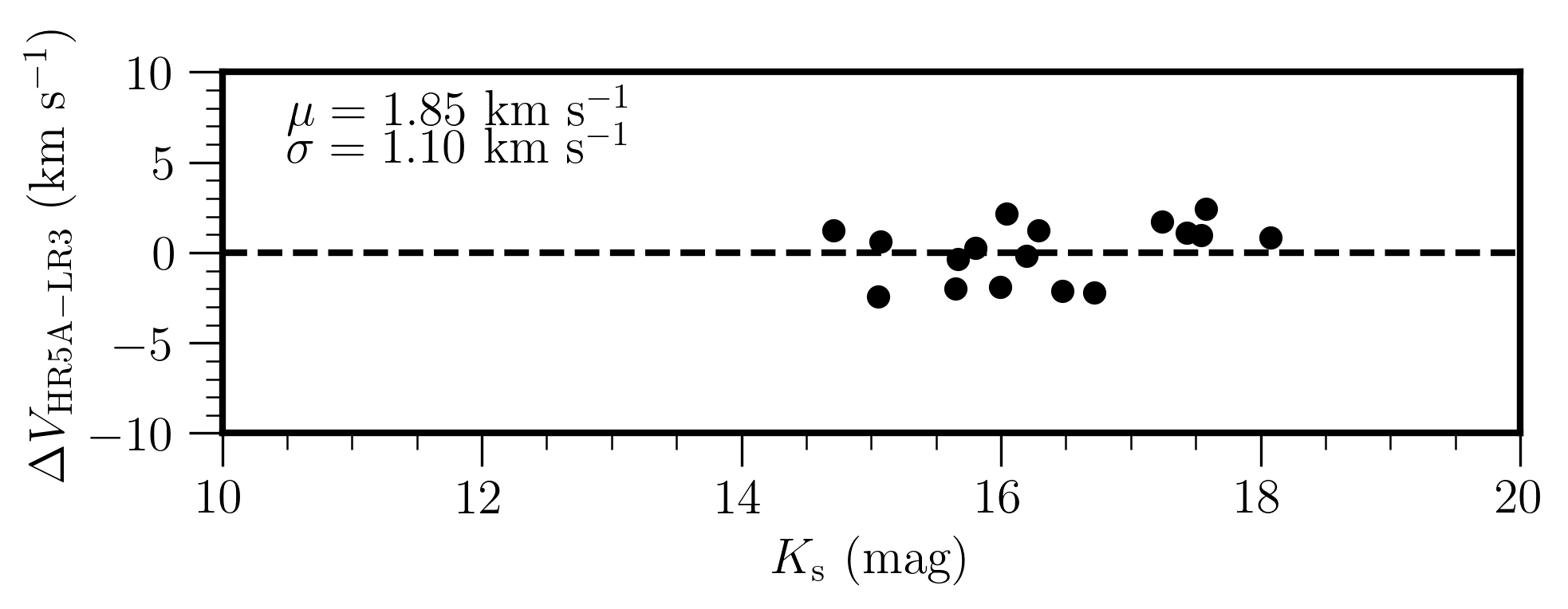}
		\includegraphics[scale=0.12]{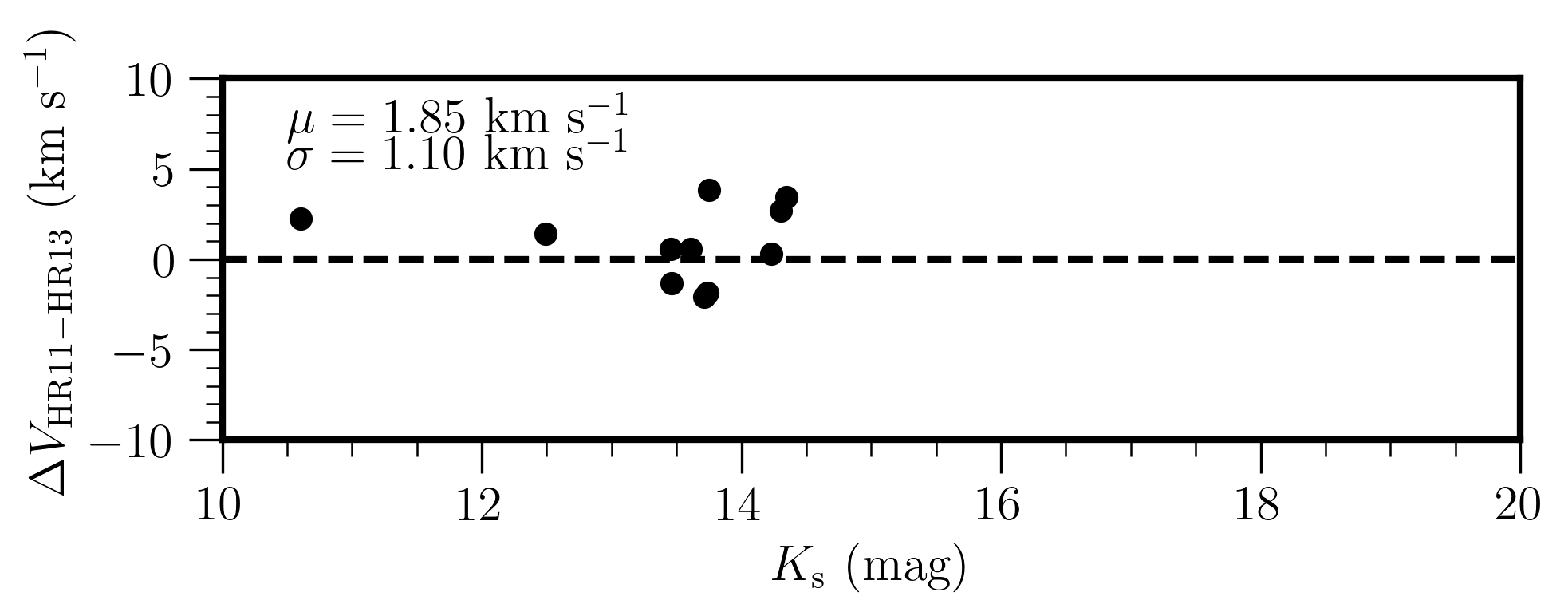}
	    \includegraphics[scale=0.12]{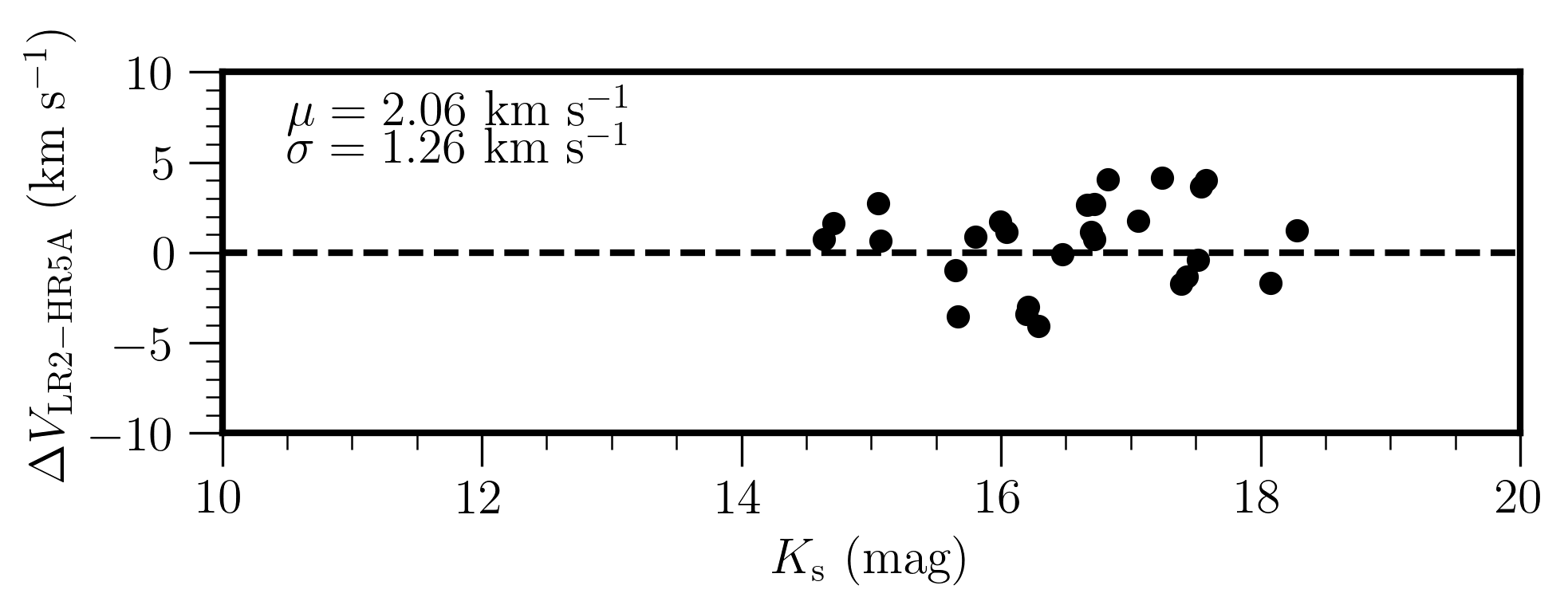}
		\includegraphics[scale=0.12]{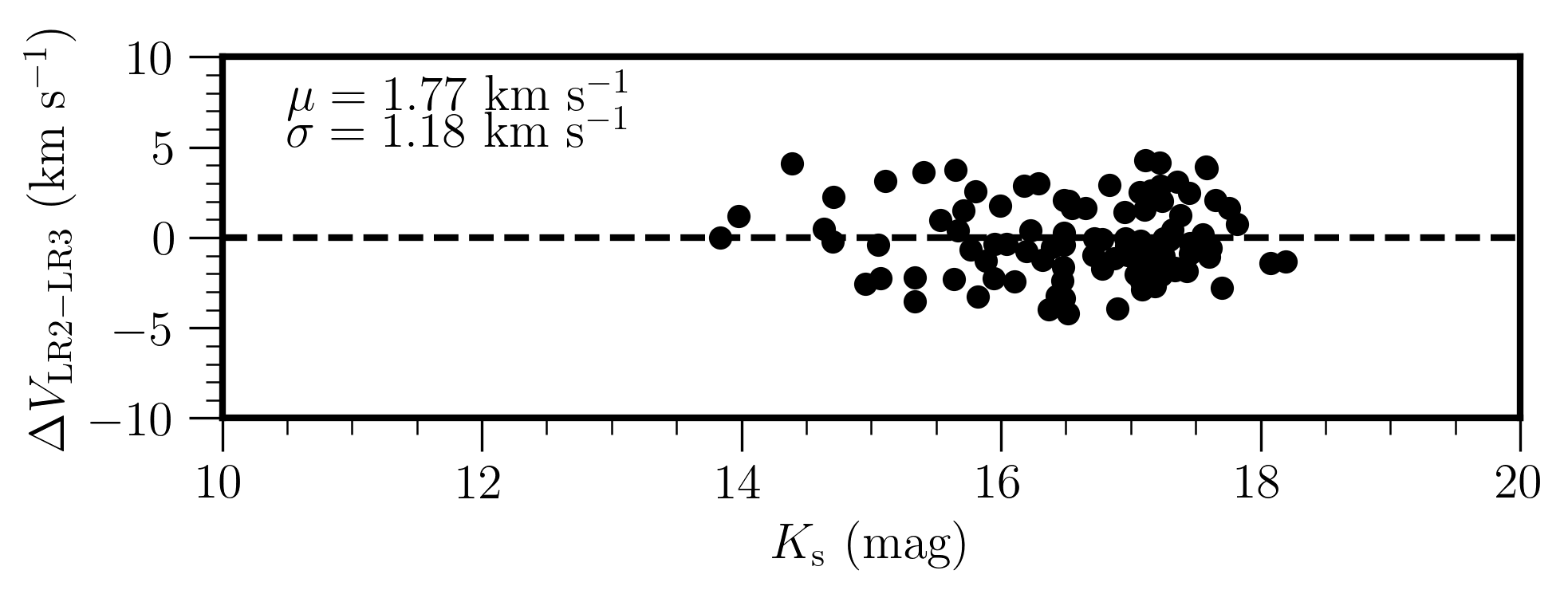}
		\includegraphics[scale=0.12]{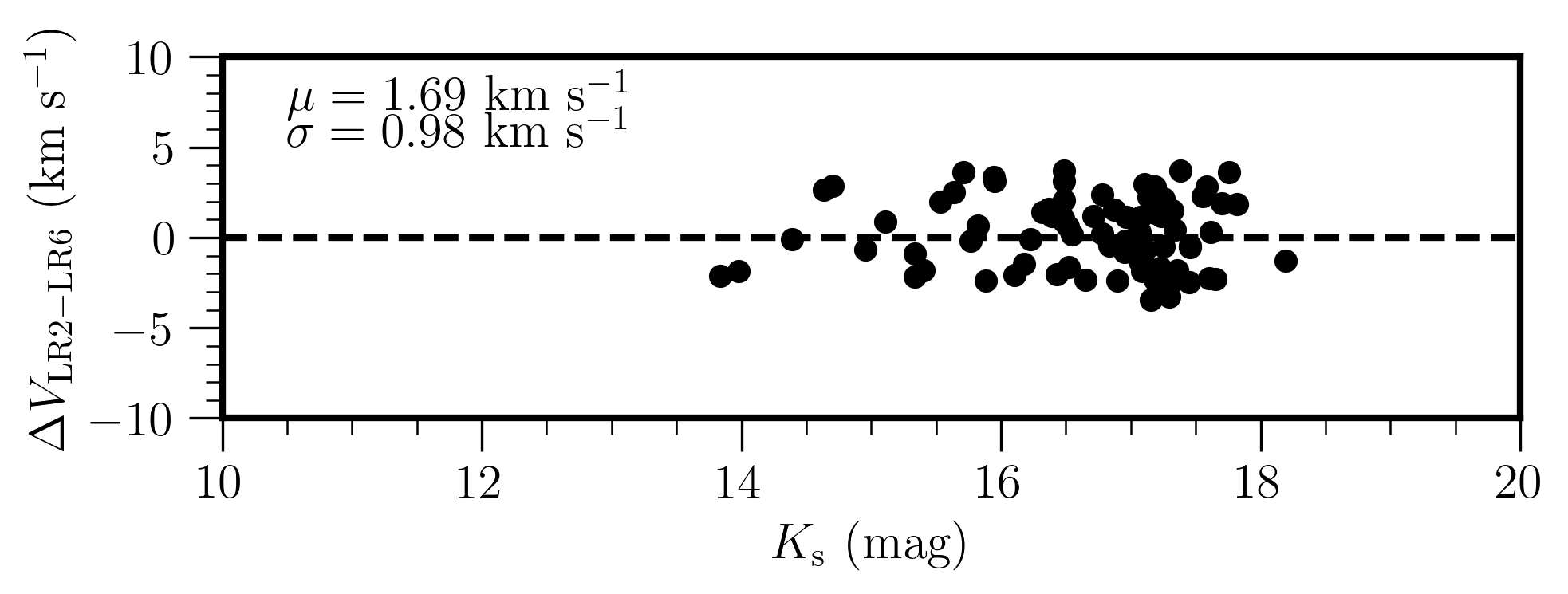}

	\caption{Comparison between the RVs estimated from spectra obtained with different GIRAFFE gratings as a function of magnitude. Means and corresponding standard deviations are indicated within each panel. The uncertainty on individual points corresponds to 2 km\,s$^{-1}$.}
	\label{fig:systemic_errors}
\end{figure*}

To evaluate the systematic uncertainties that might arise from the use of different GIRAFFE gratings, we obtain RVs for objects that have an available spectrum in at least two gratings. We found that these spectra always belong to the same Programme ID. Table \ref{table:N_common_eso} outlines the number of objects in common between these gratings. As it appears in the table, this is only available for eight combinations of gratings. These spectra have undergone the same analysis discussed in Sections \ref{sky} and \ref{method}. Figure \ref{fig:systemic_errors} shows the comparison between the RV estimates where the mean difference in terms of velocity and the respective standard deviation is indicated. We find mean differences in velocities of 1.77 $\pm$ 0.22 km\,s$^{-1}$. The dispersion within each grating is about 1 km\,s$^{-1}$.
These differences might be related to the non-uniform fibre illumination, the template mismatch in the cross-correlation procedure, as well as the resolution and SNR of the spectra. As they could not be estimated for enough combinations of gratings to homogenise the sample, we refrain from applying a correction for the systematic uncertainties.

\begin{table}
\setlength{\tabcolsep}{3pt}
	\caption{Number of sources observed in different GIRAFFE gratings.  }                      
	\label{table:N_common_eso}      
	
	\begin{tabular}{lrrrrrrrr}
		\hline
		 & HR2 & LR2 & HR3 & HR5A & LR3 & HR11 &  HR13 & LR6 \\
		\hline
	     HR2 & 66 & 0 &66 & 64 & 0 & 0 & 0 & 0\\
		 LR2  & 0 & 600&0  & 27 & 100 & 0 & 0 & 83  \\
		 HR3 & 66 &0 & 183&69 &0 &0 &0 &0 \\	    
         HR5A  & 64 & 27 & 69 &248& 17 &0 &0 &0 \\
	     LR3 & 0 &100 &0 & 17 & 591& 0 &0 &83  \\
	     HR11  & 0 & 0&0 &0 & 0 & 152 &11 &0 \\
	     HR13 & 0 &0 &0 &0 &0 &11 & 144 &0 \\
		 LR6 & 0 &83 &0 &0 &83 &0 &0 & 270 \\
		\hline
		&&&&&&&& \\			
	\end{tabular} 
\end{table}

\section{Results}\label{section4}

\subsection{Radial velocities}
\begin{figure}
 	\centering
 	\includegraphics[scale=0.1]{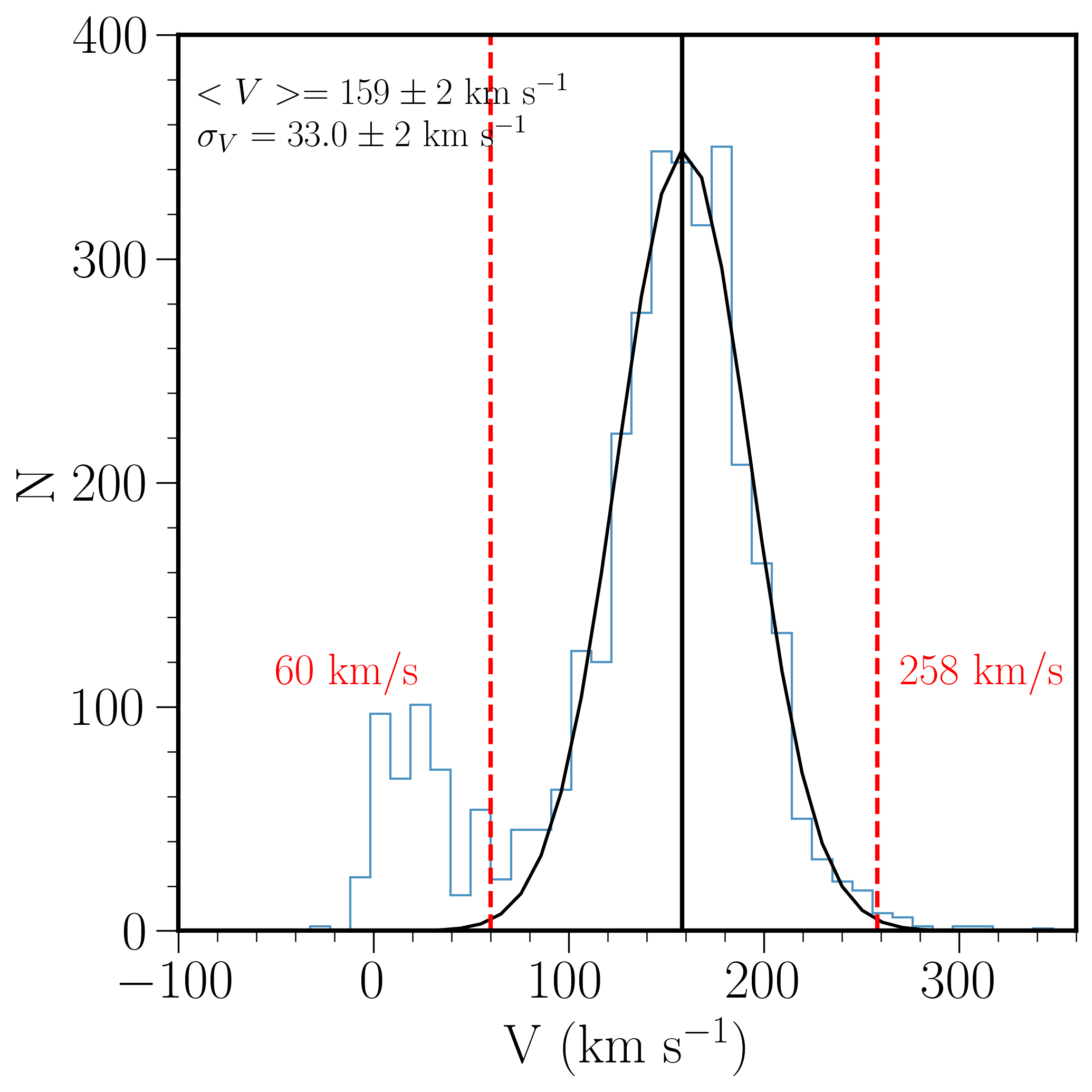}
 	\includegraphics[scale=0.1]{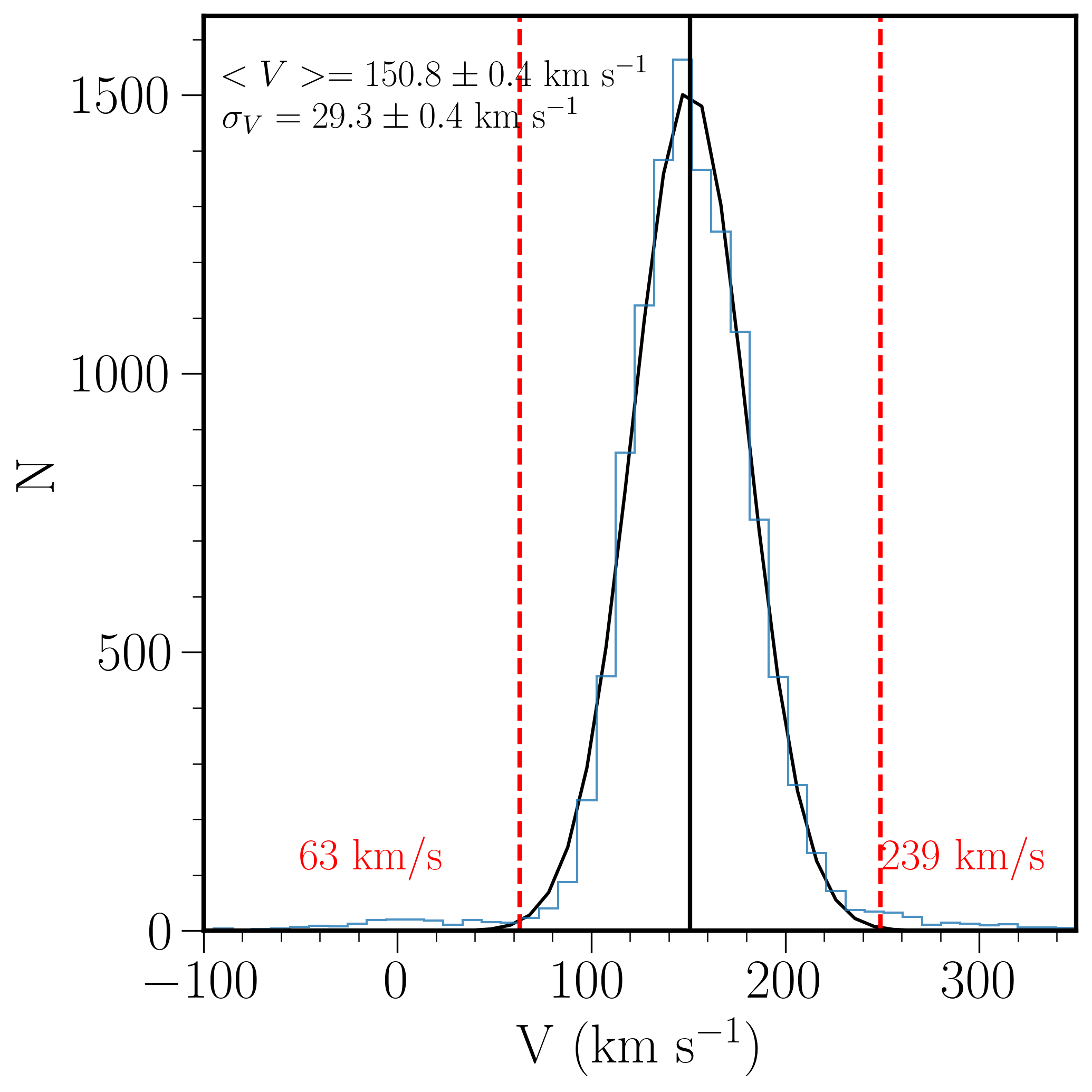}

	\caption{Distribution of RVs from our sample (top) and the homogenised literature (bottom). Black continuous lines represent the Gaussian fit of the distribution for SMC stars and the corresponding mean. The latter and its dispersion are indicated at the top left. The bin size is 10 km\,s$^{-1}$. Red vertical dashed lines mark the velocity cuts applied to reduce the influence of MW stars. These are also indicated on each side of the lines.}
    \label{fig:RV_distribution_HIstogram}
\end{figure}

Figure \ref{fig:RV_distribution_HIstogram} displays the distribution of RVs derived for the 3700 sources in the ESO SAF sample. It shows two distinct peaks with a small peak around 18 $\pm$ 2~km\,s$^{-1}$  representing the MW foreground stars in the direction of the SMC and a large peak around 159 $\pm$ 2~km\,s$^{-1}$ representing stars belonging to the SMC. The latter corresponds to a velocity dispersion ($\sigma_{V}$) of 33 $\pm$ 2 km\,s$^{-1}$ which is the best-fit sigma of the Gaussian without taking into account the velocity errors. The limits to the tails of the RV distribution representing SMC stars are set at 60~km\,s$^{-1}$ and 258~km\,s$^{-1}$  which is $\pm$ 3~$\sigma_{V}$ of the distribution. The measured velocities and their corresponding uncertainties are summarised in Table \ref{table:spectroscopic_properties}. In this table we indicate also the velocities corrected for the projected bulk proper motion, the values of the correction, the initial velocity used by the \texttt{pPXF} routine ($V_\mathrm{start}$), the reduced $\chi^2$, the formal velocity error  ($V_\mathrm{error}$) and the corrected velocity error  ($V_\mathrm{error} \times \chi^2$). Formal uncertainties provided directly by pPXF are based on the covariance matrix of the fitted parameters under the assumptions that the $\chi^2$ space is smooth, unimodal and close to unity. These errors only serve as an estimate of the order of magnitude unless $\chi^2$ $\sim$ 1. 
We find that 92\% and 99\% of our sample has a $V_\mathrm{error}$~$\leq$~5~km\,s$^{-1}$ and $V_\mathrm{error}$~$\leq$~10~km\,s$^{-1}$, respectively. 
Our mean RV is $\sim$10~km\,s$^{-1}$  higher compared to other studies such as \cite{Deleo2020} and \cite{Dobbie2014} where the mean of their distributions was found to be around 149~km\,s$^{-1}$. This difference is due to combining different type stellar populations.  \cite{Deleo2020} and \cite{Dobbie2014} observed giant stars while the mean in our study also includes main-sequence and supergiant stars. The overall distribution of RVs derived in literature studies is also shown in Fig.~\ref{fig:RV_distribution_HIstogram} and discussed in Section \ref{section42}. The RV of the different stellar populations in this work is further explored in Section \ref{section:sp_kinematics}. 
 
	\begin{table*}
	\setlength{\tabcolsep}{3pt}
		\caption{RVs of sources from the 3700 sources in our sample. The table is published in its entirety as
        supporting material with the electronic version of the article.}                        
		\label{table:spectroscopic_properties}      
		
		\begin{tabular}{crrrrrr}
			\hline
			ARCFILE & $V$ & $V_\mathrm{corr.}$ & corr. &  $V_\mathrm{error}$ & $\chi^{2}$ & $V_\mathrm{error}\times \chi^{2}$\\
			 & km s$^{-1}$ & km s$^{-1}$ & km s$^{-1}$ & km s$^{-1}$ & & km s$^{-1}$\\
			\hline
			ADP.2015-04-13T10:11:06.923 & 185.1  & 185.6 & 0.5 & 5.8 & 0.5 & 2.9 \\
            ADP.2015-04-13T10:11:07.283 & 139.5 & 142.0 & $-$2.5 & 0.9 & 0.8 & 0.7 \\
            ADP.2015-04-13T10:11:13.337 & 183.7 & 183.8 & $-$0.1 & 3.8 & 0.3 & 1.1 \\
            ADP.2015-04-13T10:11:16.837 & 164.8 & 165.3 & $-$0.5 & 8.7 & 0.4 & 3.5 \\
            ADP.2015-04-13T10:11:26.400 & 98.6 & 100.4 & $-$1.8 & 4.0 & 0.4 & 1.6 \\
            ADP.2015-04-13T10:11:53.937 & 152.5 & 152.6 & $-$0.1 & 0.7 & 5.5 & 3.8 \\
            ADP.2015-04-13T10:12:02.720 & 173.0 & 174.9 & $-$1.9 & 1.2 & 0.4 & 0.5 \\
            ADP.2015-04-13T10:12:15.860 & 165.3 & 165.5 & $-$0.2 & 0.7 & 0.8 & 0.6 \\
            ADP.2015-04-13T10:12:19.657 & 181.8 & 183.9 & $-$1.2 & 0.8 & 0.8 & 0.6 \\
			\hline
		\end{tabular} 
	\end{table*}

\subsection{Radial velocities from the literature}\label{section42}


Several spectroscopic studies have targeted the SMC using different stellar populations to characterise its internal kinematics. In our analysis, we consider data from the following literature studies and construct a comprehensive homogenised catalogue to complement the spectroscopic sample obtained in Section \ref{section2.2}.

\begin{itemize}
\item\cite{Carrera2008} presented the chemical enrichment of the SMC by providing stellar metallicities as well as RVs for 350 RGB stars distributed across the inner 4 deg of the galaxy. The observations were carried out using the visual and near-ultraviolet FOcal Reducer and low dispersion Spectrograph 2 (FORS2) at the VLT in 13 fields of about 7$\times$7 arcmin$^2$ each in size. The Spectroscopic Mask (MXU) mode was used to obtain multi-object spectra in the wavelength range 773--948 nm with $R = 2560$. This study outlined the first spectroscopic metallicity gradient in the SMC and RVs were also used to reject non SMC members from the sample. Their RV uncertainty was $\sim$ 4 km\,s$^{-1}$. 

\item\cite{Evans2008} traced the dynamics of the young stellar population of the SMC by obtaining RVs of 2045 OBA-type stars. They observed with the Two Degree Field Facility (2dF) at the Anglo-Australian Telescope (AAT) and obtained spectra that cover the wavelength range 390--480~nm. The resolving power of $R = 1600$ resulted in an accuracy of $\sim$ 11 km\,s$^{-1}$ on the RV of individual stars distributed across an area of 2 $\times$ 12.5 deg$^2$. The authors found a velocity gradient across the SMC bar of 26.3 $\pm$ 1.6 km\,s$^{-1}$\,deg$^{-1}$ at PA = 126 deg. They also noted a difference in RV between the Wing and bar as well as a $\sim$20 km\,s$^{-1}$ red-shift of their young stars compared to older stellar populations.

\item\cite{Harris2006} observed 2046 red giant stars in the central 4 $\times$ 2 kpc$^2$ of the galaxy using the Inamori-Magellan Areal Camera and Spectrograph (IMACS) at the Magellan Baade Telescope. The spectra spanned a wavelength range from 560--1000 nm with a resolving power $R\sim5000$ resulting in a RV uncertainty of $\approx$ 10 km\,s$^{-1}$. The authors found a global velocity distribution centred at 146 km\,s$^{-1}$ with a dispersion of 28 km\,s$^{-1}$ as well as evidence of a velocity gradient across the SMC.

\item\cite{Dobbie2014} provided RVs of 4172 red giant stars and 352 carbon stars across an area of 37.5 deg$^2$ with an accuracy better than 5 km\,s$^{-1}$. The spectra were acquired using the AAOmega spectrograph at the AAT. The blue and red arms of the instrument were configured with the 1500V ($R\sim 4000$) and 1700D ($R\sim 10\,000$) gratings with a wavelength coverage ranging from 425--600 nm and 845--900 nm, respectively. The authors found that a rotating disc model best represents the rest frame velocity of the youngest red giants in their sample. Beyond $\geq$ 4 deg they found signatures of tidal stripping. Their global velocity distribution is centred at 147.8 $\pm$ 0.5 km\,s$^{-1}$.

\item \cite{Lamb2016} observed 374 stars as part of the RIOTS4 Survey, a uniformly selected survey of young stars in the SMC. They found that the SMC has a systemic velocity of $\sim$150 km\,s$^{-1}$ and kinematics like those of other massive stars surveys \citep[e.g.,][]{Evans2008}. Their spectral coverage for each star varied. However every spectrum includes the wavelength range from 400--470 nm. A majority of their targets (328) were observed using IMACS at the Magellan Baade telescope with a spectral
resolving power of \textbf{$R=2600$--$3700$}. The other targets were observed with IMACS at a resolving power of \textbf{$R=1000$--$1300$} or with 
the Magellan Inamori Kyocera Echelle (MIKE) Spectrograph on the Magellan Clay telescope with a resolving power of $R\sim 28\,000$. The uncertainties on their RV measurements are $\sim$ 5 km\,s$^{-1}$ for MIKE observations and \textbf{$10$--$25$} km\,s$^{-1}$ for IMACS observations.

\item\cite{Deleo2020} targeted 2573 red giant stars within $\sim$ 4~kpc from the SMC's centre using the AAOmega spectrograph at the AAT. Observations used the grating 1700D spanning a wavelength coverage of 845--900 nm and providing a resolving power of $R\sim$8500. They achieved a velocity uncertainty better than 5~km\,s$^{-1}$ and their velocity distribution is centred at $\sim$ 150~km\,s$^{-1}$. They found an outward stellar motion in the direction towards the LMC supporting that the SMC is undergoing tidal disruption. 

\item \cite{Nidever2020} studied 3800 objects in the Magellanic Clouds using data from the Sloan Digital Sky Survey (SDSS) DR16 \citep{Ahumada2020}. Observations were obtained from the Apache Point Observatory Galactic Evolution Experiment 2 South (APOGEE-2S; \citealp{Majewski2017, Wilson2019}) on the du Pont telescope at Las Campanas Observatory (LCO) in the near-infrared (1.51--1.70 $\mu$m) with a resolving power of $R\sim20\,000$. \cite{Nidever2020} mostly focused on the RGB population, but observations included also asymptotic giant granch (AGB) stars, main-sequence stars, and supergiants distributed within 10 deg of the LMC and 6 deg of the SMC centres. They obtained a mean radial velocity of 135 km\,s$^{-1}$ and RV uncertainties of $\sim$ 0.1--0.2 km\,s$^{-1}$. In our study we use the current DR17  \citep{Abdurrouf2022} which contains observations for 3600 objects in the SMC. 

\item \textit{Gaia} DR3 \citep{Vallenari2022} contains median RVs for 33\,812\,183 stars brighter than $G$ = 14 mag and distributed throughout the entire celestial sphere. The median formal precision of the RVs is of 1.3 km\,s$^{-1}$ at $G$ = 12 and 6.4 km\,s$^{-1}$ at $G$ = 14 mag \citep{Katz2022}. The spectra are taken with the Radial Velocity Spectrograph and have a resolving power of $R=11\,500$ across a wavelength range of 845--872~nm. 
\end{itemize}

We proceed to cross-match the targets of the literature studies with our photometric sample outlined in Section \ref{section2.1} to obtain their NIR photometry and \textit{Gaia} EDR3 astrometric solution. The cross-matching facility of TOPCAT were used adopting a maximum cross-matching distance of 1 arcsec. 
Figure \ref{fig:CMD_SD_ESO_Archive} shows the NIR CMD of the sources in the literature sample and their location across the different stellar population regions outlined in \cite{ElYoussoufi2019} as well as their spatial distribution, which displays the coverage of the main body of the galaxy and of some fields in the outskirts. The location of the fields studied by \cite{Cullinane2023} using data from the Magellanic Edges Survey is also indicated.
Figure \ref{fig:BBB} shows the distribution of the sources across the PM space highlighting that most of them are enclosed within a region containing a reduced number of MW stars due to the astrometric selection criteria discussed in Section \ref{section2.1}.
 
\begin{figure}
	\centering
    \includegraphics[scale=0.45]{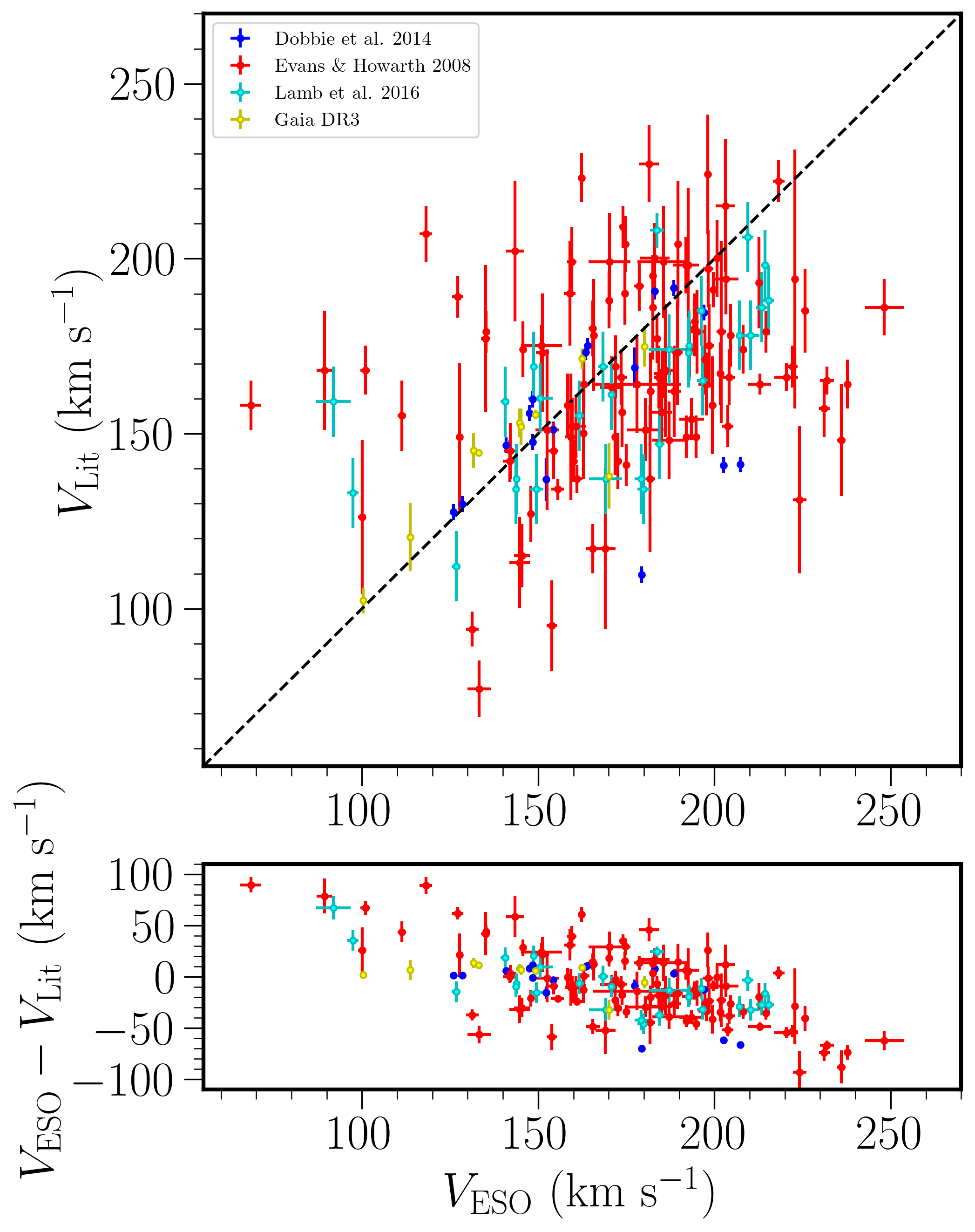}

	\caption{(top) Comparison between the RVs from our sample obtained in this study versus those derived in literature studies. The dotted line shows the one-to-one relation. (bottom) Distribution of residuals with a mean velocity of $-$6.5 km\,s$^{-1}$ and a velocity dispersion of 38.8 km\,s$^{-1}$}.
	\label{fig:RV}
\end{figure}

We then cross-match the targets among the  literature studies to find common sources. Their numbers are summarised in Table \ref{table:N_common_rv_lit}. As a consequence of combining data from different spectroscopic literature studies, we need to assess if there are any systematic offsets present in the RVs and to take them into account to produce a homogenised sample. We decided to anchor the RVs on APOGEE DR17. This catalogue has data with the highest resolving power, which span different stellar populations, and has sources in common with almost all of the other studies. Table \ref{table:offset_rv_lit} summarises the number of objects each study has in common with APOGEE DR17 and provides the median differences between the RVs \textbf{($\langle\Delta V\rangle$)} of APOGEE DR17 and the other studies as well as the standard deviation of the differences ($\sigma_{\Delta V}$). We eliminate duplicates by retaining sources with the most precise RVs and obtain a homogeneous sample by adding the median RV differences to each study respectively (for example, $V=V_\mathrm{Dobbie~et ~al.}+\langle\Delta V \rangle$). We discard the \cite{Carrera2008} sample because it does not have enough sources in common with APOGEE DR17 or other studies. Our final homogenised sample of RVs from the literature studies contains 9110 sources. 
Figure \ref{fig:RV_distribution_HIstogram} displays the distribution of RVs derived for the literature sample. It shows a single peak at 150.8 $\pm$ 0.4~km\,s$^{-1}$ representing stars belonging to the SMC, with  $\sigma_{V}=$29.3 $\pm$ 0.4 km\,s$^{-1}$. The tails of the RV distribution representing SMC stars are set at 63~km\,s$^{-1}$ and 239~km\,s$^{-1}$  which is $\pm$ 3~$\sigma_{V}$ of the distribution. We find that the velocity dispersion between the ESO SAF and literature samples agree within their dispersions. However the mean velocities differ by $\sim$ 8 km\,s$^{-1}$. A one-to-one comparison of the velocities for the sources in common is shown in Fig.~\ref{fig:RV}. The distribution appears flat because the data spread is similar within both ranges of RVs. The distribution of residuals, except for the few sources at the lowest and highest velocities, does not show any obvious trends overall and for any of the samples. We find a mean star by star offset of $V_\mathrm{ESO}-V_\mathrm{Lit}$=$-$6.5 km\,s$^{-1}$ for targets in common and a velocity dispersion of 38.8 km\,s$^{-1}$, which is larger than the errors on individual sources.

\begin{table*}

	\caption{Number of sources found in our NIR photometric sample and in common among the RV literature studies.}                        
	\label{table:N_common_rv_lit}      
	
	\begin{tabular}{lrrrrrrrr}
		\hline
		& N$_\mathrm{APOGEE~DR17}$&N$_\mathrm{Dobbie~et~al.}$&N$_\mathrm{De~Leo~et~al.}$ &N$_\mathrm{Harris~\&~Zaritsky}$ & N$_\mathrm{Evans~\&~Howarth}$&N$_\mathrm{Carrera ~et~al.}$&N$_\mathrm{Lamb~et~al.}$& N$_\mathrm{Gaia~DR3}$\\
		\hline
		N$_\mathrm{APOGEE~DR17}$       &1913 & 103 &97 &0 &9 &0 &2 & 115 \\	    
        N$_\mathrm{Dobbie~et~al.}$     &103 & 2743 & 103 &93 &1 & 2 & 0& 180 \\
		N$_\mathrm{De~Leo~et~al.}$     &97 &103 &1005 &4 &0 &2 &0 &0 \\
		N$_\mathrm{Harris~\&~Zaritsky}$&0 & 93&4 &1581 &0 &1 &0 &0 \\
		N$_\mathrm{Evans~\&~Howarth}$  &9 &1 &0 &0 &1303 &0 &35 & 18 \\
		N$_\mathrm{Carrera~et~al.}$    &0 & 2 &2 &1 &0 &215 &0 &0 \\
		N$_\mathrm{Lamb~et~al.}$       &2 &0 &0 &0 &35 &0 &196 &0 \\
		N$_\mathrm{Gaia~DR3}$         & 115 & 180 &0 &0 & 18 &0 &0 & 2833 \\
		\hline
		&&&&&&&& \\			
	\end{tabular} 
\end{table*}

\begin{table}
	\caption{RV differences between APOGEE DR17 and literature studies.}                       
	\label{table:offset_rv_lit}      
	
	\begin{tabular}{lrrr}
		\hline
		&$N$& $\langle\Delta V\rangle$&$\sigma_{\Delta V}$\\
		& & km s$^{-1}$ & km s$^{-1}$ \\
		\hline
		APOGEE DR17 -- Dobbie et al.        & 103 & 0.51 & 3.51  \\	    
		APOGEE DR17 -- De~Leo et al.        &96 &$-$0.07&3.03 \\
		APOGEE DR17 -- Evans~\&~Howarth     &9 &$-$8.86& 67.77 \\
		APOGEE DR17 -- Carrera et al.       &0 &--&-- \\
	    APOGEE DR17 -- Lamb et al.          &2 &7.14   & 99.41 \\
		APOGEE DR17 -- Gaia DR3            & 115 &  0.03  & 3.93 \\
		APOGEE DR17 -- Harris~\&~Zaritsky   & 0 &1.53$^\mathrm{a}$ &18.72$^\mathrm{a}$ \\
		Dobbie et al. -- Harris~\&~Zaritsky        &93 &1.10 &16.10  \\
        Evans~\&~Howarth -- Lamb et al.     &35 & 16.00 & 31.64 \\
		\hline		
	\end{tabular} 
    $^\mathrm{a}$This value was calculated from the combination of the results from APOGEE DR17 -- Dobbie et al. and Dobbie et al. -- Harris \& Zaritsky samples.
    
\end{table}

\subsection{Stellar kinematics of the SMC}\label{section43}

\subsubsection{Kinematics of stellar populations}\label{section:sp_kinematics}

\begin{figure*}
	\centering
   \includegraphics[scale=0.3]{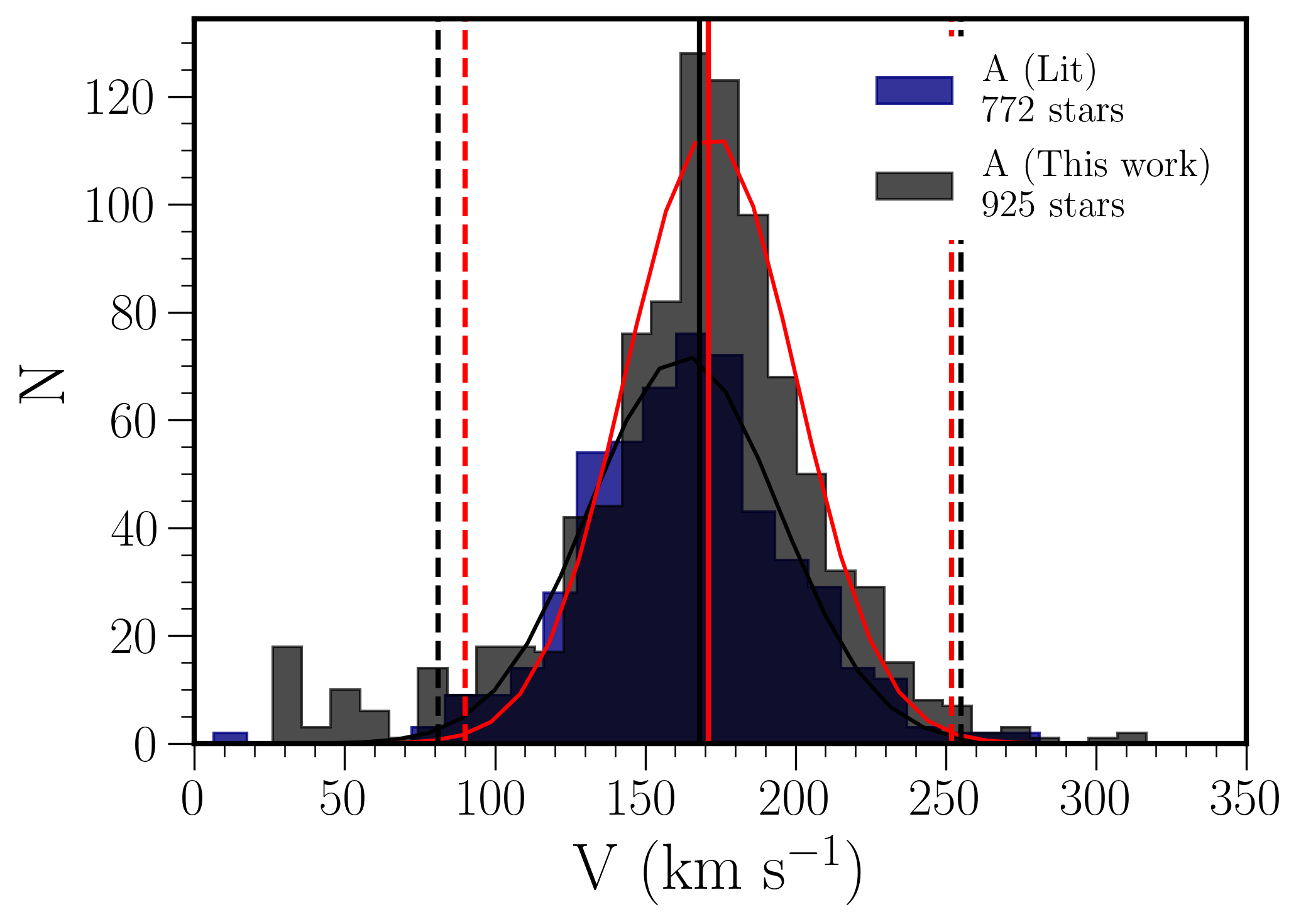}
    \includegraphics[scale=0.3]{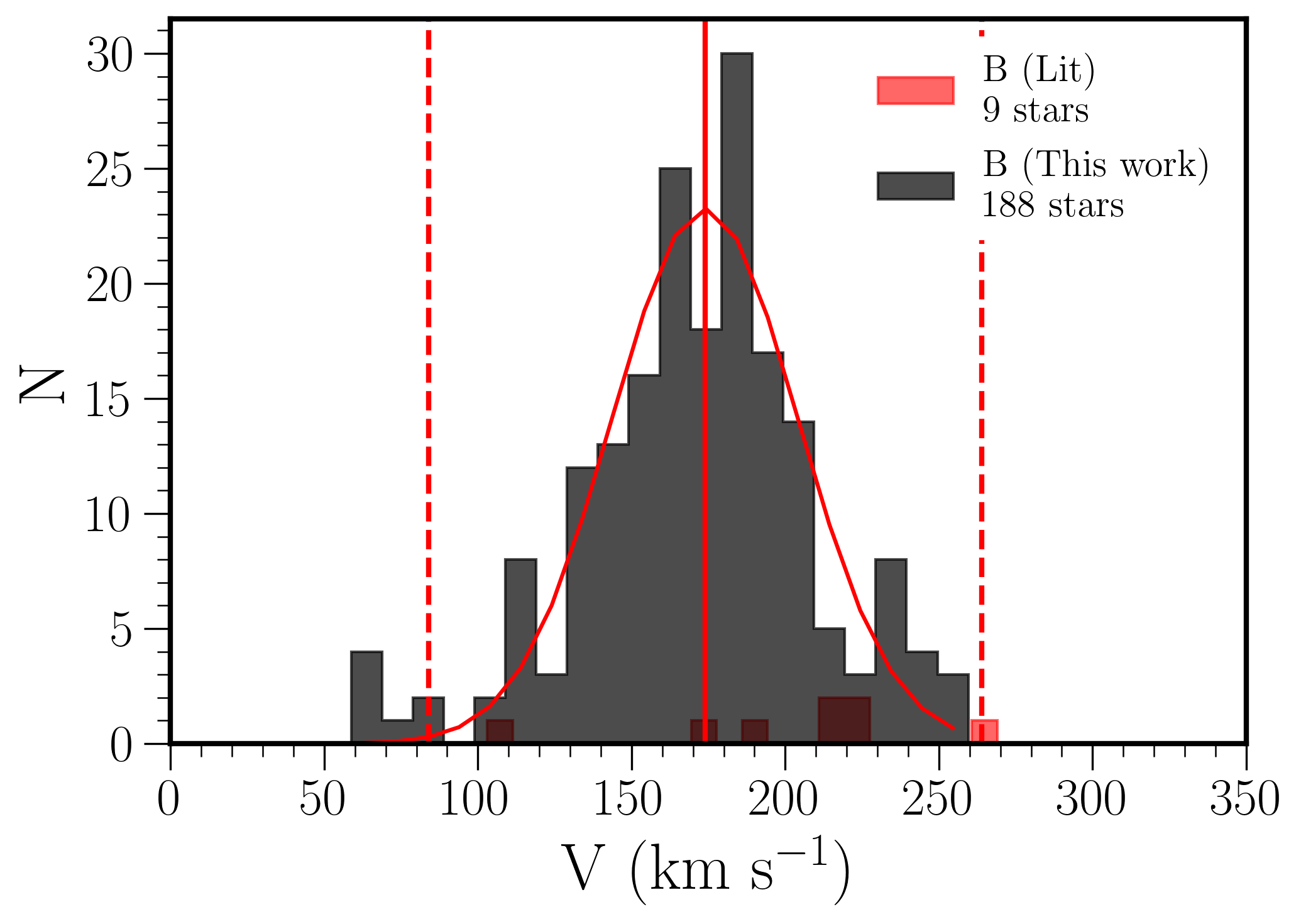}
    \includegraphics[scale=0.3]{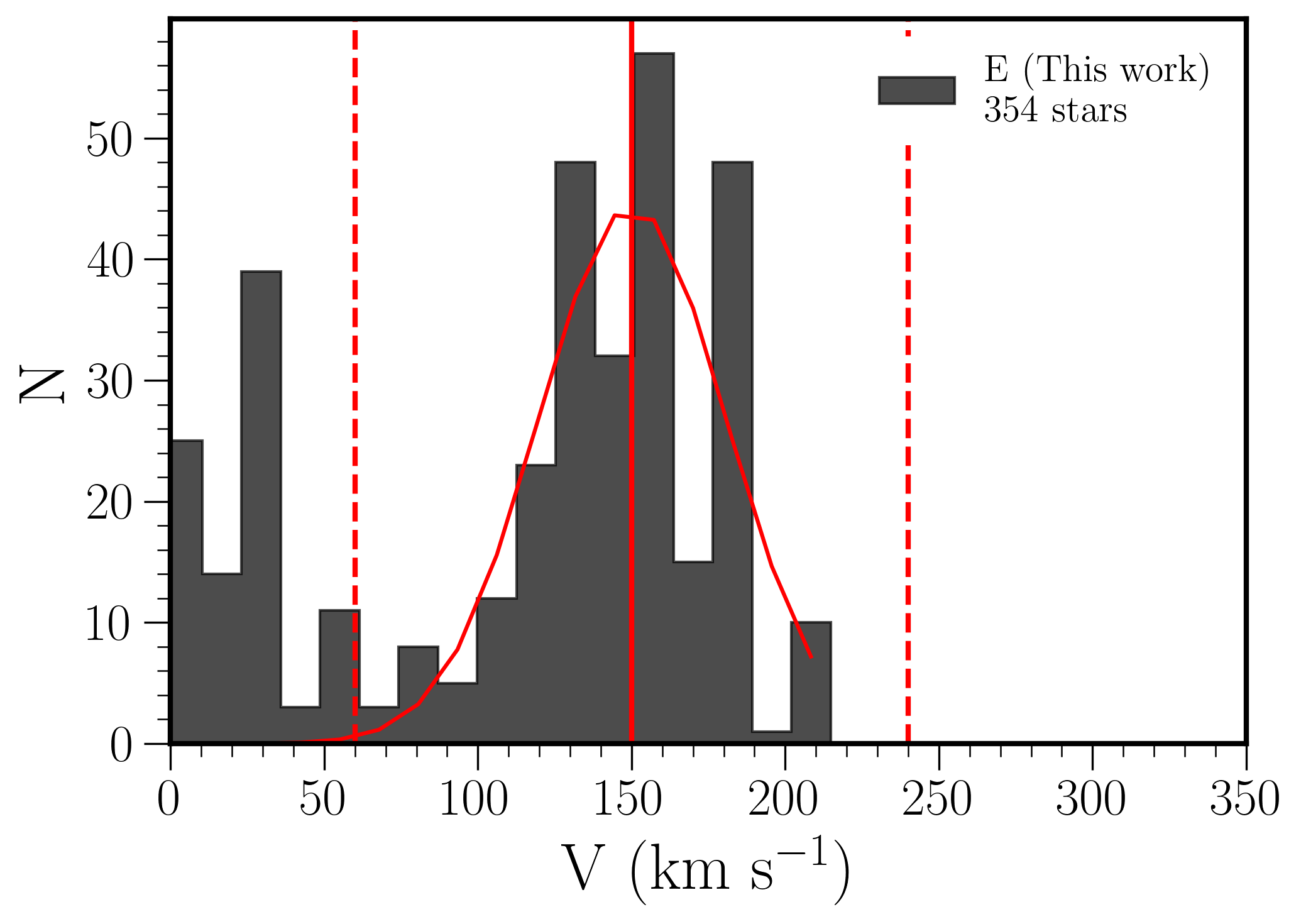}
    \includegraphics[scale=0.3]{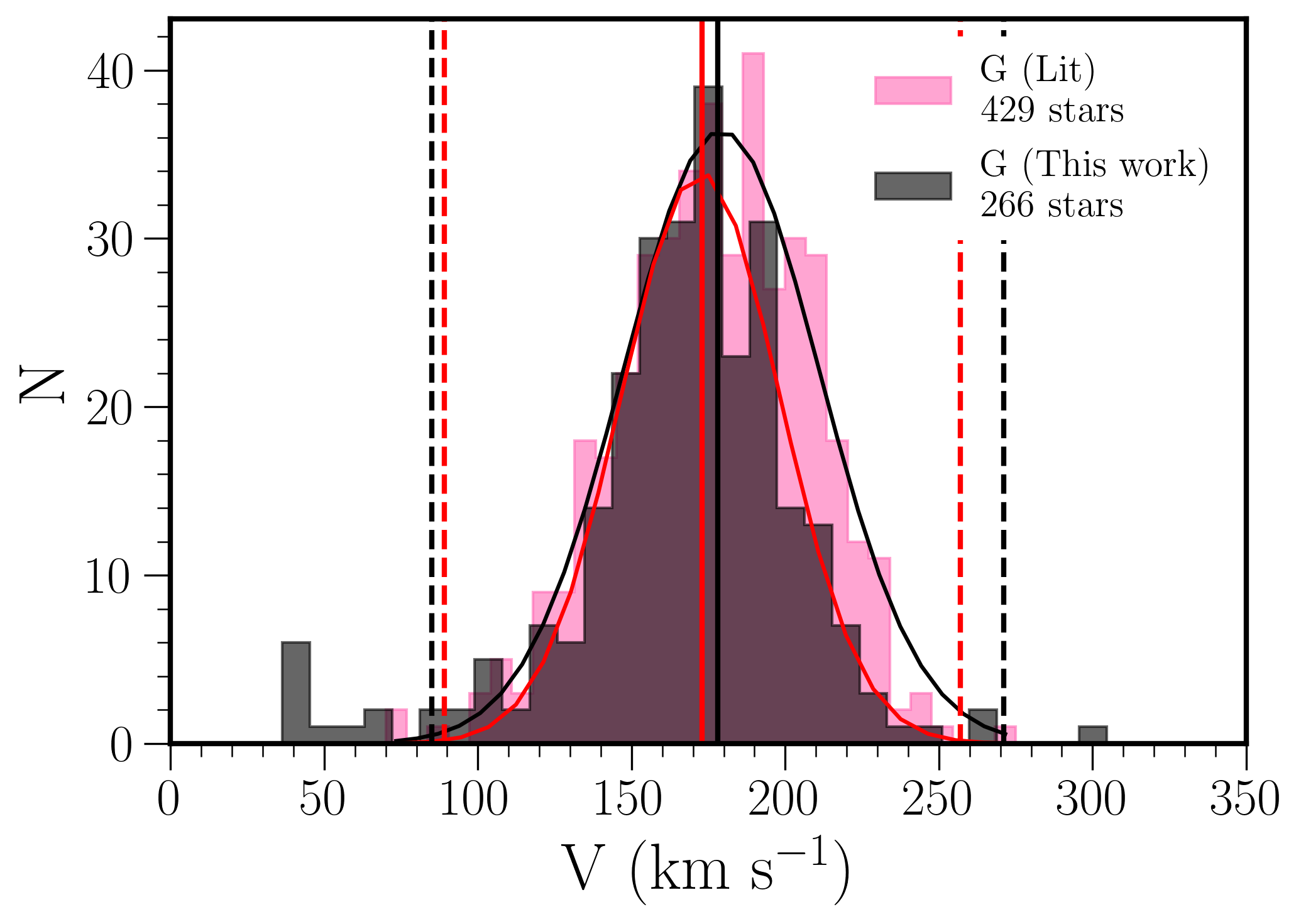}
    \includegraphics[scale=0.3]{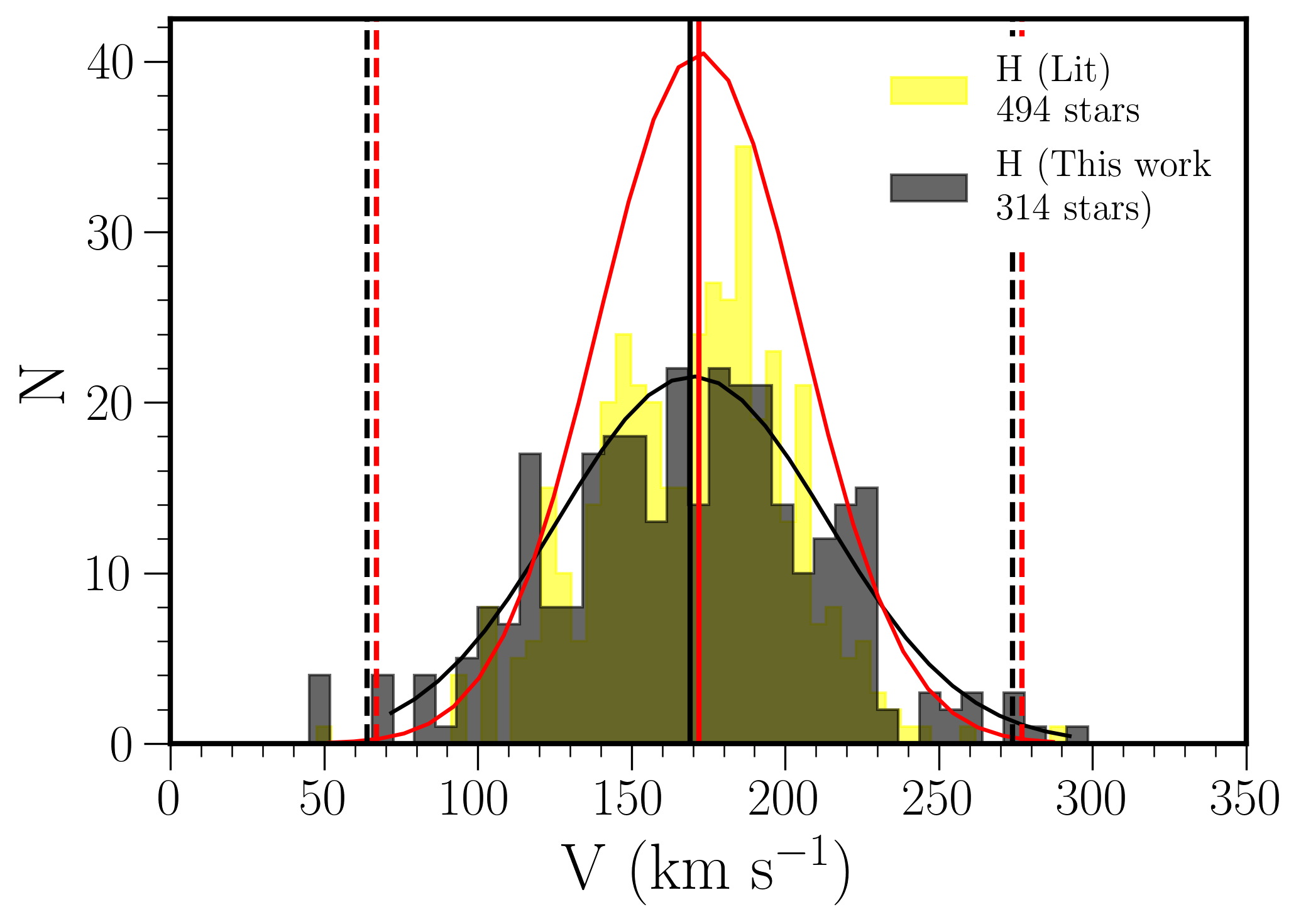}
    \includegraphics[scale=0.3]{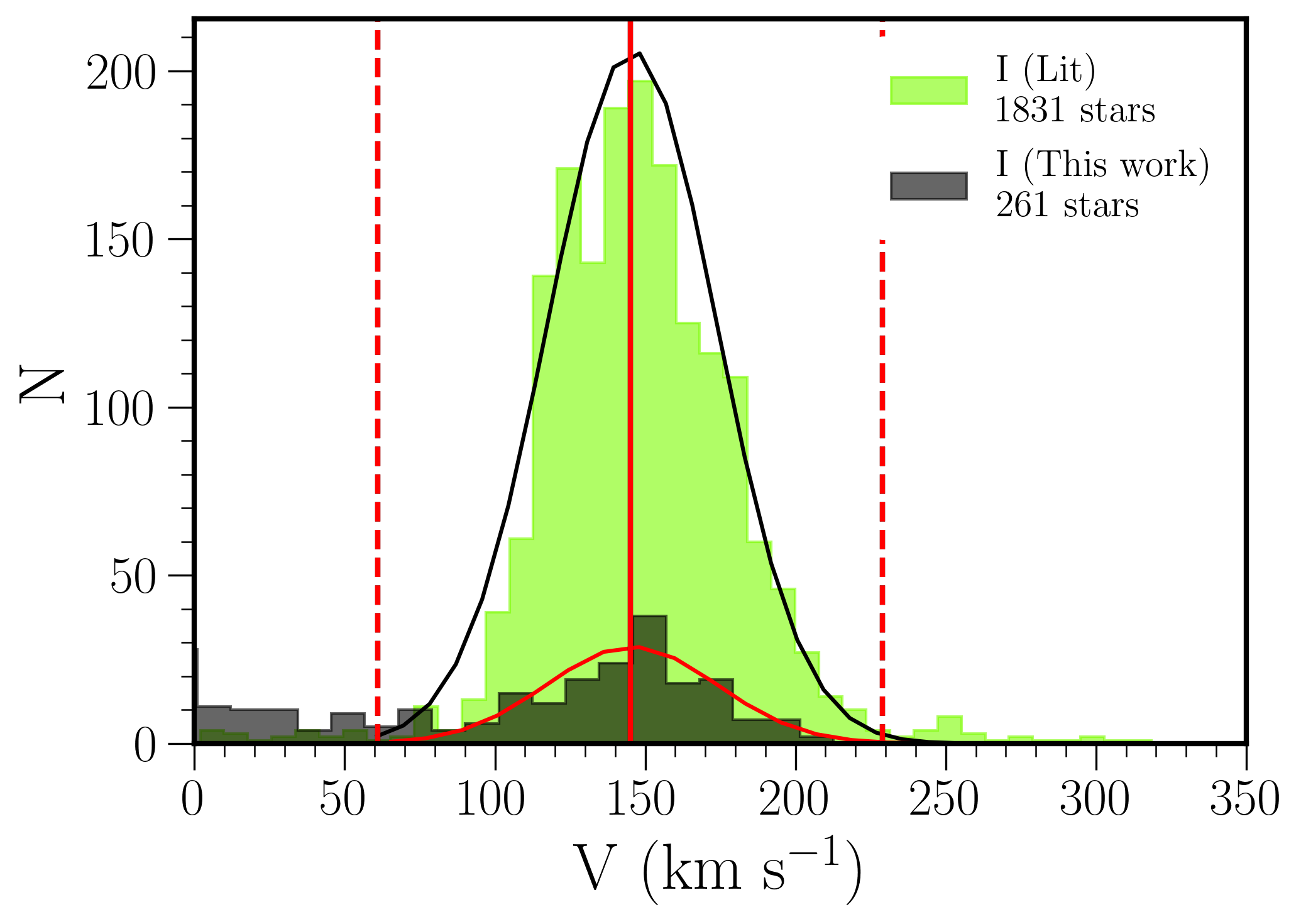}
    \includegraphics[scale=0.3]{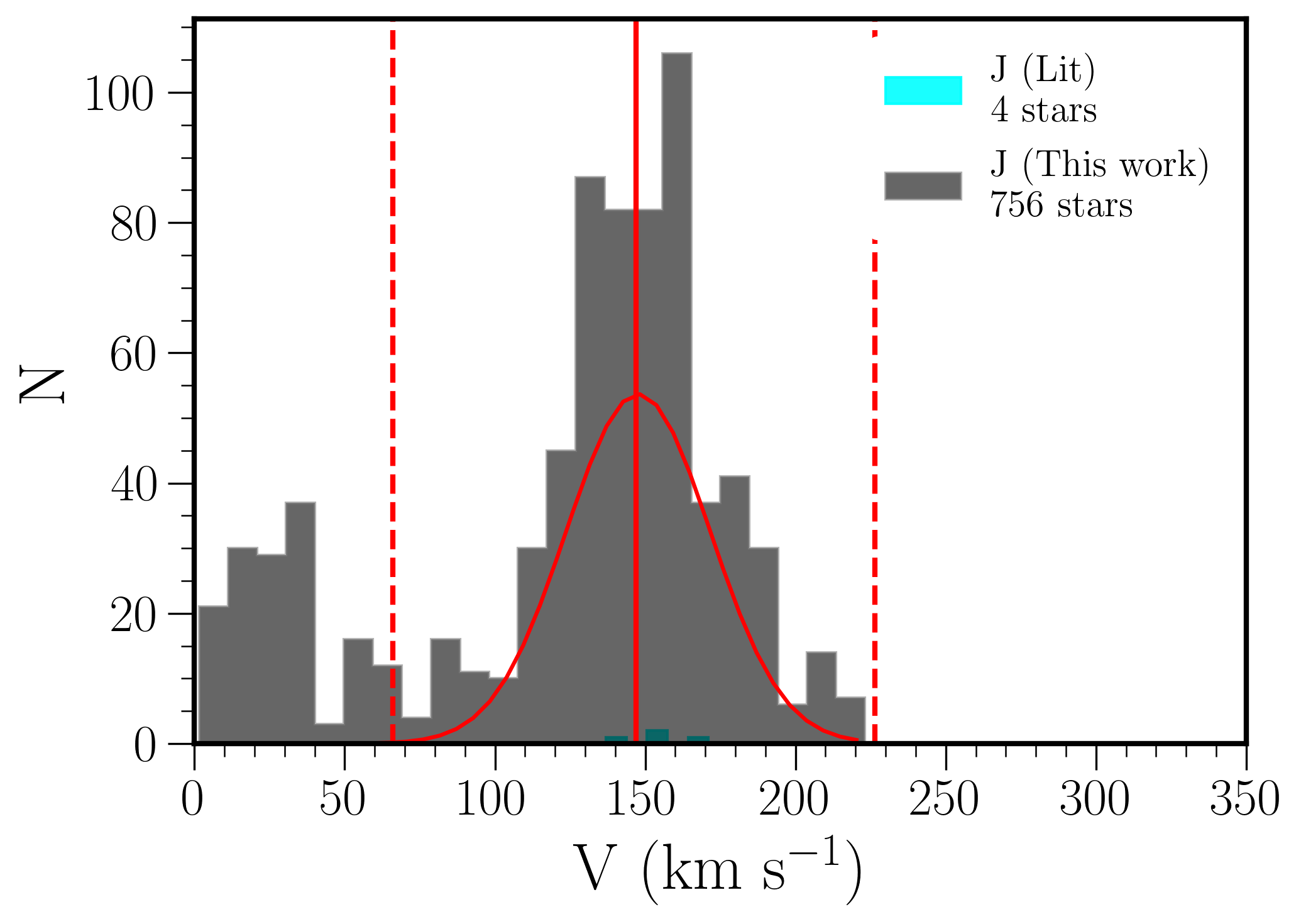}
    \includegraphics[scale=0.3]{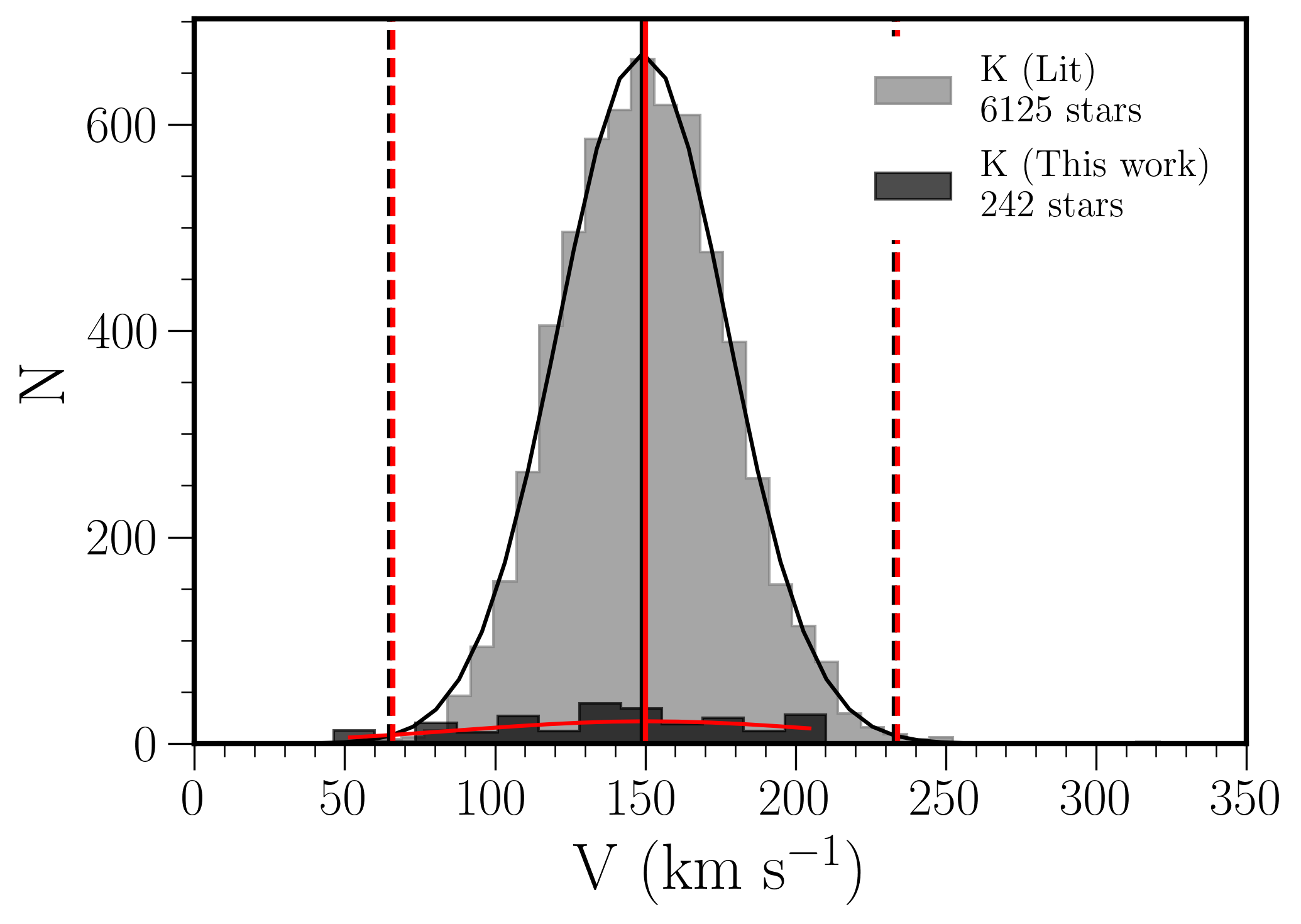}
    \includegraphics[scale=0.3]{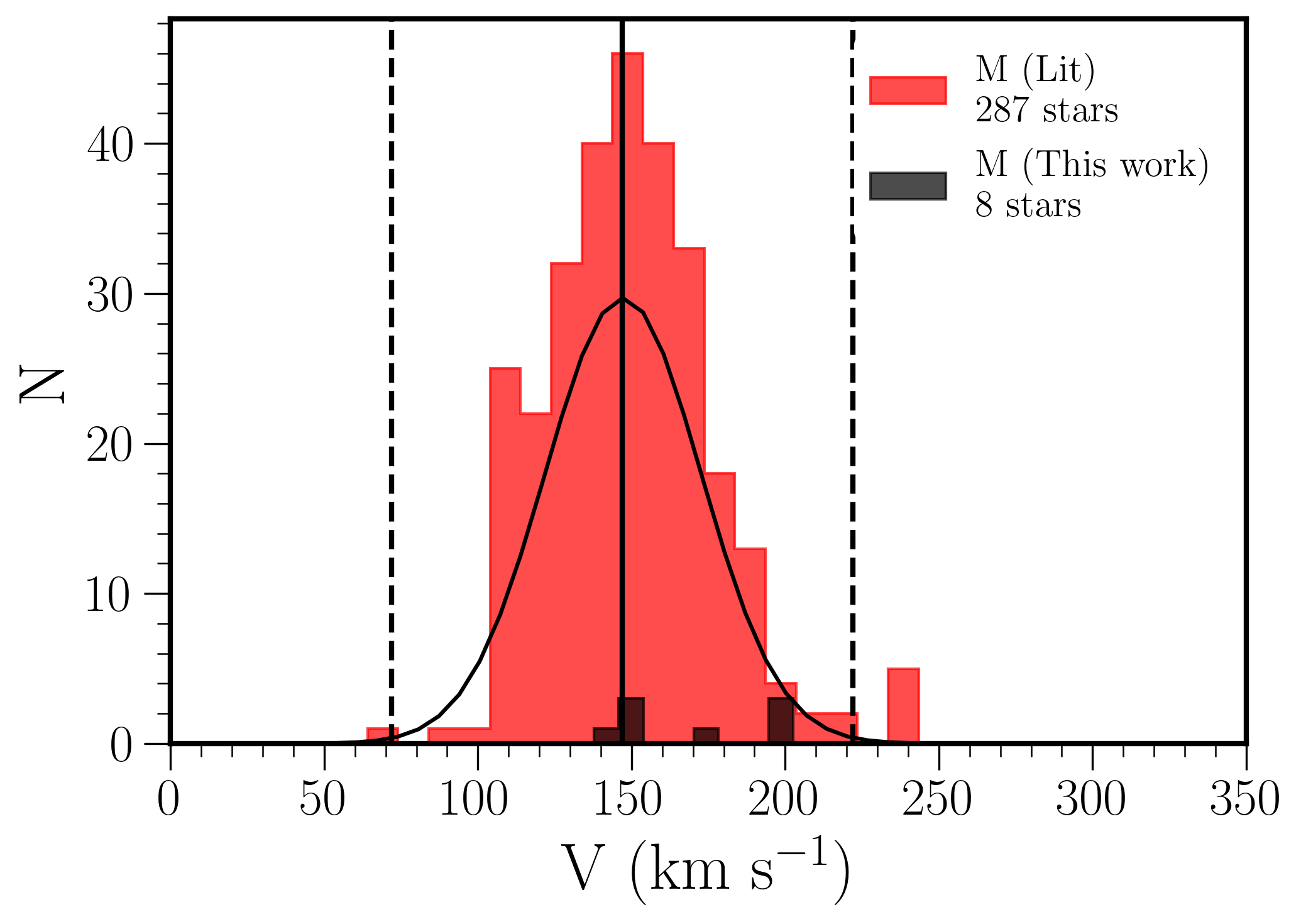}
    
    \caption{Distributions of RVs in the different CMD regions outlined in Figure \ref{fig:CMD_SD_ESO_Archive} and adapted from \protect \cite{ElYoussoufi2019}, for the literature sample (coloured histograms) and our sample (dark grey histograms). The distribution of SMC stars within each panel is fitted with a Gaussian. The vertical axes represent the number counts within bins of 12 km\,s$^{-1}$ in size. Regions C and D are not included due to the low number of sources.}

	\label{fig:HIst_all_rv}
\end{figure*}

\begin{figure}
	\centering
	\includegraphics[scale=0.65]{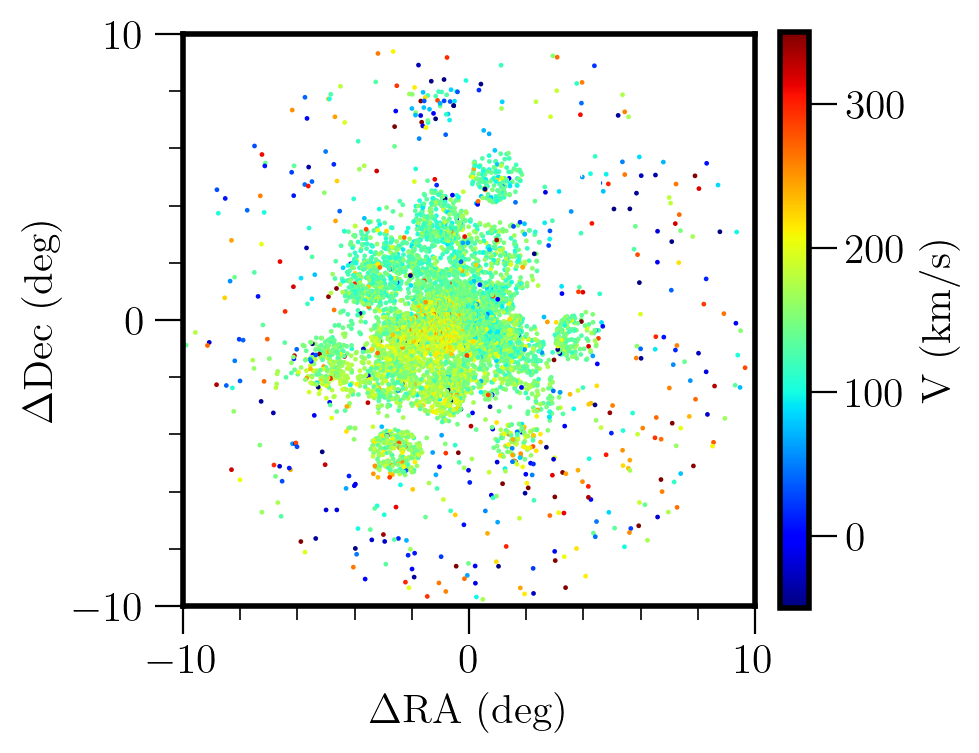}
	\includegraphics[scale=0.65]{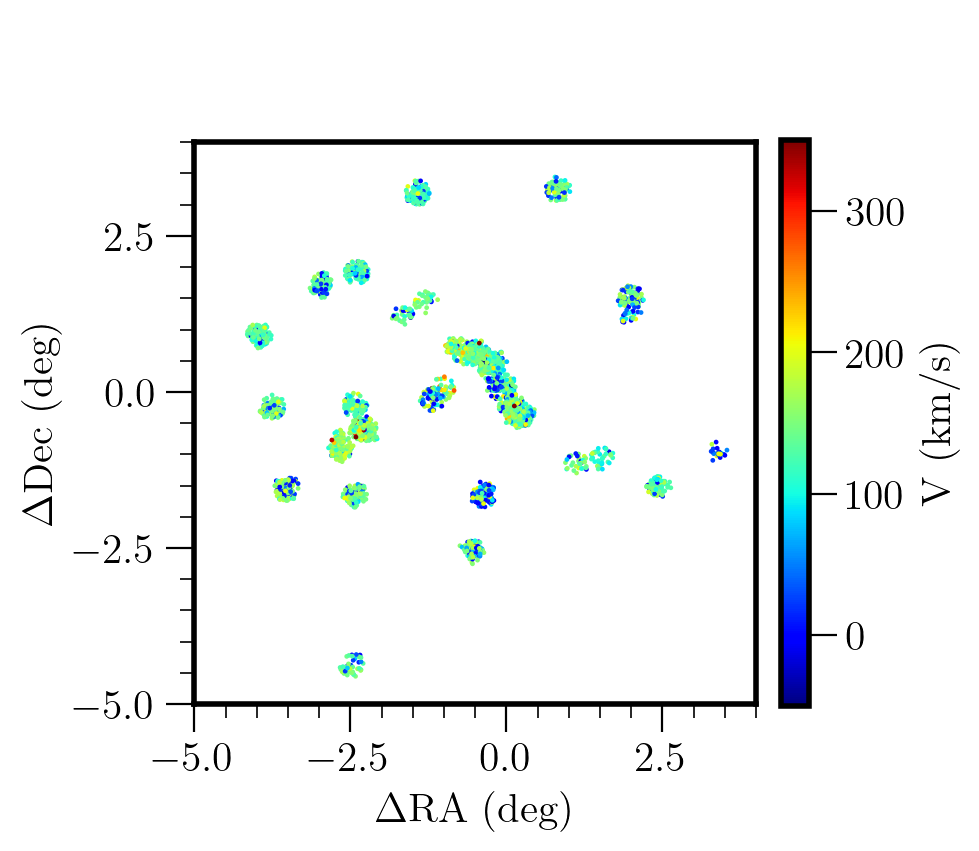}
	\caption{Distribution of RVs obtained from the literature sample (top) and our sample (bottom).}
	\label{fig:maps_liteso_rv}
\end{figure}

\begin{figure}
    \centering  
    \includegraphics[scale=0.37]{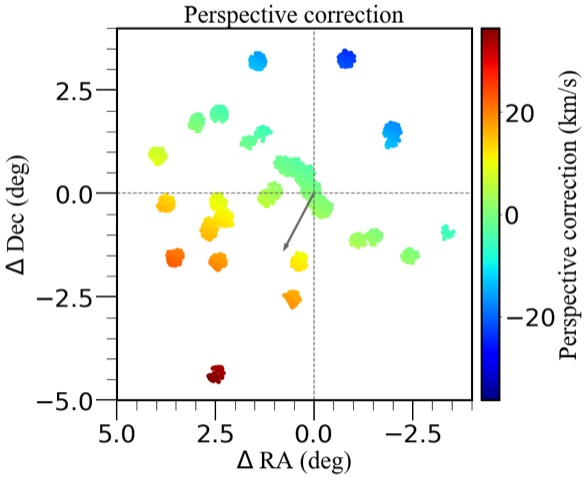}
    \caption{Distribution of the perspective correction derived from the projected bulk proper motion of the galaxy, which is indicated with an arrow departing from the centre.}
    \label{fig:RVcorr}
\end{figure}

\begin{figure*}
\centering
        \includegraphics[scale=0.05]{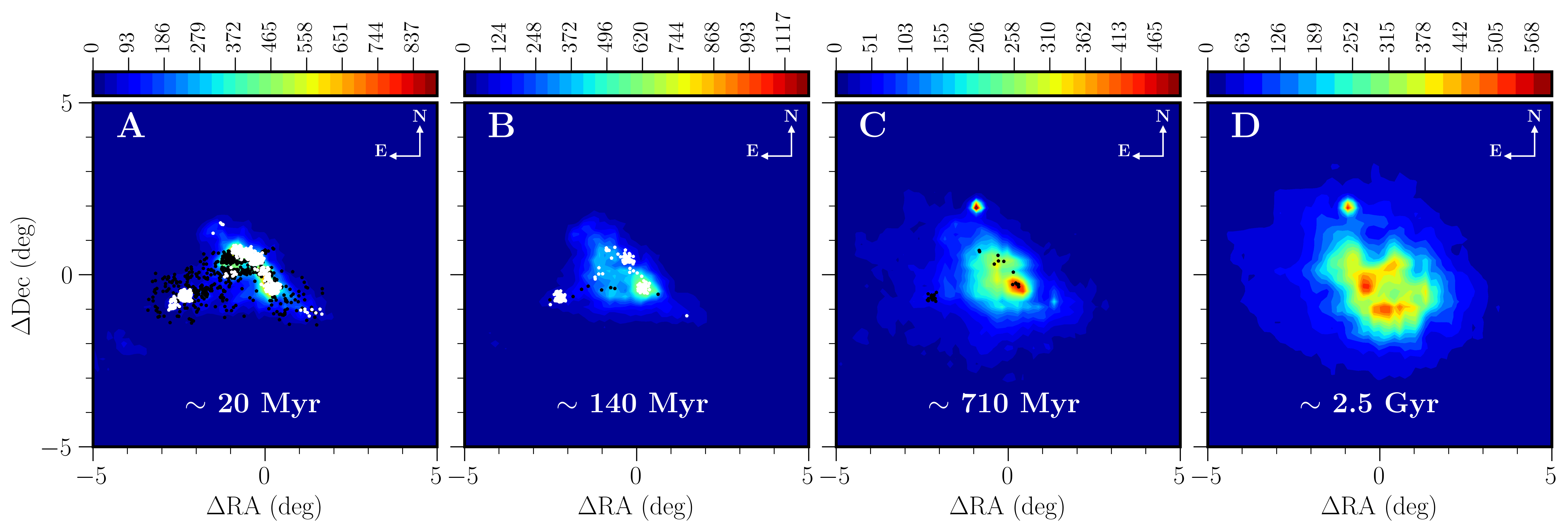}
        \includegraphics[scale=0.05]{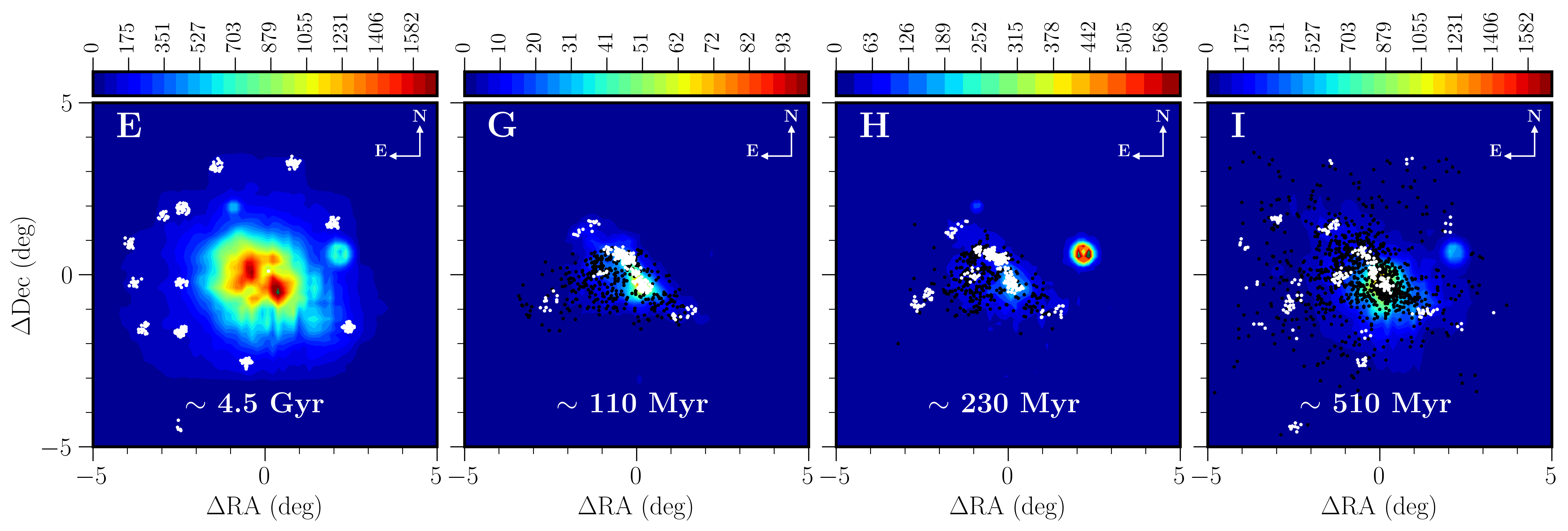}
        \includegraphics[scale=0.05]{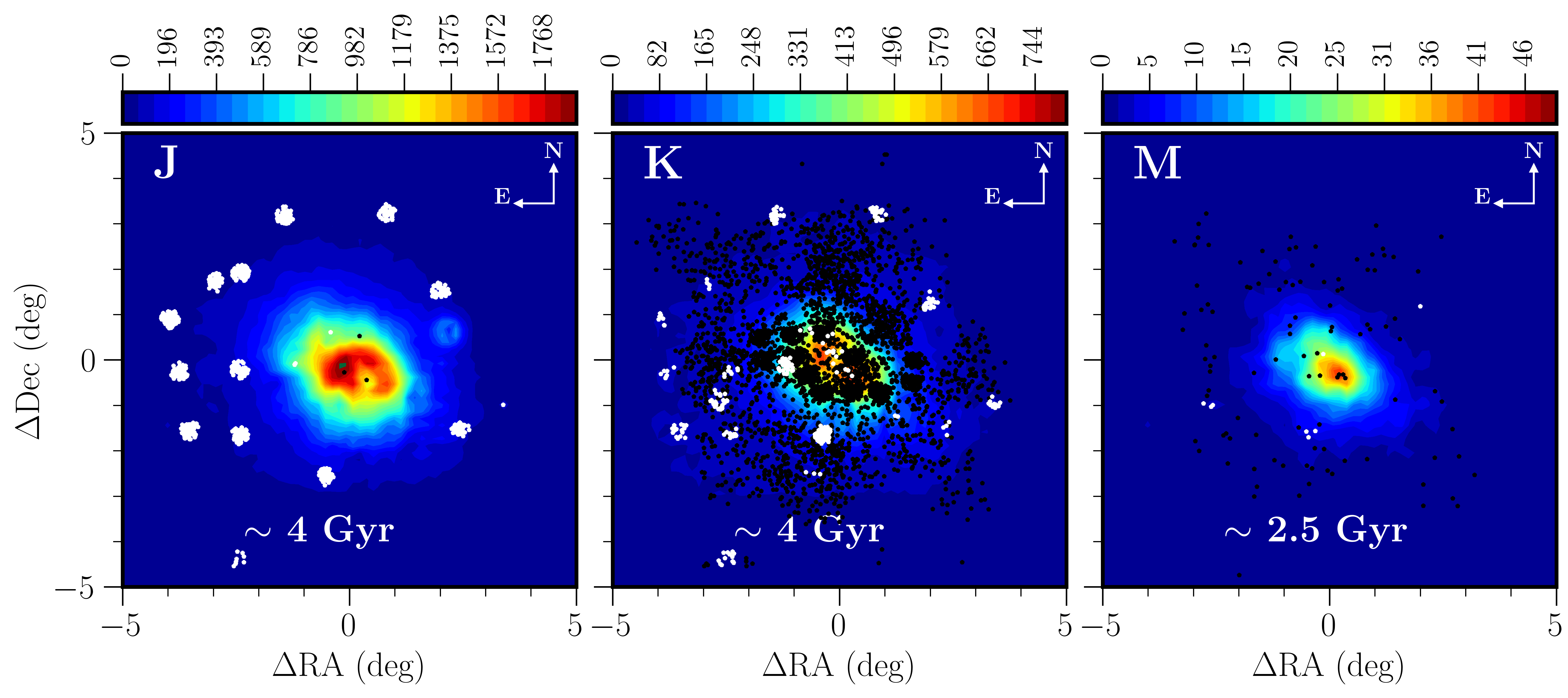}
        \caption{Distribution of RVs from our sample (white points) and the literature sample (black points) superimposed onto morphological maps of stellar populations in the SMC from \protect\cite{ElYoussoufi2019}. The bin size is 0.03 deg$^2$ and the colour bar represents the number of stars per bin. Regions A, B and C refer to main-sequence stars, D to main-sequence/subgiant stars, E to faint RGB stars, G, H and I to supergiant and giant stars, J to RC stars, K to bright RGB stars, and M to thermally pulsing AGB stars. Median ages for each stellar population, as derived by \protect\cite{ElYoussoufi2019}, are indicated in the panels.}
        \label{fig:morphology_distribution}
\end{figure*}

Figure \ref{fig:HIst_all_rv} compares the RV distributions of different CMD regions \citep{ElYoussoufi2019} between our sample  and the literature sample. The majority of the stars in our sample populate region A (main-sequence stars) and J (RC stars). There are also many stars (about 200 or more) in regions B (main-sequence stars), E (faint RGB stars), G, H and I (supergiant stars), and K (bright RGB stars). Very few stars are found in region M (AGB stars). In particular, our sample provides RVs for stars in regions B, E and J which are poorly sampled in the literature studies. It is noticeable that regions E, I and J are those mostly contaminated by MW stars which have small RVs. In the following, we consider stars with $60<V<258$ km\,s$^{-1}$ to belong to the SMC and we compute mean RVs within each region by fitting a Gaussian to their RV distributions using a non-linear least-squares minimisation technique. The number of sources in both our sample and the literature sample, the respective values of RVs and dispersions, as well as the corresponding median ages and age intervals of the stars within each CMD region are reported in Table \ref{table:rv_cmd}. This shows that there are two groups differing by ~20 km s$^{-1}$: one formed by stars in regions A, B, G, H, and M with a RV of $\sim$170 km s$^{-1}$ and another one formed by stars in regions E, I, J, and K with a RV of $\sim$150 km s$^{-1}$. The velocity dispersion of each region is comparable and corresponds to $\sim$30 km s$^{-1}$. The same dichotomy is found for stars in the literature sample with the exception of those in region M which have instead a mean RV of $\sim$150 km s$^{-1}$. However, in our sample regions B, J and M contain only a few stars and it may not be appropriate to compare their mean values with those from regions that are instead well populated. Figure \ref{fig:maps_liteso_rv} shows the spatial distribution of the RVs across the SMC for stars in the two samples whereas Figure \ref{fig:morphology_distribution} shows similar distributions but for each CMD region. The literature sample, which encompasses a rather homogeneous wide area of the galaxy, shows  larger RVs to the SE of the SMC centre and lower RVs in the other area along and west of the bar. A large RV to the SE is also present in our sample, but it is more difficult to discern an overall RV trend because our sample is concentrated on the bar with only sparse fields located around it. We computed the perspective correction to the RVs from the projected bulk proper motion of the galaxy (Fig. \ref{fig:RVcorr}). We used the coordinates for the centre ($\alpha=13.05$ deg, $\delta=-72.03$ deg) as given in Sect. \ref{section2}, the proper motions ($\mu_\alpha$cos$\delta$=0.686 mas yr$^{-1}$, $\mu_\delta$=$-$1.237 mas yr$^{-1}$) from \cite{Luri2021} and the distance modulus (m-M)=18.88 mag from \cite{Muraveva2018}. Figure \ref{fig:RVcorr} shows that the correction on the eastern side of the galaxy is larger than on its western side suggestive of a rotation pattern, but it is constant along the bar region.


\begin{table*}
	\caption{Mean radial velocity for each CMD region.}                       
	\label{table:rv_cmd}      
	
	\begin{tabular}{crccrccrr}
	    \hline
	    CMD Region & \multicolumn{3}{c}{ESO SAF sample} & \multicolumn{3}{c}{Literature sample} & Median age$^\mathrm{a}$ & Age range$^\mathrm{a}$\\
	     & N & $\langle V\rangle$ & $\sigma_{V}$ & N & $\langle V\rangle$ & $\sigma_{V}$ & & \\
	     & & km s$^{-1}$ & km s$^{-1}$ & & km s$^{-1}$ & km s$^{-1}$ & Myr & Myr\\
		\hline
		A & 925 & 171 & 28 & 772 & 168 & 30 & 20 & 10--50 \\
		B & 188 & 174 & 30 & 9 & 201 & 38 & 150 & 50--410 \\
		C & 28 & 32 & 8 & -- & -- & -- & 700 & 250--2000 \\
		D & -- & -- & -- & -- & -- & -- & 2500 & 1500--5000 \\
		E & 354 & 150 & 30 & -- & -- & -- & 4500 & 2500--8000 \\
		G & 266 & 173 & 28 & 429 & 179 & 31 & 110 & 50--230 \\ 
  	    H & 314 & 169 & 36 & 494 & 172 & 33 & 230 & 150--360 \\
		I & 261 & 145 & 28 & 1831 & 146 & 28 & 510 & 90--3000 \\
		J & 756 & 148 & 27 & 4 & 152 & 12 & 4000 & 2000--9000 \\
		K & 242 & 151 & 28 & 6125 & 149 & 28 & 4000 & 2000--8000 \\
		M & 8 & 170 & 25 & 287 & 147 & 25 & 2500 & 1000--5000 \\
		\hline
	\end{tabular} 

 $^\mathrm{a}$ Ages are from \cite{ElYoussoufi2019}.
 
\end{table*}

Regions A and B represent both main-sequence populations and their morphology delineates the asymmetric nature of the galaxy \citep{ElYoussoufi2019}. In these regions, both samples are of a comparable size and populate the bar and the Wing. The literature sample is also present between the two structures, but only for stars 10--50 Myr old, and our sample provides stars 50--410 Myr old \citep{ElYoussoufi2019}.
Regions G and H represent supergiant populations of  30--200 Myr old and 130--340 Myr old, respectively and the two samples show similar distributions to those in regions A and B despite having about twice as many sources in the literature sample than in our sample.
In contrast in region I, which is also populated by supergiant stars with a median age of $\sim$510 Myr \citep{ElYoussoufi2019}, we find that both samples probe the extended body of the galaxy. Region I includes fainter stars than those in regions G and H. The literature sample has also a large concentration of stars in the SW part of the bar compared to the part in the NE.
Regions E and K represent the faint and bright RGB populations, respectively with an approximate range of 2--8 Gyr \citep{ElYoussoufi2019}. Our sample in region E occupies sparse fields in the outskirts of the galaxy with more fields being present in the east than in the west. There are no stars in the literature sample from this region. Our sample in region K  has a similar spatial distribution as that of region E. Region K encompasses the largest number of stars in the literature sample and the widest coverage of the galaxy. The comparison between region K and regions populated by younger stars highlights that the NE of the bar is on average younger than its SW as well as other regions around it. The RVs of the old populations are on average lower than those of the young populations. This is perhaps due to the young stars having formed out of gas already influenced by tidal stripping.
Region J represents the RC population with an age in the range from 2--9 Gyr \citep{ElYoussoufi2019} for which the spatial distribution and average RV from our sample is similar to that of regions E and K; no stars in the literature sample populate this region. Region M represents the AGB population with an approximate age of 1--5 Gyr \citep{ElYoussoufi2019}. The literature sample is sparsely distributed mostly across the outer region of the galaxy. Our sample only has 8 stars in this region. In addition, regions I and M show a small overdensity in the RV histograms around 250 km\,s$^{-1}$. However, these stars are sparsely distributed across the galaxy and do not trace a particular substructure. Due to their low number they could be fluctuations of the histogram distributions, as it appears in other CMD regions, or perhaps members of a stellar stream. 


We conclude that the RVs obtained from our sample and the literature sample for stellar populations within each CMD region are in good agreement within the uncertainties. Existing differences for regions B and M can be attributed to the small size of the literature sample. We chose to proceed continuing to keep both samples separated, in order to avoid introducing systematic effects and uncertainties that may be relevant at the level of fewer sources and their spatial distribution compared to the mean values.

\begin{figure}
	\centering
	\includegraphics[scale=0.055]{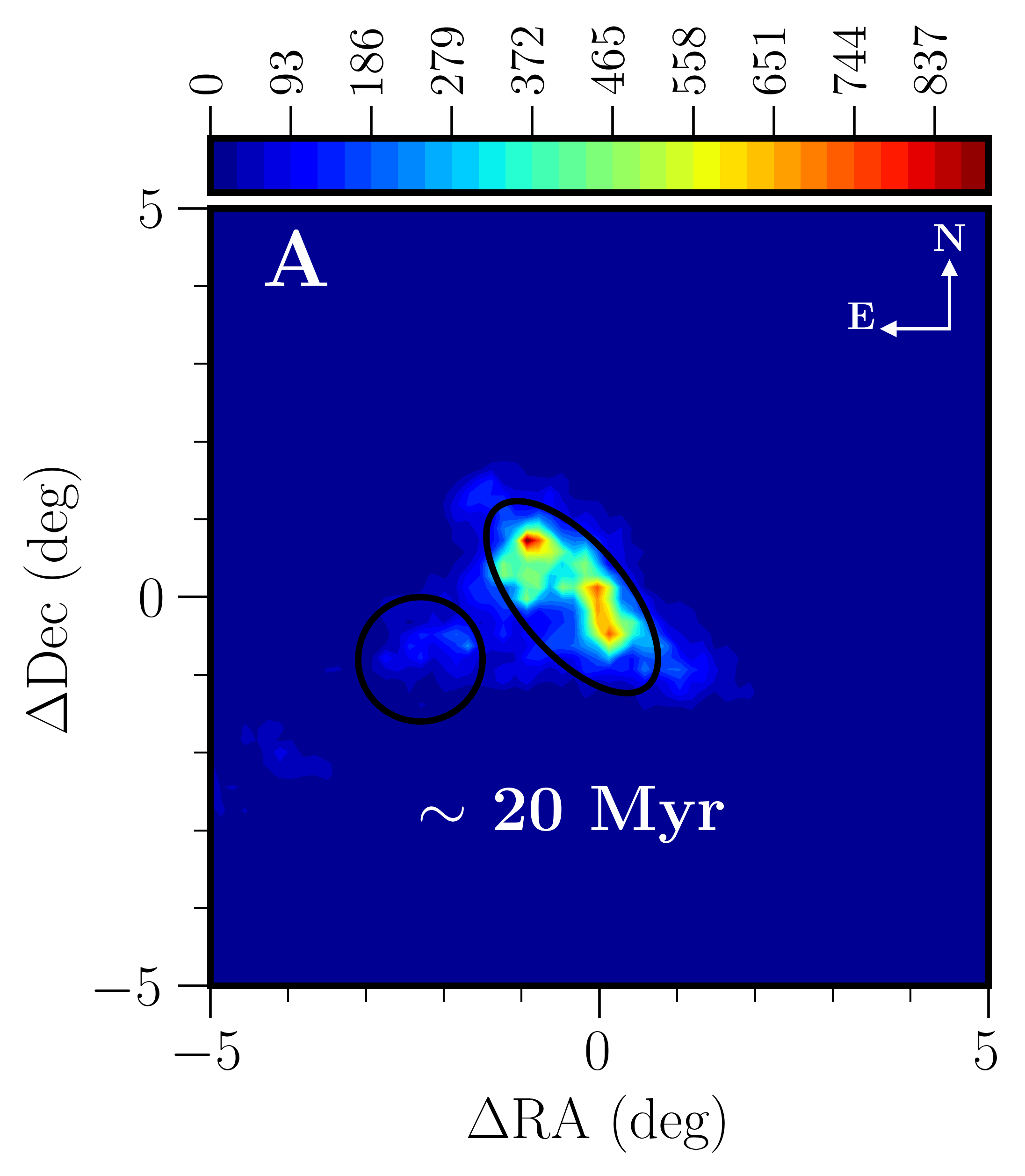}
	\includegraphics[scale=0.055]{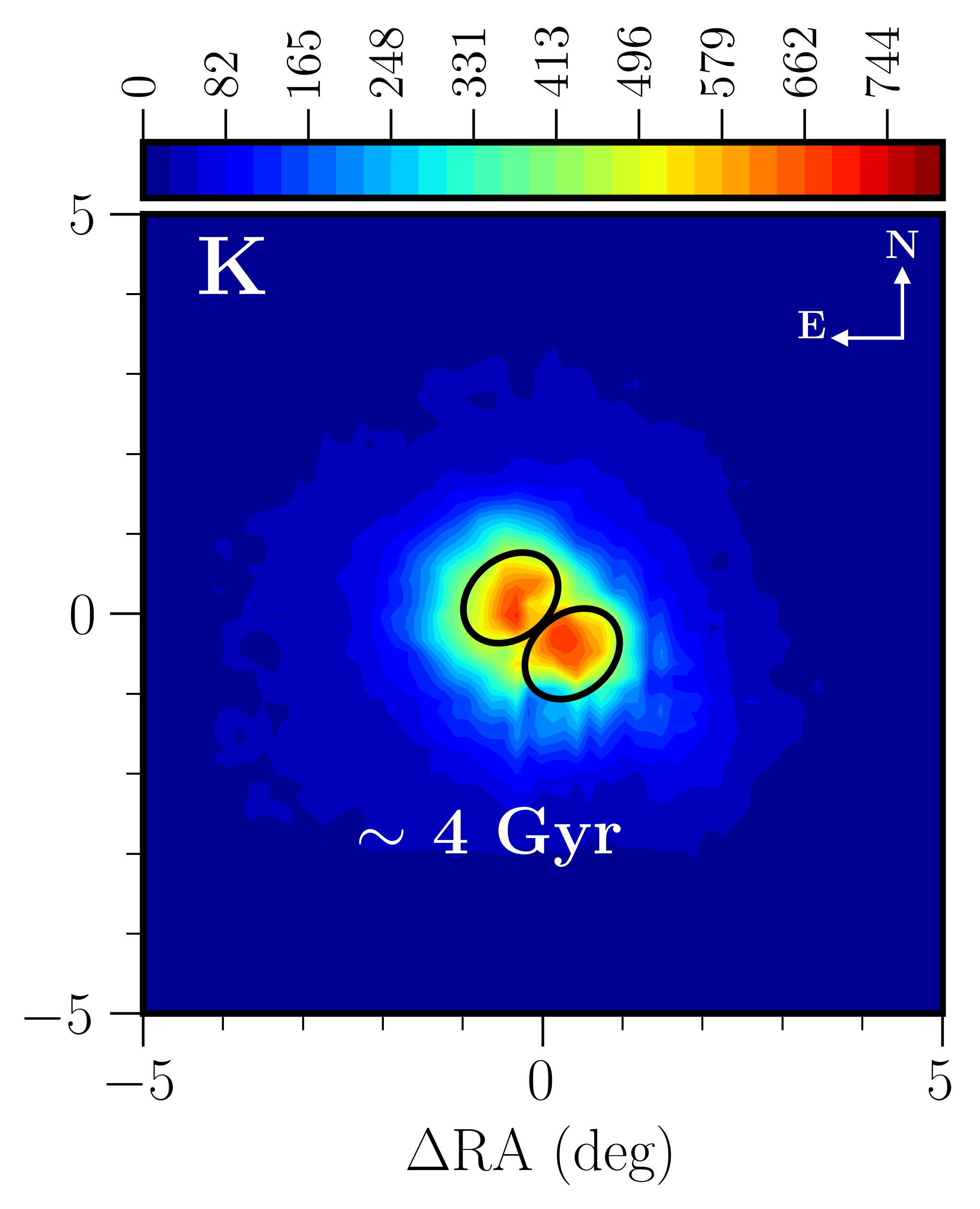}

	\caption{(Left) Morphological map of young main-sequence stars (region A) with  superimposed circular and elliptical masks encompassing the Wing and bar of the SMC, respectively. (Right) Morphological map of bright RGB stars (region K) with elliptical masks encompassing two apparent overdensities. The bin size is 0.03 deg$^2$ and the colour bars represent the number of stars per bin. Median ages are as in Fig. \ref{fig:morphology_distribution}.}
	\label{fig:substructures_mask}
\end{figure}


\subsubsection{Kinematics of the bar and Wing}
The distribution of sources with available RV estimates from the ESO SAF and literature samples encompasses different morphological features of the SMC. In the following, we examine the kinematics of the bar and Wing regions which are prominent features in our sample. To study the kinematics of these two regions, we define two masks to enclose them \textbf{--} see Figure \ref{fig:substructures_mask}.

The SMC bar spans $\sim$2.5\,deg in length  \citep{ElYoussoufi2019} and is also known to be elongated across the line of sight reaching up to 23 kpc \citep{Gardiner1995,Haschke2012,Subramanian2012,Nidever2013,Jacyszyn-Dobrzeniecka2017,Muraveva2018}. The elliptical mask defined to enclose the bar is centred at ($\Delta$RA=$-$0.35\,deg, $\Delta$Dec=0 \,deg), has a semi-minor axis of 3\,deg, a semi-major axis of 1.4\,deg, and a PA of 40\,deg in the NE direction.
We find that in the bar region, the mean RV estimates decrease by about 20~km\,s$^{-1}$ with increasing stellar population age, from young main-sequence stars ($\sim$170 km\,s$^{-1}$) to old supergiant stars ($\sim$150~km\,s$^{-1}$) corresponding to an age difference of about 100 Myr. Differences in kinematics between the NE ($\Delta$Dec>0 deg) and SW ($\Delta$Dec<0 deg) bar correspond to 10 km\,s$^{-1}$ in the youngest main-sequence stars, with the NE bar having the highest velocity (177 km\,s$^{-1}$). In supergiants, there is a similar velocity difference (8 km\,s$^{-1}$), but only for stars in region H (yellow supergiants) and it is the SW bar that instead has the highest velocity (172 km\,s$^{-1}$). Uncertainties on the RV measurements are $\sim$1 km\,s$^{-1}$.

At intermediate-ages, traced by bright RGB stars in region K, the inner SMC is also dominated by two overdensities which are located along the bar, but that differ in shape compared to those traced by younger stars (Fig.~\ref{fig:substructures_mask}). Data for these substructures are only available in the literature sample. We calculate the mean RV of these overdensities using two masks defined as follows. The elliptical mask defined to encompass the overdensity in the NE is centred at  ($\Delta$RA=$-$0.4\,deg, $\Delta$Dec=0.2\,deg), has a semi-minor axis of 1\,deg, a semi-major axis of 1.3\,deg, and a PA of 40\,deg whereas the elliptical mask defined to encompass the overdensity in the SW is centred at ($\Delta$RA=0.37\,deg, $\Delta$Dec=$-$0.50\,deg), has a semi-minor axis of 1.0\,deg, a semi-major axis of 1.3\,deg, and a PA of 40\,deg. Using only RVs from bright RGB stars we obtain that the SW overdensity has a mean RV of 152 $\pm$ 1 km s$^{-1}$ whereas the NE overdensity has a mean RV of 144 $\pm$ 1 km s$^{-1}$. These velocities are similar to the velocity of old supergiant stars. 

The Wing is a morphological substructure discovered by \cite{Shapley1940}, located at the SE of the SMC, connecting the galaxy to the Magellanic Bridge. The circular mask defined to cover the Wing is centred at ($\Delta$RA=$-$2.30\,deg, $\Delta$Dec=$-$0.80\,deg) and has a radius of 0.80 deg. We find that both the youngest main-sequence stars and supergiant stars show the same mean RV of $\sim$180 km s$^{-1}$ which is on average larger than the RV derived for stellar populations within the bar, but for young main-sequence stars in the NE. 

\section{Conclusions}\label{section6}

In \cite{ElYoussoufi2019} we provided an updated view of the morphology of the SMC through a comprehensive age tomography. In this study, we complement their work through a homogeneous and extensive kinematic study of the different resolved stellar populations. The SMC is known to have a complicated interaction history, a complex 3D structure and an eventful star formation history; all of these factors can have an impact on the kinematics of stellar populations within the galaxy. To this end we proceed as follows:

(i) We search the ESO SAF for the best available spectra of objects belonging to the SMC, making use of near-infrared photometry from the VMC, VHS, 2MASS and OGLE IV projects. Parallaxes and PMs from {\it Gaia} EDR3 are used to reduce the influence of MW foreground stars.  The selected stars form a sample of 3700 sources. The spectra used in this study are taken with the FLAMES/GIRAFFE instrument, have a SNR $\geq$ 10 and a resolving power from 6500 to 38\,000.  The spectra are sky subtracted and a full spectrum fitting method is used to obtain RVs; 
the stellar spectra and their associated uncertainties were re-sampled onto the arbitrary wavelength grid of the templates which were taken from the X-shooter spectral library. Systematic uncertainties among the FLAMES gratings were found to be $\sim$ 2 km\,s$^{-1}$, but were not applied  to correct the spectra due to not having enough combinations of gratings to homogenise the sample. 

(ii) We obtain a RV distribution which shows two distinct peaks: a small peak around 18 $\pm$ 2 km\,s$^{-1}$ representing the MW foreground stars in the direction of the SMC and a large peak around 159 $\pm$ 2 km\,s$^{-1}$ representing stars belonging to the SMC and corresponding to a velocity dispersion ($\sigma_V$ ) of 33 $\pm$ 2 km\,s$^{-1}$. Furthermore, we compare these values to those obtained from a homogenised RV sample of several spectroscopic studies that have previously observed SMC stars and that we anchor to the APOGEE DR17 spectra. 

(iii) We find that stars occupying regions of the CMD, as defined by  \cite{ElYoussoufi2019}, that have been poorly sampled by spectroscopic studies in the literature such as regions B (main-sequence stars 50--410 Myr old),  E (faint RGB stars 2.5--8 Gyr old) and J (RC stars 2--9 Gyr old) have RVs resembling those of stellar populations of similar median ages: regions G (supergiant stars 50--230 Myr old) and K (bright RGB stars 2--8 Gyr old). On average, the RV of stellar populations younger than 500 Myr is $\sim$ 20 km s$^{-1}$ larger than that of stellar populations older than 2 Gyr. A similar RV dichotomy has recently been found by \cite{Mucciarelli2023} from an independent analysis of 206 RGB stars, which are also included in our sample, located on the eastern side of the galaxy around the globular cluster NGC 419. The authors characterised a metal-rich component with a large RV ([Fe/H]$\sim$-0.9 dex and RV$\sim$172 km s$^{-1}$) and a metal-poor component with a low RV ([Fe/H]$\sim$-1.1 dex and RV$\sim$154 km s$^{-1}$), suggesting that they could result from separate bursts of star formation \citep{Massana2020}.

(iv) We measure a RV difference between the Wing and bar, dominating the morphology of the SMC as traced by our sample, confirming that these two substructures are kinematically distinct. This RV difference was already noticed by \cite{Evans2008} from their analysis of the kinematics of OBA type stars. PM studies corroborate this result, with the Wing kinematics following the transverse motion along the Magellanic Bridge towards the LMC (e.g., \citealp{Zivick2019,Schmidt2020,Niederhofer2021}). Consequently, the increased star formation rate in the Wing region following the interaction between the LMC and the SMC about 200 Myr ago \citep{Harris2004,Rubele2015} imprinted a large RV on the newly formed stars compared to those in the bar. The Wing substructure is found to lie at a distance closer to us than the main body of the SMC, further contributing to these discrepancies \citep{Tatton2021}. The differences in kinematics between the bar and Wing are more pronounced in supergiants (150--360 Myr old; $\sim$ 30 km s$^{-1}$) than in young main-sequence stars (10--50 Myr old; $\sim$10 km s$^{-1}$) where the RV of stars in the bar is smaller in the former than in the latter population.
Moreover, the RVs of stars in the bar are found to be larger in young main-sequence stars than in red supergiants; we attribute this to the peak of star formation in the SMC bar at $\sim$ 40 Myr \citep{Rubele2015} from gas already influenced by tidal stripping. 
This is consistent with a cold gas outflow, which is stronger in the Northern bar, that originated 33--56 Myr ago \citep{McClure-Griffiths2018}. SMC debris with large RVs has been found at the extreme southern disc of the galaxy (6--9 deg from the centre beyond the Wing) using data from the Magellanic Edges Survey \cite{Cullinane2023}.

(v) We measure different RVs between the NE and SW portions of the bar. These differences are \textbf{$\sim$} 10 km\,s$^{-1}$ and are found to be present in young populations such as main-sequence stars (10--50 Myr old) and yellow supergiants (150--360 Myr old) as well as in old populations like bright RGB stars (2--8 Gyr old). This also shows consistency between our (main-sequence and supergiant stars) and literature (bright RGB stars) studies. The NE bar region has a larger RV in young main-sequence stars than the SW region; the reverse is true for the supergiants and RGB populations. These gradients are not influenced by the projected bulk proper motion. However, we confirm a previously found rotation pattern (\citealp{Dobbie2014, DiTeodoro2019, Deleo2020, Abdurrouf2022}). A prominent episode of star formation $\sim$25 Myr ago is evident at the NE extremity of the bar according to \cite{Rubele2015,Rubele2018} which coincides with the NE bar structure. In this region there is also a predominance of young (<140 Myr old) Cepheids in front of the main body of the SMC \citep{Ripepi2017}. The larger RV of stars in this region corroborates their formation subsequent to the gas stripping resulting from the dynamical interaction with the LMC. The largest outflow gas velocities are also found in this region \citep{McClure-Griffiths2018}. Furthermore, numerical simulations including interstellar gas dynamics and star formation processes showed that the observed structural, kinematic and stellar properties of the SMC are predominantly of tidal origin \citep{Yoshizawa2003}.


(vi) \cite{Soszynski2010} and \cite{Haschke2012} noticed that RR Lyrae stars, which are distributed fairly homogeneously over the whole body of the SMC, trace two overdensities at the centre similar to those we detect from bright and faint RGB stars.  \cite{Jacyszyn-Dobrzeniecka2017} found similar features in the on-sky projection of the OGLE-IV data. However the two overdensities are not present in the 3D Cartesian density maps, deducing that they might be due to a projection effect. The RR Lyrae distribution does not show any substructure or asymmetry along the line-of-sight. Furthermore, no discrepancies in the line-of-sight depth are found between the locations of the two overdensities from the analysis of RC stars \citep{Tatton2021}. In the plane of the sky, several studies based on PMs \citep{Deleo2020,Niederhofer2018,Zivick2019} found that stars follow a non-uniform velocity structure at the location of the SW-bar overdensity which might indicate a stretching or tidal stripping of the SMC.\\


In our future work, we plan to carry out a similar investigation of the stellar populations of the LMC. The upcoming large-scale spectroscopic facilities and their planned survey projects: the One Thousand and One Magellanic Fields survey (1001MC; \citealp{cioni2019}) using the  4-metre Multi-Object Spectroscopic Telescope (4MOST; \citealp{dejong2019}) facility at VISTA and the Survey of the Milky Way and its Satellites \citep{gonzalez2020} using the Multi-Object Optical and Near-infrared spectrograph (MOONS; \citealp{cirasuolo2020}) facility at the VLT will significantly enlarge spectroscopic stellar sample sizes, cover large areas encompassing also the halo of the Magellanic Clouds, and increase the sensitivity to enable us to investigate substructures with kinematics, metallicity and chemistry in great detail.

\section*{Acknowledgements} 
This project received funding from the European Research Council (ERC), under the European Union’s Horizon 2020 research and innovation programme (grant agreement no. 682115), and was supported in part by the Australian Research Council Centre of Excellence for All Sky Astrophysics in 3 Dimensions (ASTRO 3D), through project number CE170100013. We thank the Cambridge Astronomy Survey Unit (CASU) and the Wide Field Astronomy Unit (WFAU) in Edinburgh for providing the necessary data products under the support of the Science and Technology Facility Council (STFC) in the UK. This study is based on observations obtained with VISTA at the Paranal Observatory under programmes 179.B-2003 and 179.A-2010. 
This work has made use of data products from the Two Micron All Sky Survey, which is a joint project of the University of Massachusetts and the Infrared Processing and Analysis Center/California Institute of Technology, funded by the National Aeronautics and Space Administration and the National Science Foundation. This work has made use of data from the European Space Agency (ESA) mission \textit{Gaia} (\url{http://www.cosmos.esa.int/gaia}), processed by the \textit{Gaia} Data Processing and Analysis Consortium (DPAC; \url{ http://www.cosmos.esa.int/web/gaia/dpac/consortium}). Funding for the DPAC has been provided by national institutions, in particular the institutions participating in the \textit{Gaia} Multilateral Agreement.
This project has made extensive use of the Tool for OPerations on Catalogues And Tables (TOPCAT) software package \citep{Taylor2005} as well as the following open-source Python packages: matplotlib \citep{Hunter2007}, NumPy \citep{Oliphant2015}, pandas \citep{McKinney2010}, SciPy \citep{Jones2001}.
We thank Marica Valentini and Salvatore Taibi for the fruitful discussions concerning the sky subtraction.

\section*{Data Availability}
The data underlying this study are available in the ESO Archive (http://archive.eso.org). This concerns both catalogues to select the stellar sample and spectra to obtain the radial velocities. The Gaia data to establish the memberships to the SMC are also publicly available. All other data resulting from our study are made available with the publication.


\bibliographystyle{mnras} 
\bibliography{ref-Kinematics}

\appendix

\section{Examples of spectral fits}

\begin{figure}
    \centering
    \includegraphics[scale=0.5]{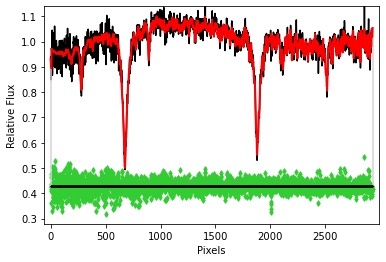}
    \includegraphics[scale=0.5]{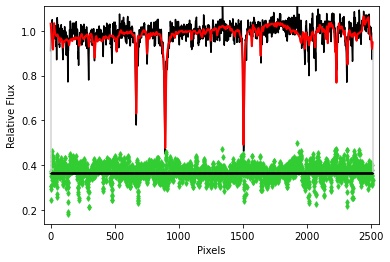}
    \caption{Spectra, spectral fits and residuals for sources ADP.2019-02-01T01\_00\_32.982 (top) and ADP.2019-10-23T14\_43\_58.789 (bottom).}
    \label{fig:spectra}
\end{figure}

Figure \ref{fig:spectra} shows the spectra of two  randomly-selected sources, their spectral fit obtained using pPXF, and the residuals. Each fit is the combination of a number of templates at the same radial velocity of which the first five, with their associated parameters, are listed in Table \ref{tab:spectra} in order of weight. The sum of the weights of the templates participating to a fit is about 1.0, which means that the individual weights can be interpreted as percentage contributions to the spectrum. However, the range of stellar parameters should not be over-interpreted because first, they are not set on a regular grid and second because we have not applied regularisation of the template weights, which is necessary to attach a  physical interpretation to the results. Nevertheless, the 1.2 arcsec aperture of the MEDUSA fibres would, in dense stellar regions like the Magellanic Clouds, collect the flux from more than just one source. This might be captured by the pPXF method similarly to what happens in the observation of nearby galaxies with integral field spectrographs for which the pPXF method was originally designed.

\begin{figure}
    \centering
    \includegraphics[scale=3]{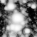}
    \includegraphics[scale=3]{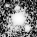}
    \caption{Post-stamp 0.2$\times$0.2 arcmin$^2$ $K_\mathrm{s}$-band images of sources ADP.2019-02-01T01\_00\_32.982 (left) and ADP.2019-10-23T14\_43\_58.789 (right).}
    \label{fig:stamps}
\end{figure}

Source ADP.2019-02-01T01\_00\_32.982 is represented by a hot stellar spectrum. It is located in CMD region B and it is most likely a main-sequence star. The resulting RV is of 112.9 km s$^{-1}$. Source ADP.2019-10-23T14\_43\_58.789 is represented by a rather heterogeneous set of template spectra. It is located in CMD region E and it is most likely a red giant branch star. The resulting RV is of 146.6 km s$^{-1}$. Figure \ref{fig:stamps} shows that both sources have neighbouring sources and/or nebulous emission suggesting a possible contamination of their spectra.

\begin{table}
    \caption{{Spectral templates and their parameters contributing to the pPXF-based fits of two sources.}
    \label{tab:spectra}}
    \centering
    \begin{tabular}{lrccc}
        \hline
        Template & T$_\mathrm{eff}$ & log(g) & [Fe/H] & weight \\
         & K & cm s$^{-2}$ & dex & \\
         \hline
         \multicolumn{5}{l}{Source\,ADP.2019-02-01T01\_00\_32.982:}\\
         HD\,38856\,(X0817) & 15659 & 4.18 & -0.04 & 0.23 \\
         HD\,149382\,(X0692) & 34165 & 5.54 & -1.16 & 0.20 \\
         HD\,176301\,(X0676) & 14552 & 3.77 & 0.26 & 0.11 \\
         HD\,170783\,(X0470) & 15500 & 3.63 & 0.29 & 0.10 \\
         HD\,43286\,(X0539) & 16006 & 4.16 & -0.11 & 0.06 \\
          & & & &  \\
         \multicolumn{5}{l}{Source\,ADP.2019-10-23T14\_43\_58.789:}\\
         HE\,1201-1512\,(X0622) & 5729 & 3.59 & -2.61 & 0.40 \\
         HD\,4359\,(X0448) & 17340 & 4.86 & -0.04 & 0.39 \\
         NGC\,1904\,223\,(X0523) & 4226 & 0.69 & -1.42 & 0.25 \\
         HE\,1207-3108\,(X0630) & 5435 & 3.24 & -2.65 & 0.23 \\
         HD\,4539\,(X0463) & 17303 & 4.84 & -0.04 & 0.18 \\
         \hline
    \end{tabular}
\end{table}

\end{document}